\documentclass[10pt,a4paper]{article}
\pdfoutput=1
\usepackage{jheppub}

\usepackage{bbding}
\usepackage{pifont}
\usepackage{wasysym}
\usepackage{wrapfig}
\usepackage{amsmath}
\usepackage{verbatim}
\usepackage{amssymb}
\usepackage{inputenc}
\usepackage{textcomp}
\usepackage{appendix}
\usepackage{mathrsfs}
\usepackage{float}
\restylefloat{table}
\newcommand{\be}[1]{\begin{equation}\label{#1} }
\newcommand{\ee}{\end{equation}}
\newcommand{\bea}[1]{\begin{eqnarray}\label{#1} }
\newcommand{\eea}{\end{eqnarray}}
\newcommand{\p}{\partial}
\newcommand{\refb}[1]{(\ref{#1})}

\renewcommand{\>}{\rangle}
\newcommand{\<}{\langle}

\renewcommand{\a}{\alpha}

\renewcommand{\b}{\beta}
\renewcommand{\t}{\tau}
\newcommand{\s}{\sigma}

\newcommand{\bes}{\begin{subequations}}
\newcommand{\ees}{\end{subequations}}

\newcommand{\n}{\nu}
\newcommand{\m}{\mu}

\newcommand{\normord}[1]{\xcentcolon\mathrel{#1}\xcentcolon}
\newcommand{\xcentcolon}{%
\mathrel{\vbox{\hbox{$:$}\kern.2ex}}%
}

\title{Tensionless Tales: Vacua and Critical Dimensions}

\author{Arjun Bagchi,} \author{Mangesh Mandlik,} \author{and Punit Sharma.} \author{\\}

\affiliation{Indian Institute of Technology Kanpur, Kanpur 208016, INDIA.\\} 

\emailAdd{abagchi@iitk.ac.in, mangeshm@iitk.ac.in, spunit@iitk.ac.in}

\abstract{Recently, a careful canonical quantisation of the theory of closed bosonic tensionless strings has resulted in the discovery of three separate vacua and hence three different quantum theories that emerge from this single classical tensionless theory. In this note, we perform lightcone quantisation with the aim of determination of the critical dimension of these three inequivalent quantum theories. The satisfying conclusion of a rather long and tedious calculation is that one of vacua does not lead to any constraint on the number of dimensions, while the other two give $D=26$. This implies that all three quantum tensionless theories can be thought of as consistent sub-sectors of quantum tensile bosonic closed string theory.}

\begin{document}
\maketitle

\vfill 

\section{Introduction}
Over the past several decades, string theory has emerged as the leading framework in which to formulate a theory of quantum gravity. As is well known, the point particle limit of string theory is the limit where the string length goes to zero or equivalently the tension of the fundamental string goes to infinity. This leads back to classical Einstein gravity (and its supersymmetric cousins). The opposite limit \cite{Schild:1976vq}, where the tension goes to zero offers a possible window into the deep quantum mechanical regime of gravity. 

\medskip

This very high temperature limit of string theory is also supposed to carry hints of a possible phase transition to the ever-mysterious Hagedorn phase of string theory \cite{Pisarski:1982cn, Olesen:1985ej, Atick:1988si}. Again, through the seminal work of Gross and Mende \cite{Gross:1987kza, Gross:1987ar, Gross:1988ue}, this sector has been shown to have interesting simplifications for string scattering. One also expects emergent higher spin symmetry to appear here. It is also envisioned that the tensionless limit would be instrumental in understanding singularities in spacetime and how string theory perceives them. So, in order to address a whole host of interesting problems, a better look at the theory of tensionless strings is important. 

\medskip 

The classical theory of tensionless or null strings \cite{Schild:1976vq} is by now well understood. Building on the seminal work \cite{Isberg:1993av}, we now know that the worldsheet of the null string is characterised by a degenerate metric, which enables a rewriting in terms of vector densities. The action becomes 
\be{lst}
S_{\text{ILST}} = \int d^2 \xi \,\ V^\alpha V^\beta \p_\alpha X^m \p_\beta X^n \eta_{mn}.
\ee
There is still diffeomorphism invariance, which is fixed by fixing a gauge for $V^\a$. One can choose the analogue of the conformal gauge, {\it{i.e.}} $V^\a = (1,0)$. But even with this, there is some residual gauge symmetry left over which now leads to the symmetry algebra 
\be{bms} 
[L_m,L_n]=(m-n)L_{m+n}, \quad [L_m,M_n]=(m-n)M_{m+n}, \quad [M_m,M_n]=0. 
\ee
This algebra, which replaces the two copies of the Virasoro algebra on the tensionless worldsheet, is known as the 3d Bondi-Metzner-Sachs (BMS$_3$) algebra \cite{Bagchi:2013bga, Bagchi:2015nca}. BMS algebras have made their appearance previously in the asymptotic symmetries of Minkowski spacetimes on the null boundary \cite{Bondi:1962px, Sachs:1962wk}. Specifically the algebra \refb{bms} appears as asymptotic symmetries of 3d flat spacetimes \cite{Barnich:2006av}. This has recently been heavily used in attempts of a holographic description of flatspace \cite{Bagchi:2010eg, Bagchi:2012xr,Bagchi:2012yk,Bagchi:2014iea}.  
  
\medskip
  
A formulation of the tensionless string with worldsheet BMS symmetries at its heart has been recently pursued in \cite{Bagchi:2013bga, Bagchi:2015nca, Bagchi:2019cay, Bagchi:2020fpr, Bagchi:2020ats} (see \cite{Bagchi:2016yyf, Bagchi:2017cte, Bagchi:2018wsn} for progress on tensionless superstrings). This paper is another step in this direction. Here we address the rather important question of critical dimensions of the tensionless string. The past literature on this subject has been very confusing with various sets of authors finding various different answers to this question \cite{Karlhede:1986wb, Lizzi:1986nv, Gamboa:1989zc, Gamboa:1989px, Gustafsson:1994kr, Lindstrom:2003mg}. In this paper, we provide a way out of this confusion. 

\medskip

It has recently been shown \cite{Bagchi:2020fpr}, through a careful analysis of canonical quantisation, that from a single classical tensionless theory, governed by the ILST action \refb{lst}, there are three different and distinct tensionless theories that can appear. These are theories built on three distinct vacua, which were called the Flipped, Induced and Oscillator vacua. The Flipped vacuum gives rise to the bosonic version of the Ambitwistor string \cite{Casali:2016atr}. The Induced vacuum is one which naturally arises when one takes a high energy limit on a particular string theory. The Oscillator vacuum is perhaps the most non-standard among the three and this arises when one considers Bogoliubov transformations on the worldsheet in an attempt to link tensile and tensionless oscillators. The induced and the oscillator vacua are intimately related through these Bogoliubov transformations and can be thought in terms of worldsheet analogues of inertial and accelerated Rindler observers \cite{Bagchi:2020ats}. 

\medskip 

In our paper, with a lightcone analysis where we focus on the closure of the background Lorentz algebra, we find that the consistent dimensions for these three different quantum theories. We find that while the Induced vacuum does not give us any constraints on the dimensions (and hence can live in any dimension), the Flipped vacuum and the Oscillator vacuum both lead us to $D=26$. This means that all of the three quantum tensionless theories can be thought of as sectors of usual relativistic string theory{\footnote{The theory built on the induced vacuum does not have any restrictions on dimensions, and hence there is no contradiction to this arising as a limit of a parent tensile string theory. There would be contradictions only if we found dimensions other than $D=26$}}.  

\medskip

To the uninitiated reader, our results may a priori look trivial, so let us pause to explain why this is far from obvious. Firstly, we seem to have three distinct quantum tensionless theories from the canonical quantisation techniques. Let us assume that these tensionless theories form sectors of a consistent quantum tensile string theory. If the lightcone quantisation results in anything other than $D=26$ (or no further constraints on $D$ as we have for the induced vaccuum), these tensionless theories would cease to be a sub-sector in the tensile theory and can be ruled out. We are here of course thinking of tensionless theories as consistent sub-sectors of a usual tensile theory. It could very well be that a quantum tensionless theory does not have a tensile parent and the tensionless strings are fundamental objects by themselves. In this case, where the tensionless strings are fundamental, these bosonic objects can live in dimension other than 26. But these theories cannot be connected to any relativistic tensile string theory. 

\medskip
We have seen that on the tensionless worldsheet, the metric vanishes and the (pseudo-)Riemannian  structure is replaced by a Carrollian worldsheet structure. This is indicated by the emergence of the BMS algebra \refb{bms} on the worldsheet. Recently, strings have been studied on non-relativistic Newton-Cartan backgrounds \cite{Harmark:2017rpg, Harmark:2018cdl, Harmark:2019upf, Bergshoeff:2019pij, Gomis:2019zyu}. Some of these theories also yield a worldsheet 2d Galilean conformal algbebra \cite{Harmark:2018cdl, Harmark:2019upf}, which is isomorphic to the BMS$_3$ algebra \refb{bms} \cite{Bagchi:2010eg}. Hence the appearance of \refb{bms} does not a priori guarantee that these strings can propagate in flat spacetimes for any dimension. It may so happen, as it does for these non-relativistic string theories, that the background obeys equations of Newton-Cartan or Carrollian geometry instead of Einstein equations. Indeed the vanishing of the beta-functions on the worldsheet gives results consistent with this line of thought \cite{Gomis:2019zyu, Gallegos:2019icg}. Hence our attempt to close the Lorentz algebra may not have yielded any positive result for any dimension at all. That would mean that the loss of the relativistic structure of the worldsheet necessarily means that the spacetime has to be deformed and not obey Einstein's equations. We have taken a singular limit on the worldsheet by sending the tension to zero. This manifesting itself as a singular limit on the spacetime, thereby modifying Einstein's equations to non-Lorentzian backgrounds would not be very surprising.  

\medskip
Having made an argument each in favour and against the plausibility our results, let us finally tilt the scale back in our favour. Consider the analogy with point particles. Massless point particles trace out null trajectories in any spacetime. If strings are thought of as natural generalisations, tensionless strings are the analogues of massless point particles. It is then not expected that these strings would need a completely different formulation from usual tensile strings, and hence we expect that the tensionless theories to appear as limits of tensile theories and therefore be consistent in the same dimensions as the tensile parents. Our results in this paper happily support such expectations.

\medskip

In what follows, we will first revisit the canonical quantisation of the tensionless bosonic closed string briefly before going on to explain how the three different quantum theories arise out of a single classical theory. Then we move onto lightcone quantization for the tensionless theories. We consider each vacua separately and show how the closing of the Lorentz algebra of the background spacetime leads to consistent dimensions for the three different theories. The calculations involved are very long and tedious, so much of it is excluded from the main body of the paper. For the interested reader, we however include appendices which has details of the calculations.

\bigskip

\section{Tensionless strings: Canonical quantisation}
We start with the ILST action\footnote{We now prefix \refb{lst} by a constant $\frac{1}{4\pi c'}$ to match dimensions in mode expansions that follow.}
\begin{equation}\label{TSAc}
S_{\text{ILST}} = \frac{1}{4\pi c'}\int d^2\sigma V^aV^b\partial_aX^\mu\partial_bX_\mu~,
\end{equation}
where $X^\m$ are space-time co-ordinates and the string worldsheet coordinates are
$\sigma^{0,1} \equiv (\tau, \sigma).$ $V^\alpha$'s are vector densities that have replaced the tensile structure $\frac{1}{\alpha '}\sqrt{g}g^{\alpha\beta}$ with the tensionless structure $\frac{1}{c'}V^\alpha V^\beta$. This action can be systematically derived from the tensile action \cite{Isberg:1993av}. We are however free to take \refb{TSAc} as the starting point of the analysis of tensionless strings where we treat these null strings as fundamental objects which may or may not descend from a tensile parent. 

\paragraph{Residual symmetries:} The above action \refb{TSAc},  like its tensile counterpart, has gauge symmetries on the worldsheet, which have to be fixed. $V^\alpha$ has the following transformation under the worldsheet reparametrization:
\begin{equation}\label{Vtransform}
\s^\alpha\rightarrow\s^\alpha+\epsilon^\alpha: \quad \delta V^\alpha=-V^\beta\partial_\alpha\epsilon^\alpha+\epsilon^\beta\partial_{\beta}V^\alpha+\frac{1}{2}(\partial_\beta\epsilon^\beta)V^\alpha.
\end{equation}
This allows us to fix the gauge
\be{v}
V^0 = 1,~~~V^1 = 0.
\ee
In this gauge the $X^\mu$ equation of motion (EOM) becomes
\begin{equation}\label{EOM}
\ddot{X}^\mu = 0, 
\end{equation}
while the variation of \eqref{TSAc} with respect to $V^\alpha$ followed by usage of the gauge gives the constraints
\begin{eqnarray}
\label{constr}
\dot{X}^\mu\dot{X}_\mu = 0, \quad \dot{X}^\mu X'_\mu = 0.
\end{eqnarray}
In the above equations, dot represents differentiation with respect to $\t$ while prime represents differentiation with respect to $\s$. It is worth noting at this point that reparameterization symmetry is still not completely fixed. In fact, for $V^\alpha=(1,0)$, the residual symmetry transformation that still keeps the gauge fixed action invariant are 
$$(\tau,\sigma)\rightarrow (\tau f'(\sigma) + g(\sigma), f(\sigma)),$$
where the functions $f(\sigma),g(\sigma)$ can be arbitrary. These can be obtained by fixing the gauge in \eqref{Vtransform}. The effect of such a transformation on a generic function of $\s, \t$ $(F(\s,\t))$ is 
\be{} \delta F(\s,\t)=[f'(\sigma)\tau\p_\tau+f(\sigma)\p_\sigma+g(\sigma)\p_\tau]F(\s,\t) =[L(f)+M(g)]F. \ee
We can now read off the generators of the residual gauge symmetry:
\bea{LM} 
&& L(f)=f'(\sigma)\tau\p_\tau+f(\sigma)\p_\sigma=\sum_n a_n e^{in\sigma}(\p_\sigma+in\tau\p_\tau)=-i\sum_n a_n L_n, \\
&& M(g)=g(\sigma)\p_\tau=\sum_n b_n e^{in\sigma} \p_\tau=-i\sum_n b_n M_n 
\eea
where we have expanded $f, g$ in terms of Fourier modes in $\s$. The algebra of the modes reads
\be{bms1} 
[L_m,L_n]=(m-n)L_{m+n}, \quad [L_m,M_n]=(m-n)M_{m+n}, \quad [M_m,M_n]=0. 
\ee
This is the classical part of BMS algebra \refb{bms}, {\it i.e.} without central extensions. 

\medskip

\paragraph{Energy-Momentum tensor:} We now proceed to find the energy-momentum (EM) tensor for the tensionless string, that would be central to understanding quantisation. Let $$\s^\a\rightarrow\s'^\a=\s^\a+\delta\s^\a$$ be an infinitesimal transformation of \refb{TSAc}. The EM tensor constructed from the Noether current $J^\a=T^\a_{\ \b}\delta\s^\b$ is: 
\be{} 
T^\a_{\ \beta}=V^\a V^\rho \p_\rho X^\mu \p_\beta X_\mu-\frac{1}{2}V^\lambda V^\rho \p_\lambda X^\mu\p_\rho X_\mu \delta^\a_{\ \beta}. 
\ee
In our gauge of choice, $V^\a=(1,0)$, $\delta\xi^\a=(f'\tau+g,f)$ and the non-trivial components of $T^\a_{\ \beta}$ are
\be{conT} 
T^0_{\ 1} = \dot X\cdot X' \equiv T_1 (\s, \t),  \quad  T^0_{\ 0}=-T^1_{\ 1}=\frac{1}{2} \dot X^2 \equiv T_2 (\s, \t). 
\ee
The associated Noether current $Q=\int d\sigma J^0 = \int d\sigma \left[T_1 f + T_2 (f'\tau+g) \right]$ can be expressed as
\be{}  
Q=\sum_n \left( a_n \int d\sigma \ (T_1+ in\tau T_2)e^{in\sigma}+ b_n \int d\sigma \ T_2 e^{in\sigma} \right) = \sum_n \left(a_n L_n + b_n M_n \right). 
\ee
by expanding $f$ and $g$ in fourier modes as before. Thus we have:
\be{} 
L_n=\int d\sigma (T_1+in\tau T_2)e^{in\sigma}, \quad M_n=\int d\sigma\ T_2  \ e^{in\sigma}. 
\ee
We can invert the above expressions to find an expansion of the EM tensor in terms of the modes of the BMS algebra: 
\be{Tmode}
T_1 (\s, \t) = \frac{1}{2\pi}\sum_{n} (L_n - in\tau M_n) e^{-i n\s}, \quad T_2 (\s, \t)= \frac{1}{2\pi}\sum_{n} M_n e^{-in\s}.
\ee

\medskip

\paragraph{Mode expansions:} Returning to the EOM, the most general solution of \eqref{EOM} is given by
\begin{equation}
X^\mu = x^\mu + c'p^\mu\tau + A^\mu\sigma  +i \sqrt{\frac{c'}2}\sum_{n\neq 0}\frac{1}{n}\left(A^\mu_n-i n\tau B^\mu_n\right)e^{-i n\sigma}.
\end{equation}
where $x^\mu$ and $p^\mu$ are real, and the mode coefficients $A^\mu_n$ and $B^\mu_n$ have to satisfy the reality condition ($X^\mu = \left(X^\mu\right)^*$)
\begin{equation}\label{reality}
A^\mu_n = \left(A^\mu_{-n}\right)^*,~~~B^\mu_n = \left(B^\mu_{-n}\right)^* .
\end{equation}
for $X^\mu$ to be real. In what follows, we will restrict our attention to closed strings. Hence this implies $A^\mu_0=0$. So we have
\begin{equation}\label{solmode}
X^\mu = x^\mu + c'p^\mu\tau  +i \sqrt{\frac{c'}2}\sum_{n\neq 0}\frac{1}{n}\left(A^\mu_n-i n\tau B^\mu_n\right)e^{-i n\sigma}.
\end{equation}
The mean position of the string is
\begin{equation}\label{pos}
x^\mu(\tau) = \frac{1}{2\pi}\int_0^{2\pi}d\sigma X^\mu = x^\mu+c' p^\mu \tau,
\end{equation}
which implies that $x^\mu$ is the initial mean position, whereas the canonical momentum conjugate to $X^\mu$ is given by
\begin{equation}
P^\mu \equiv \eta^{\mu\nu}\frac{\delta S}{\delta\dot{X}^\nu} =  \frac{1}{2\pi c'}\dot{X}^\mu = \frac{p^\mu}{2\pi} + \frac{1}{2\pi}\sqrt{\frac{1}{2c'}}\sum_{n\neq 0}B^\mu_n e^{-i n\sigma},
\end{equation}
resulting in the interpretation of $p^\mu$ as the mean momentum of the string,
\begin{equation}
p^\mu = \frac{1}{2\pi}\int_0^{2\pi}d\sigma P^\mu.
\end{equation}
We have already seen that the constraints \refb{constr} are expressible in terms of vanishing of the EM tensor \refb{conT}: 
\be{}
T_1= \dot X\cdot X' = 0, \quad T_2 (\s, \t) = \frac{1}{2} \dot X^2 = 0,
\ee
where $P\cdot Q \equiv P^\mu Q_\mu$. Using the mode expansion \refb{solmode} and the expression \refb{Tmode} of the EM tensor in terms of its modes, we see that 
\be{lmab}
L_m = \frac{1}{2} \sum_n A_{n} \cdot B_{m-n}, \quad M_m = \frac{1}{2} \sum_n B_{n} \cdot B_{m-n}
\ee

\bigskip

\paragraph{Quantization:} The tensionless action in the gauge \refb{v} takes the simplified form  
\be{}
S = \frac{1}{4\pi c'}\int d^2\sigma \partial_\t X^\mu\partial_\t X_\mu~. 
\ee
We will proceed to quantise this in the usual canonical way, keeping in mind that there are constraints \refb{constr} that we also need to impose. 
The Poisson bracket structure is given by
\begin{equation}\label{canPB}
\{X^\mu(\tau,\sigma),P_{\nu}(\tau,\sigma')\} = \delta(\sigma-\sigma')\delta^\mu_\nu.
\end{equation}
which gives rise to the symplectic structure
\begin{equation}
\Omega = \int_0^{2\pi}d\sigma\left(\delta P_\mu\wedge\delta X^\mu\right).
\end{equation}
Using the mode expansion \eqref{solmode}, we obtain 
\begin{equation}
\Omega = \delta p_\mu\wedge\delta x^\mu+\sum_{n\neq 0}\frac{i}{2n}\delta\left(B_{\mu}\right)_{-n}\wedge\delta A^\mu_{n},
\end{equation}
from which we can read off the nonvanishing Poisson brackets between the modes
\begin{equation}\label{modePB}
\{A^\mu_m,B^\nu_n\} = -2ni\eta^{\mu\nu}\delta_{m,-n}, \quad \{x^\mu,p^\nu\} = \eta^{\mu\nu}.
\end{equation}
which translate into the commutators
\begin{equation}\label{modecomm}
\left[A^\mu_m,B^\nu_n\right] = 2n\eta^{\mu\nu}\delta_{m,-n},\quad
\left[x^\mu,p^\nu\right] = i\eta^{\mu\nu}.
\end{equation}
We note here that that the algebra of $A, B$ is not that of harmonic oscillators. As a sanity check, we can use \refb{modecomm} and \refb{lmab} to rederive the BMS algebra \refb{bms1}. 

\medskip

Now to impose the constraints quantum mechanically, we impose 
\be{qT}
\<\text{phy}'|T_1|\text{phy}\> = 0, \quad \<\text{phy}'|T_2|\text{phy}\> = 0
\ee
on the total Hilbert space to pick out the physical states. Here $ |\text{phy}\>, |\text{phy}'\>$ are two arbitrary physical states. Using \refb{Tmode}, this condition \refb{qT} is equivalent to 
\be{qLM}
\<\text{phy}'|L_n|\text{phy}\> = 0, \quad \<\text{phy}'|M_n|\text{phy}\> = 0, \qquad \forall \, n \in \mathbb{Z}
\ee
In order to implement this, we shall make the assumption that the vacuum is a physical state:
\be{}
\<0|L_n|0\> = 0, \quad \<0|M_n|0\> = 0, \qquad \forall \, n \neq 0
\ee
where the zero mode will come with normal ordering ambiguities and is hence excluded. Following the analysis of \cite{Bagchi:2020fpr}, one can then find that there are three distinct choices of vacuum (and hence three distinct quantum mechanical theories) compatible with the conditions above:
\begin{subequations}
\bea{}
&& \text {(A) Flipped:}  \qquad  L_n|\text{phy}\> = 0, \quad M_n|\text{phy}\> = 0 \qquad \forall \, n>0, \\
&& \text {(B) Induced:}  \qquad  L_n|\text{phy}\> \neq 0, \quad M_n|\text{phy}\> = 0 \qquad \forall \, n\neq0, \\
&& \text {(C) Oscillator:} \quad  L_n|\text{phy}\> \neq 0, \quad M_n|\text{phy}\> \neq 0, \quad \forall \, n \quad {\text {but \refb{qLM} satisfied}}.
\eea
\end{subequations}
\begin{figure}[t]
\centering
\includegraphics[width=10cm]{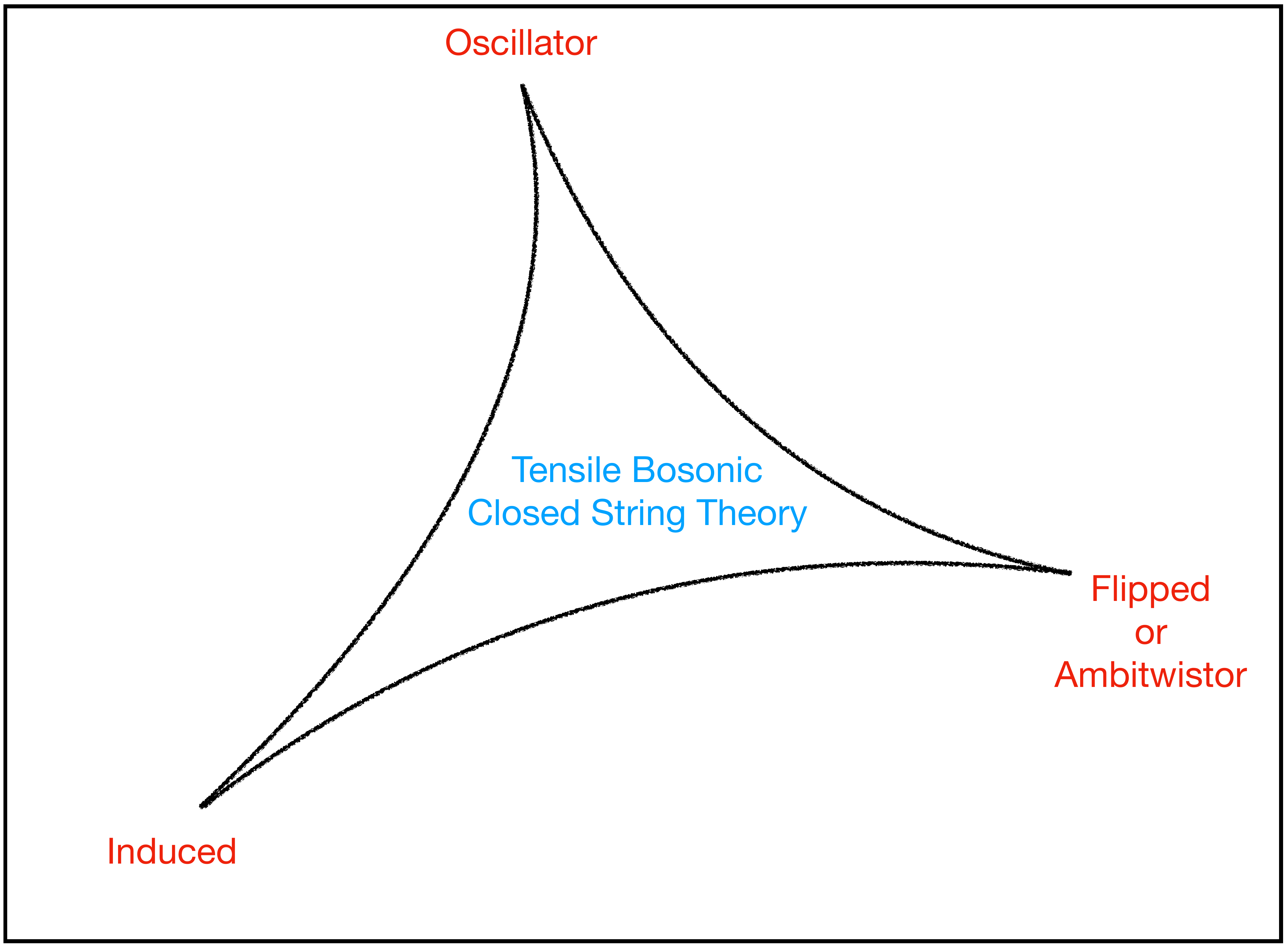}
  \caption{Tensionless corners of Bosonic String Theory.}
\end{figure}

The flipped vacuum imposes the conditions in the familiar highest weight manner. The resulting theory is actually the bosonic version of the Ambitwistor string \cite{Casali:2016atr}, which has been put forward to explain the Cachazo-He-Yuan formulae for tree-level scattering amplitudes. 

\medskip

The induced vacuum is named from the induced representations of the BMS algebra under which it transforms. This can be thought of as the limit of the tensile vacuum. A lot of interesting phenomena occur here like the emergence of an open string from the condensation of all the closed string modes \cite{Bagchi:2019cay}.

\medskip

The oscillator vacuum seems to be the most intimately intrinsic tensionless vacuum and hence has the maximal chance of not satisfying a spacetime Lorentz algebra in any dimensions. However the oscillator and the induced vacua are closely related by worldsheet Rindler transformations \cite{Bagchi:2020ats} and our present analysis would show that this too is consistent in $d=26$. 

\medskip

Figure 1 above depicts these three tensionless corners of the closed bosonic string.

\section{Lightcone quantisation}
We now turn our attention to lightcone (LC) quantisation. In order to implement this, the spacetime coordinates \{$X^\mu$\} are converted in to the lightcone coordinates $(X^+,X^-,\{X^i\})$, where
\be{}
X^\pm = \frac{1}{\sqrt{2}}\left(X^0\pm X^{D-1}\right)
\ee
and $i = 1,\dots,D-2$ denote the ``transverse directions". In these coordinates the nonzero components of the flat metric of the string background becomes
\begin{equation}\label{flatmetLC}
g_{+-} = -1, \, g_{ij} = \delta_{ij}.
\end{equation}
We now reconsider the mode expansions \refb{solmode} and introduce the $n=0$ extensions of $A^i_n$, $A^+_n$, $B^i_n$ and $B^+_n$
\begin{equation}\label{ABnaught}
B^i_0 \equiv\sqrt{2c'}p^i,~~~~B^+_0 \equiv\sqrt{2c'}p^+,~~~~ A^i_0 = 0 = A^+_0~.
\end{equation}
when acting of physical states. The mode expansion of the first of the constraints in \eqref{constr} gives
\begin{equation}\label{Constr1LC}
\begin{split}
2\sum_{n\neq m}B^+_nB^-_{m-n}+2\sqrt{2c'}p^- B^+_m &= \sum_{n}B^i_nB^i_{m-n}~~~\forall m\neq 0~,\\
2c'p^2 &= \sum_{n\neq 0}\left(2B^+_nB^-_{-n}-B^i_nB^-_{-n}\right)
\end{split}
\end{equation}
The second equation of \eqref{Constr1LC} is called the mass shell condition. Here, $p^2 = -2p^+p^- + p^ip^i$. Whereas the second constraint in \eqref{constr} similarly gives
\begin{equation}\label{Constr2LC}
\begin{split}
\sum_{\substack{n\neq 0\\n\neq m}}A^+_nB^-_{m-n}+\sum_{n\neq 0}A^-_nB^+_{m-n}+\sqrt{2c'}p^- A^+_m &= \sum_{n\neq 0}\left(A^i_nB^i_{m-n}\right)~~~\forall m\neq 0,\\
\sum_{n\neq 0}\left(A^+_nB^-_{-n}+A^-_nB^+_{-n}\right) &= \sum_{n\neq 0}\left(A^i_nB^i_{-n}\right)
\end{split}
\end{equation}
The second equation of \eqref{Constr1LC} is called the level matching condition.\\
Then fixing the residual gauge by choosing $A^+_n = 0 = B^+_n~\forall n \neq 0$ reduces the $m\neq 0$ parts of \eqref{Constr1LC} and \eqref{Constr2LC} to
\begin{equation}\label{minusdef}
A^-_m = \frac{1}{B_0^+}\sum_{n\neq 0}\normord{A^i_n B^i_{m-n}}~,\quad B^-_m = \frac{1}{2B_0^+}\sum_n \normord{B^i_n B^i_{m-n}}~.
\end{equation}
These constraint equations tell us that the modes in the ``$-$" direction are not independent degrees of freedom, but they depend on the modes in the orthogonal directions. In these equations, the free index $m$ runs over nonzero integers. But they also allow us to define the extensions $A^-_0$ and $B^-_0$. The notation $\normord{X}$ indicates that a given operator product $X$ is normal ordered according to the choice of the vacuum. The $A^-_n$ and $B^-_n$ operators themselves satisfy the algebra
\begin{equation}\label{cent}
\begin{split}
\left[A^-_m,A^-_n\right] &= \frac{2(m-n)}{B^+_0}A^-_{m+n}+\frac{c_L}{12}\delta_{m+n,0},\\
\left[A^-_m,B^-_n\right] &= \frac{2(m-n)}{B^+_0}B^-_{m+n}+\frac{c_M}{12}\delta_{m+n,0},\\
\left[A^-_m,A^-_n\right] &= 0.
\end{split}
\end{equation}
$c_L$ and $c_m$ are called ``central charges", so the terms involving them are called the ``central terms", and this algebra is called the ``central algebra". These charges depend on the choice of normal ordering (i.e. the vacuum).

\medskip

Classically, if we use the definition \eqref{minusdef} in the mass shell condition and the level matching condition, we get $A^-_0 = 0$ and $B^-_0 = \sqrt{2c'}p^-$. This is analogous to the definitions in \eqref{ABnaught}. However, due to normal ordering ambiguity, $A^-_0$ and $B^-_0$ don't follow the same definitions as the $i$ and $+$ counterparts as in \eqref{ABnaught} but they receive quantum corrections
\be{} 
A^-_0 = \frac{a}{B^+_0},~~~~B^-_0 = \sqrt{2c'}p^- + \frac{b}{B^+_0}.
\ee
$a$ and $b$ are constants the values of which will be fixed using the Poincare algebra. When we separate out the $\pm$ indices from the other spacetime indices, we can write the basic commutators \eqref{modecomm} as
\begin{equation}\label{modecommij}
\left[A^i_m,B^j_n\right] = 2n\delta^{ij}\delta_{m,-n}~, \quad \left[x^i,p^j\right] = i\delta^{ij}~,\quad \left[x^-,p^+\right] = -i~.
\end{equation}
These can be used to find some other useful commutations

\begin{equation}\label{ABij}
\begin{split}
\left[A^i_m,A^-_n\right] &= \frac{2}{B^+_0}A^i_{m+n},~~~\forall m\neq -n, \quad \left[A^i_{-m},A^-_m\right] =0. \\
\left[B^i_m,A^-_n\right] &= \left[A^i_m,B^-_n\right] = \frac{2}{B^+_0}B^i_{m+n}~,\quad \left[B^i_m,B^-_n\right] = 0.
\end{split}
\end{equation}
and
\begin{equation}\label{xAB}
\begin{split}
\left[x^-,B^j_0\right] &= i\sqrt{2c'}\delta^{ij}~,\quad \left[x^-,B^+_0\right] = -i\sqrt{2c'}~,\\
\left[x^\mu,A^-_n\right] &= \frac{i\sqrt{2c'}}{B^+_0}A^\mu_n \quad 
\left[x^\mu,B^-_n\right] = \frac{i\sqrt{2c'}}{B_0^+}B^\mu_n~~{\rm for}~\mu\neq +~.
\end{split}
\end{equation}
The expressions in \eqref{ABij} and \eqref{xAB} are independent of the choice of the vacuum.
{\footnote{Naively the equations \eqref{ABij} and \eqref{xAB} may seem to be contradictory to \eqref{modecomm}. However, note that \eqref{modecomm} arose from covariant quantization where $A^-_n$ and $B^-_n$ were also treated as independent variables and constraints were imposed afterwards on the physical states, while here $A^-_n$ and $B^-_n$ are treated as dependent on the transverse oscillators.}}

\subsection*{Lorentz algebra and problematic commutators}\label{Lormain}
We now figure out which of the commutators of the background Lorentz algebra may be problematic going forward. The below discussion mirrors that of the usual string theory in e.g. \cite{Green:1987sp}, but since we work with a different action \refb{lst}, we revisit and work these details out carefully. 

\medskip

As mentioned in the previous section, the reparametrization gauge choice here is
$$X^+ = x^+ + c'p^+\tau.$$
The theory in this gauge has clear structure for the variables: $D$ zero mode positions $x^\mu$ and momenta $p^\mu$, and $2(D-2)$ sets of transverse oscillators namely $A^i_n$ and $B^i_n$. The $A^-_n$ and $B^-_n$ oscillator modes are dependent on the transverse oscillator modes.

\medskip

For the theory to be Lorentz invariant, this gauge choice needs to be preserved by arbitrary Lorentz transformation, or else this structure is spoiled. Let's first consider an infinitesimal Lorentz transformation
$$\delta X^\mu = {\omega^\mu}_\nu X^\nu.$$
The variation of the action \eqref{TSAc} produced by this transformation is 
\begin{equation}
\begin{split}
\delta S &= \frac{1}{2\pi c'}\int d^2\sigma\left(V^aV^b{\omega^\mu}_\nu\partial_aX^\nu\partial_bX_\mu\right) = -\frac{1}{2\pi c'}\int d^2\sigma~\partial_b\left(V^b\omega_{\mu\nu} X^\mu V^a\partial_aX^\nu\right)\\
&= -\frac{1}{2\pi c'}\int d^2\sigma~\partial_\tau\left(\omega_{\mu\nu} X^\mu \partial_\tau X^\nu\right) = -\int d^2\sigma~\partial_\tau\left(\omega_{\mu\nu} X^\mu P^\nu\right) = -\frac{1}2\int d\tau~\partial_\tau\left(\omega_{\mu\nu}J^{\mu\nu}\right)
\end{split}
\end{equation}
The last equality defines $J^{\mu\nu}$ which is the generator for the Lorentz transformation $\omega_{\mu\nu}$. Considering the antisymmetry of $\omega_{\mu\nu}$:
\be{}
J^{\mu\nu} \equiv  \int d\sigma \left(X^\mu P^\nu-X^\nu P^\mu\right).
\ee
For $\delta S$ to vanish between two constant $\tau$ slices for arbitrary $\omega_{\mu\nu}$, $J^{\mu\nu}$ needs to be conserved.

For the theory to be Lorentz invariant, the algebra generators have to obey:
\begin{equation}
\left[J^{\mu\nu},J^{\rho\sigma}\right] = -i\left(J^{\mu\rho}\eta^{\nu\sigma}-J^{\nu\rho}\eta^{\mu\sigma}-J^{\mu\sigma}\eta^{\nu\rho}+J^{\nu\sigma}\eta^{\mu\rho}\right)
\end{equation}

Now for the gauge choice to be preserved under an arbitrary Lorentz transformation, we need to accompany it with a compensating reparametrization gauge transformation as follows. Let $\sigma^a\to\sigma^a+\xi^a$ be the compensating gauge transformation for ${\omega^\mu}_\nu$. i.e.
\begin{equation}\label{comptrans}
\delta X^+ = {\omega^+}_\mu X^\mu + \xi^a\partial_aX^+={\omega^+}_\mu X^\mu + c'\xi^0 p^+.
\end{equation}
For the gauge to be preserved, we need
\begin{equation}\label{transreq}
\delta X^+ = {\omega^+}_\mu\left(x^\mu+c'p^\mu\tau\right).
\end{equation}
Comparing \eqref{comptrans} and \eqref{transreq} evaluates $\xi^0$:
\begin{equation}\label{compgt}
\xi^0 = \frac{{\omega^+}_\mu}{c'p+}\left(x^\mu+c'p^\mu\tau-X^\mu\right)= -\frac{i{\omega^+}_i}{\sqrt{2c'}p^+}\sum_{n\neq 0}\frac{1}{n}\left(A^i_n-i n\tau B^i_n\right)e^{-i n\sigma},
\end{equation}
having used ${\omega^+}_- = \omega_{--} = 0$. The analysis so far tells us that the Lorentz transformations ${\omega^+}_i = \omega_{-i}$ require a nontrivial compensating gauge transformation, so the algebra elements involving the corresponding generators, namely $J^{i-}$, shouldn't necessarily satisfy Lorentz algebra. We have used reparametrization invariance to set $V^a = (1,0)$. Under the transformation 
\eqref{compgt},
\begin{equation}
\delta V^a = \xi^b\partial_bV^a - V^b\partial_b\xi^a +\frac{1}2V^a\partial_b\xi^b = -\partial_\tau\xi^a+\frac{1}2\delta^a_\tau\partial_b\xi^b
\end{equation}
First, for $\delta V^1 = 0$, we need $\partial_\tau\xi^1 = 0$, so $\xi^1$ is a function of only $\sigma$. Then, for setting $\delta V^1 = 0$, we need
\begin{equation}\label{xi1}
\partial_\sigma\xi^1 = \partial_\tau\xi^0 = -\frac{{\omega^+}_i}{\sqrt{2c'}p^+}\sum_{n\neq 0}B^i_n e^{-i n\sigma}
\end{equation}
Thus a ${\omega^+}_i$ Lorentz transformation should be accompanied by a compensating gauge transformation obeying \eqref{compgt} and \eqref{xi1} in order to maintain the gauge choice $V = (1,0)$ and $X^+ = x^+ + c'p^+\tau$. This Lorentz transformation is generated by $J^{i-}$, hence the algebra elements involving $J^{i-}$ may have anomalies that have to be checked for.

\section{Closing the Lorentz algebra in different vacua}
As shown in the appendix \ref{Lortriv}, all of the components of the Lorentz algebra are obeyed in all the vacua, except the $\left[J^{i-},J^{j-}\right]$ components which are nontrivial, so they have to be imposed as operator equations. This gives rise to interesting constraints on the theory, the most important being the determination of the critical number of dimensions.

\medskip

The goal of this section is to evaluate $\left[J^{i-},J^{j-}\right]$ in different vacua of the tensionless strings. The Lorenzian ``boost" components $J^{i-}$ are given by
\begin{equation}\label{Jsplit}
J^{i-} = \int_0^{2\pi}d\sigma\normord{\left(X^iP^--X^-P^i\right)}= L^{i-}+S^{i-}.
\end{equation}
The expression is split between two contributions, namely the zero mode contribution\footnote{Actually, for neither of the terms in $L^{i-}$ the ordering is fixed by the choice of the vacuum. But it's easy to see that for calculating neither $\left[L^{i-},L^{j-}\right]$ nor $\left[L^{i-},S^{j-}\right]$ does the ordering matter.}
\begin{equation}\label{Ldef}
L^{i-} = x^ip^- - x^-p^i.
\end{equation}
and the oscillator contribution
\begin{equation}\label{Sdef}
S^{i-} = -\frac{i}2\sum_{m\neq 0}\frac{1}m \left(\normord{A^i_{-m}B^-_m} + \normord{A^-_m B^i_{-m}}\right)~.
\end{equation}
First, it is easy to see that $\left[L^{i-},L^{j-}\right] = 0,$ which means
\begin{equation}\label{JJprim}
\left[J^{i-},J^{j-}\right] = \left[L^{i-},S^{j-}\right]+\left[S^{i-},L^{j-}\right]+\left[S^{i-},S^{j-}\right]
\end{equation}
The section has three subsections dedicated to three vacua of tensionless strings (induced, flipped and oscillator). These subsections have roughly the same structure. First, the annihilators of the vacuum are reported. Then the explicit normal ordered structures are written, followed by the calculation of the ``central terms". Then $\left[L^{i-},S^{j-}\right]+\left[S^{i-},L^{j-}\right]$ and $\left[S^{i-},S^{j-}\right]$ are computed, and combined to get $\left[J^{i-},J^{j-}\right]$ according to \eqref{JJprim}. In the end we impose the Lorentz algebra
$$\left[J^{i-},J^{j-}\right]=0.$$

The results of the main steps of this calculation are reposted in this section. The detailed calculation is done in the appendix which interested readers can refer to.

\subsection{Induced Vacuum}\label{Indsec}
This vacuum, as mentioned earlier, is defined by 
\be{}
L_n|0\>_I \neq 0, \quad M_n|0\>_I = 0 \qquad \forall \, n\neq0
\ee
In the oscillator language, this translates into the vacuum being annihilated by all $B^i_m$ except $m=0$:
\be{}
B^i_m |0\>_I = 0 ~~\forall~m\neq0.
\ee
We will be concerned with this oscillator definition going forward, not only for this, but corresponding oscillator definitions for all other vacua. The above fixes the normal ordering in \eqref{minusdef},
\begin{equation}\label{minusdefind}
A^-_m = \frac{1}{B_0^+}\sum_{n\neq 0}A^i_n B^i_{m-n},\quad
B^-_m = \frac{1}{2B_0^+}\sum_n B^i_n B^i_{m-n}.
\end{equation}
Their commutation gives
\begin{equation}
\left[A^-_m,A^-_n\right] = \frac{2(m-n)}{B^+_0}A_{m+n},\quad
\left[B^-_m,A^-_n\right] = \frac{2(m-n)}{B^+_0}B_{m+n}.
\end{equation}
i.e. both the central charged $c_L$ and $c_M$ vanish. This is in keeping with earlier findings in \cite{Bagchi:2020fpr} and also makes sense when we consider the tensionless limit on the algebra where the two copies of the Virasoro algebra turns into the BMS$_3$. The normal ordering in \eqref{Sdef} is explicitly written as
\begin{equation}
S^{i-} = -\frac{i}2\sum_{m\neq 0}\frac{1}m \left(A^i_{-m}B^-_m + A^-_mB^i_{-m}\right).
\end{equation}
Then we evaluate
\begin{equation}\label{LStot}
\begin{split}
\left[L^{i-},S^{j-}\right]+\left[L^{i-},S^{j-}\right]
=& -\sqrt{2c'}\sum_{m\neq 0}\frac{1}{mB^+_0}\left(A^i_{-m}B^j_m + B^i_{-m}A^j_m\right)p^-\\
&+\frac{\sqrt{2c'}}{2}\sum_{m\neq 0}\frac{1}{mB^+_0}\left(A^i_{-m}p^j-p^iA^j_{-m}\right)B^-_m\\
&+\frac{\sqrt{2c'}}{2}\sum_{m\neq 0}\frac{1}{mB^+_0}\left(B^i_{-m}p^j-p^iB^j_{-m}\right)A^-_m~.
\end{split}
\end{equation}
The details for this and the subsequent calculations are collected in Appendix \ref{Ivac}. Continuing with our calculations, we proceed to find
\begin{equation}\label{SS}
\begin{split}
\left[S^{i-},S^{j-}\right]
=& \sum_{m\neq 0}\frac{1}{mB^+_0}\left(A^i_{-m}B^j_m+B^i_{-m}A^j_m\right)B^-_0 -\frac{\sqrt{2c'}}2\sum_{m\neq 0}\frac{1}{mB^+_0}\left(A^i_{-m}p^j-p^iA^j_{-m}\right)B^-_m\\
&-\frac{\sqrt{2c'}}2\sum_{m\neq 0}\frac{1}{mB^+_0}\left(B^i_{-m}p^j-p^iB^j_{-m}\right)A^-_m + \sum_{m\neq 0}\frac{1}{mB^+_0}B^i_{-m}B^j_mA^-_0.
\end{split}
\end{equation}
and finally using \eqref{LStot} and \eqref{SS},
\begin{equation}
\begin{split}
\left[J^{i-},J^{j-}\right] &= \left[L^{i-},S^{j-}\right]+\left[S^{i-},L^{j-}\right]+\left[S^{i-},S^{j-}\right]\\
=& \sum_{m\neq 0}\frac{1}{mB^+_0}\left(A^i_{-m}B^j_m+B^i_{-m}A^j_m\right)\left(B^-_0-\sqrt{2c'}p^-\right)+\sum_{m\neq 0}\frac{1}{mB^+_0}B^i_{-m}B^j_mA^-_0.
\end{split}
\end{equation}

Now imposing $\left[J^{i-},J^{j-}\right] = 0$ as an operator equation on the physical states requires
\begin{subequations}
\bea{}
A^-_0|phys\rangle &=& 0, \\
B^-_0|phys\rangle &=& \sqrt{2c'}p^-|phys\rangle.
\eea
\end{subequations}
which is what we expect classically. Also, notice that there is no constraint on the dimensionality of the theory. This means that this vacuum is consistent with the Lorentz algebra in any number of spacetime dimensions. To stress something that we have already mentioned in the introduction, this means that if we start off with a consistent relativistic string theory, this quantum tensionless theory, build over the induced vacuum, can emerge as a consistent quantum mechanical sector of the quantum tensile theory. 

\subsection{Flipped Vacuum}\label{Flsec}
The flipped vacuum, which gives rise to the bosonic version of the Ambitwistor string, is given by 
\be{}
L_n|0\>_F=0, \quad M_n|0\>_F = 0 \qquad \forall \, n>0.
\ee
In terms of oscillators, this translates into 
\be{ABFv}
A^i_m |0\>_F = 0, B^i_m|0\>_F = 0 ~~~\forall~m > 0.
\ee
This is a familiar highest weight conditions for a vacuum, similar to the case for usual strings. But unlike usual strings, the algebra for the oscillators $A, B$ are not the same as simple harmonic oscillator. The above condition \refb{ABFv} fixes the normal ordering in \eqref{minusdef}:
\begin{equation}\label{minusdeffl}
\begin{split}
A^-_m &= \frac{1}{B_0^+}\left(\sum_{n < 0}A^i_n B^i_{m-n}+\sum_{n > 0} B^i_{m-n}A^i_n\right),~~~\forall m \neq 0\\
A^-_0 &= \frac{1}{B_0^+}\left(\sum_{n < 0}A^i_n B^i_{-n}+\sum_{n > 0} B^i_{-n}A^i_n\right),\\
B^-_m &= \frac{1}{2B_0^+}\sum_n B^i_n B^i_{m-n}.
\end{split}
\end{equation}
Their commutation gives
\begin{equation}
\begin{split}
\left[A^-_m,A^-_n\right] &= \frac{2(m-n)}{B^+_0}A_{m+n}+\frac{2(D-2)}{3(B_0^+)^2}(m^3-m)\delta_{m+n,0}~,\\
\left[B^-_m,A^-_n\right] &= \frac{2(m-n)}{B^+_0}B_{m+n}.
\end{split}
\end{equation}
According to the normal ordering suited to this vacuum,
\begin{equation}\label{Sflip}
S^{i-} = -\frac{i}2\sum_{m> 0}\frac{1}m \left(A^i_{-m}B^-_m + B^i_{-m}A^-_m\right)-\frac{i}2\sum_{m< 0}\frac{1}m \left(B^-_mA^i_{-m} + A^-_mB^i_{-m}\right)~.
\end{equation}

The calculation of $\left[L^{i-},S^{j-}\right]+\left[S^{i-},L^{j-}\right]$ follows a path very similar to the induced vacuum case, and we have given the details of this computation in Appendix \ref{FVac}. The calculation doesn't have any nontrivial shift of indices as in the case of $\left[S^{i-},S^{j-}\right]$. We just have to keep in mind the ordering of \eqref{Sflip}. We get

\bea{LSflip}
\left[L^{i-},S^{j-}\right]+\left[L^{i-},S^{j-}\right]&
=& -\sqrt{2c'}\sum_{m\neq 0}\frac{1}{mB^+_0}\left(A^i_{-m}B^j_m + B^i_{-m}A^j_m\right)p^- +\frac{\sqrt{2c'}}2\sum_{m > 0}\frac{1}{mB^+_0}\left(A^i_{-m}p^j-p^iA^j_{-m}\right)B^-_m\nonumber\\
&&+\frac{\sqrt{2c'}}2\sum_{m < 0}\frac{1}{mB^+_0}B^-_m\left(A^i_{-m}p^j-p^iA^j_{-m}\right) +\frac{\sqrt{2c'}}2\sum_{m > 0}\frac{1}{mB^+_0}\left(B^i_{-m}p^j-p^iB^j_{-m}\right)A^-_m\nonumber\\
&&+\frac{\sqrt{2c'}}2\sum_{m < 0}\frac{1}{mB^+_0}A^-_m\left(B^i_{-m}p^j-p^iB^j_{-m}\right).
\eea
As shown in Appendix \ref{FVac}, we also get: 
\bea{SSflip}
\left[S^{i-},S^{j-}\right]
&=& \sum_{m\neq 0}\frac{1}{mB^+_0}\left(A^i_{-m}B^j_m + B^i_{-m}A^j_m\right)B^-_0 -\frac{\sqrt{2c'}}2\sum_{m > 0}\frac{1}{mB^+_0}\left(A^i_{-m}p^j-p^iA^j_{-m}\right)B^-_m\\
&&-\frac{\sqrt{2c'}}2\sum_{m < 0}\frac{1}{mB^+_0}B^-_m\left(A^i_{-m}p^j-p^iA^j_{-m}\right) -\frac{\sqrt{2c'}}2\sum_{m > 0}\frac{1}{mB^+_0}\left(B^i_{-m}p^j-p^iB^j_{-m}\right)A^-_m\nonumber\\
&&-\frac{\sqrt{2c'}}2\sum_{m < 0}\frac{1}{mB^+_0}A^-_m\left(B^i_{-m}p^j-p^iB^j_{-m}\right) -4\sum_{m>0}\frac{m}{(B^+_0)^2}B^i_{-m}B^j_m-4\sum_{m<0}\frac{m}{(B^+_0)^2}B^i_{-m}B^j_m\nonumber\\
&&+\frac{(D-2)}{6(B^+_0)^2}\left(\sum_{m>0}\left(m-\frac{1}m\right)B^i_{-m}B^j_m+\sum_{m<0}\left(m-\frac{1}m\right)B^i_{-m}B^j_m\right) + \sum_{m\neq 0}\frac{1}{mB^+_0}B^i_{-m}B^j_mA^-_0~. \nonumber
\eea
Finally, using \eqref{LSflip} and \eqref{SSflip},
\bea{JJflip}
&&\left[J^{i-},J^{j-}\right] = \left[L^{i-},S^{j-}\right]+\left[S^{i-},L^{j-}\right]+\left[S^{i-},S^{j-}\right] \nonumber\\
&& =\sum_{m\neq 0}\frac{1}{mB^+_0}\left(A^i_{-m}B^j_m + B^i_{-m}A^j_m\right)\left(B^-_0-\sqrt{2c'}p^-\right) +\left(\frac{D-2}{6}-4\right)\sum_{m\neq 0}\frac{m}{(B^+_0)^2}B^i_{-m}B^j_m \nonumber \\
&& \hspace{1cm} + \left(A^-_0 - \frac{D-2}{6B^+_0}\right)\sum_{m\neq 0}\frac{1}{mB^+_0}B^i_{-m}B^j_m~.
\eea
Now imposing $\left[J^{i-},J^{j-}\right] =0$ on the physical states requires
\begin{subequations}
\bea{}
D &=& 26, \\
A^-_0|phys\rangle &=& \frac{4}{B^+_0},~~{\rm \text{and}} \\
B^-_0|phys\rangle &=& \sqrt{2c'}p^-|phys\rangle.
\eea
\end{subequations}
This means, unlike the induced vacuum, this vacuum is consistent with Lorentz algebra only when the number of spacetime dimensions $D=26$.

\subsection{Oscillator Vacuum}\label{Oscsec}
The third and perhaps the most interesting of the vacua is the oscillator vacuum. This, as stated earlier is given 
\be{}
L_n|0\>_c \neq 0, \quad M_n|0\>_c \neq 0, \quad {\text {but }} \<L_n\>_c = \<M_n\>_c = 0, \quad \forall \, n
\ee
Here $\<L_n\>_c = \, _c\<0|L_n|0\>_c$ and so on. In order to express the above in oscillator modes, we now define some new oscillators: 
\begin{equation}\label{Cdef}
C^\mu_n = \frac{1}{2} (B^\mu_n+A^\mu_n), \quad \tilde{C}^\mu_n = \frac{1}{2}(B^\mu_{-n}-A^\mu_{-n})~~~~\forall n\neq 0.
\end{equation}
or equivalently, 
\begin{equation}\label{Cinv}
A^\mu_n = C^\mu_n-\tilde{C}^\mu_{-n},\quad B^\mu_n = C^\mu_n+\tilde{C}^\mu_{-n}~~~~\forall n\neq 0~.
\end{equation}
These definitions make the algebra of the $C$ oscillators like that of standard simple harmonic oscillators. From \eqref{modecommij} we evaluate
\begin{equation}\label{Ccomm}
\left[C^i_m,C^j_n\right] = \delta^{ij}\delta_{m,-n}, \, \left[\tilde{C}^i_m,\tilde{C}^j_n\right] = \delta^{ij}\delta_{m,-n}, \, \left[\tilde{C}^i_m,C^j_n\right] = 0~.
\end{equation}
The original motivation in defining these sets of oscillators was an attempt to link the usual string theory $\a$ oscillators to these $C$ oscillators by a worldsheet Bogoliubov transformation. There are very interesting consequences of this and for a detailed account the reader is pointed to the recent work \cite{Bagchi:2019cay, Bagchi:2020ats}. 

In terms of these new $C$ oscillators, the oscillator vacuum is defined in the usual way: 
\be{}
C^i_m |0\>_c = 0 \quad {\rm and}~~~\tilde{C}^i_m|0\>_c=0 ~~~\forall m>0.
\ee

Here it turns out that the central algebra is more conveniently and usefully expressed in therms of commutators involving $C^-_n$ and $\tilde{C}^-_n$ than those involving $A^-_n$ and $B^-_n$.
\begin{equation}\label{CC-}
\begin{split}
\left[C^-_{m},C^-_{n}\right] =& \frac{m-n}{2B^+_0}\left(3C^-_{m+n}+\tilde{C}^-_{-m-n}\right) +\frac{(D-2)}{6(B^+_0)^2}(m^3-m)\delta_{n,-m}~,\\
\left[C^-_m,\tilde{C}^-_{-n}\right] =& -\frac{(m-n)}{2B^+_0}\left(C^-_{m+n}-\tilde{C}^-_{-m-n}\right)~,\\
\left[\tilde{C}^-_{-m},\tilde{C}^-_{-n}\right] =& -\frac{(m-n)}{2B^+_0}\left(C^-_{m+n}+3\tilde{C}^-_{-m-n}\right)-\frac{(D-2)}{6(B^+_0)^2}(m^3-m)\delta_{n,-m}.
\end{split}
\end{equation}

Again, we provide the details of this admittedly laborious and tedious computation in Appendix \ref{Cvac}. Here we provide some of the major steps. According to the normal ordering suited to this vacuum,
\begin{equation}\label{Sosc}
S^{i-} = -i\sum_{m> 0}\frac{1}m \left(C^i_{-m}C^-_m - \tilde{C}^-_{-m}\tilde{C}^i_m\right)-i\sum_{m< 0}\frac{1}m \left(C^-_mC^i_{-m} - \tilde{C}^i_m\tilde{C}^-_{-m}\right)~.
\end{equation}
Then we evaluate
\begin{equation}\label{LStotosc}
\begin{split}
\left[L^{i-},S^{j-}\right]+\left[S^{i-},L^{j-}\right] &
= -\frac{2\sqrt{2c'}}{B^+_0}\sum_{m\neq 0}\frac{1}m\left(C^i_{-m}C^j_m - \tilde{C}^i_m\tilde{C}^j_{-m}\right)p^-\\
&-\frac{\sqrt{2c'}}{B^+_0}\sum_{m > 0}\frac{1}m \left(\left(C^j_{-m}C^-_m - \tilde{C}^-_{-m}\tilde{C}^j_m\right)p^i- \left(C^i_{-m}C^-_m - \tilde{C}^-_{-m}\tilde{C}^i_m\right)p^j\right)\\
&-\frac{\sqrt{2c'}}{B^+_0}\sum_{m < 0}\frac{1}m \left(\left(C^-_mC^j_{-m} - \tilde{C}^j_m\tilde{C}^-_{-m}\right)p^i - \left(C^-_mC^i_{-m} - \tilde{C}^i_m\tilde{C}^-_{-m}\right)p^j\right).
\end{split}
\end{equation}
and
\begin{equation}\label{SStotosc}
\begin{split}
\left[S^{i-},S^{j-}\right] &=\frac{1}{\left(B^+_0\right)^2}\left(\frac{(D-2)}{6}-4\right)\sum_{m<0}m\left(C^i_{-m}C^j_m-\tilde{C}^i_m\tilde{C}^j_{-m}\right) -\frac{(D-2)}{6\left(B^+_0\right)^2}\sum_{m\neq 0}\frac{1}m\left(C^i_{-m}C^j_m-\tilde{C}^i_m\tilde{C}^j_{-m}\right)\\
&+\frac{1}{B^+_0}\sum_{m \neq 0}\frac{1}m B^i_{-m}B^j_mA^-_0+\frac{2}{B^+_0}\sum_{m \neq 0}\frac{1}m\left(C^i_{-m}C^j_m - \tilde{C}^i_m\tilde{C}^j_{-m}\right)B^-_0\\
&+\frac{\sqrt{2c'}}{B^+_0}\sum_{m > 0}\frac{1}m \left(\left(C^j_{-m}C^-_m - \tilde{C}^-_{-m}\tilde{C}^j_m\right)p^i- \left(C^i_{-m}C^-_m - \tilde{C}^-_{-m}\tilde{C}^i_m\right)p^j\right)\\
&+\frac{\sqrt{2c'}}{B^+_0}\sum_{m < 0}\frac{1}m \left(\left(C^-_mC^j_{-m} - \tilde{C}^j_m\tilde{C}^-_{-m}\right)p^i - \left(C^-_mC^i_{-m} - \tilde{C}^i_m\tilde{C}^-_{-m}\right)p^j\right).
\end{split}
\end{equation}
In the end, using \eqref{LStotosc} and \eqref{SStotosc} we have
\begin{equation}\label{JJosc}
\begin{split}
\left[J^{i-},J^{j-}\right] &= \left[L^{i-},S^{j-}\right]+\left[S^{i-},L^{j-}\right]+\left[S^{i-},S^{j-}\right]\\
&=\frac{1}{\left(B^+_0\right)^2}\left(\frac{(D-2)}{6}-4\right)\sum_{m<0}m\left(C^i_{-m}C^j_m-\tilde{C}^i_m\tilde{C}^j_{-m}\right)\\
&+\frac{1}{B^+_0}\sum_{m\neq 0}\frac{1}m\left(C^i_{-m}C^j_m-\tilde{C}^i_m\tilde{C}^j_{-m}\right)\left(2B^-_0- 2\sqrt{2c'}p^- -\frac{(D-2)}{6B^+_0}\right)\\
&+\frac{1}{B^+_0}\sum_{m \neq 0}\frac{1}m B^i_{-m}B^j_mA^-_0~.\\
\end{split}
\end{equation}
where we have used \eqref{LStotosc} and \eqref{SStotosc}. Note that the $p^i$ and $p^j$ terms cancel out. Thus, for $\left[J^{i-},J^{j-}\right]$ to vanish on physical states, we need
\begin{subequations}
\bea{}
D &=& 26, \\
A^-_0|phys\rangle &=& 0,~~{\text{ and}} \\
B^-_0|phys\rangle &=& \left(\sqrt{2c'}p^- + \frac{2}{B^+_0}\right)|phys\rangle.
\eea
\end{subequations}
Hence this vacuum also needs to live in $26$ dimensions to be consistent with Lorentz (Poincare) algebra.

\newpage

\section{Conclusions}
\subsection*{Summary}
In this paper,  through the process of lightcone quantisation and the closure of the Lorentz algebra in the background, we asked in which dimensions the quantum tensionless bosonic string theory was consistent. A previous detailed analysis in terms of canonical quantisation of the tensionless theory had revealed that the quantum tensionless theory had three, not one, different manifestations. In other words, there were three different allowed vacua allowed by canonical quantisation, from which three different quantum tensionless theories were built. So three different quantum theories emerged from a single classical one. Our investigations in this paper revealed that two of these theories, the ones build over the Flipped and the Oscillator vacua, were only consistent in $d=26$, while there were no such constraints on the dimension arising from the Induced vacuum which thus could be consistent in any dimension. This was in keeping with the idea that these quantum tensionless theories could arise as corners of the quantum tensile theory, envisioned in Fig 1. We believe that our present work clarifies a lot confusion in existing literature about the dimensionality of tensionless string theory.    

\subsection*{Discussions}
Moving forward, there are a number of immediate directions of research that are being currently pursued. First and foremost is the question of the tensionless string spectra around these different vacua. An initial analysis of this was carried out in \cite{Bagchi:2020fpr}. The lightcone gauge is better suited to this and should yield definitive answers which would help us concretise the claims of the preliminary analysis in \cite{Bagchi:2020fpr}. We are currently working on this. 

\medskip

It would be good to have a cross-check of our results in this paper by an independent analysis. BRST quantisation appears to be very suited for such a verification. There are some older papers which have performed BRST quantisation for the tensionless string, e.g. \cite{Gamboa:1991nj, Bozhilov:1997xq}. It would be good to re-examine these papers from our recent understanding of the three inequivalent vacua.

\medskip

Another very immediate generalisation is to the theory of tensionless superstrings. Here the classical theory itself has two different avatars, where the homogeneous and inhomogeneous Super BMS$_3$ arise as the worldsheet symmetries. It is of interest to understand what canonical quantisation leads to here and how many consistent tensionless quantum theories arise. It would then be important to do the light-cone analysis generalising the work in this paper to determine the critical dimensions of the obtained quantum theories. 

\medskip

Finally, returning to the bosonic analysis, we would like to address an important point we made in the introduction. The tensionless limit on the worldsheet is a rather singular limit. It is then perhaps a valid expectation that this can lead to a similar singular limit on the spacetime on which the string moves. Our present analysis seems to indicate that this is not so, and we can think of these in close analogy with massless point particles which move on null geodesics, but in the same spacetime as the massive particles. It is perhaps also possible that tensionless strings can consistently live in Carrollian spacetimes in lower dimensions. It has been shown Carrollian spacetimes can be embedded in higher dimensional relativistic spacetimes by the so called Eisenhart lift \cite{Duval:2014uoa, Cariglia:2016oft}. Hence it is possible that tensionless strings are consistent in both relativistic and ultra/non-relativistic spacetimes. 

\bigskip

\section*{Acknowledgements}
It is a pleasure to thank Aritra Banerjee, Shankhadeep Chakrabortty, and Pulastya Parekh for discussions and for a continued collaboration on aspects of tensionless strings.  

\smallskip 

\noindent AB's research is supported by a Swarnajayanti fellowship of the Department of Science and Technology and the Science and Engineering Research Board (SERB), India. AB is further supported by the following grants from SERB: EMR/2016/008037, ERC/2017/000873, MTR/2017/000740. MM would like to acknowledge the support provided by the Max Planck Partner Group grant MAXPLA/PHY/2018577.

\newpage
\section*{APPENDICES}
\appendix
\section{Trivial components of Lorentz algebra}\label{Lortriv}
In this appendix, we justify the claim in Sec.~\ref{Lormain} that the only nontrivial components of the Lorentz algebra are the commutations between the boosts, i.e. $\left[J^{i-},J^{j-}\right]$.

\medskip

We begin by noting that $J^{\mu +}$ doesn't have any oscillator contribution analogous to \eqref{Sdef}, i.e. $S^{\mu +}$ = 0, thus
$$J^{\mu +} = L^{\mu +} = x^\mu p^+-x^+p^\mu.$$
With this simple observation, it is very easy to see that the Lorentz algebra components involving $J^{\mu +}$ are obeyed in any of the three vacua. So we are left with following potentially nontrivial ones:
$$\left[J^{ij},J^{kl}\right], \, \left[J^{i-},J^{kl}\right], \,\left[J^{i-},J^{j-}\right].$$
We take care of the first two in this appendix. while the last one is truly nontrivial, leading to interesting Physics that is at the heart of this paper.

\subsection{Induced and flipped vacua}

\subsubsection*{Computation of $[J^{ij},J^{kl}]$}
Similar to $J^{i-}$, the $J^{ij}$ can be split as
\begin{equation}
J^{ij}=L^{ij}+S^{ij}.
\end{equation}
The zero mode contribution is
\begin{equation}\label{Lijdef}
L^{ij}=x^ip^j-x^jp^i,
\end{equation}
and the oscillator contribution is
\begin{equation}\label{Sijdef}
S^{ij} = \frac{i}2\sum_{n\neq 0}\frac{1}n\normord{\left(A^i_{n}B^j_{-n}-A^j_{n}B^i_{-n}\right)}
\end{equation}
First, we have a pretty simple commutator
\begin{equation}\label{LLrr}
\begin{split}
\left[L^{ij},L^{kl}\right]=&\left[\left(x^ip^j-x^jp^i\right),\left(x^kp^l-x^lp^k\right)\right],\\
=&i\left(L^{li}\delta^{jk}+L^{kj}\delta^{il}+L^{ik}\delta^{jl}+L^{jl}\delta^{ik}\right)
\end{split}
\end{equation}
Then we trivially have
\begin{equation}\label{LSindflrr}
\left[L^{ij},S^{kl}\right] = 0
\end{equation}
Since $x^i$ and $p^i$ commute with all $A^i_n$ and $B^i_n$. Now we compute $\left[S^{ij},S^{kl}\right]$. It comprises of four terms, each taking the form
$$\sum_{\substack{m\neq 0\\n \neq 0}}\frac{-1}{4mn}\left[\normord{A^i_{m}B^j_{-m}},\normord{A^k_{n}B^l_{-n}}\right]$$
with some permutation of the spacetime indices. Now, the commutation ``clicks" only when $m=n$. But when $m=n$, $\normord{A^i_{m}B^j_{-m}}$ and $\normord{A^k_{n}B^l_{-n}}$ have the same ordering\footnote{For the induced vacuum the ordering is the same for all $m$ and $n$, so the statement is trivially true when $m=n$.}, and their commutation after eating up one $A$ and one $B$ from either side has the same ordering for the remaining $AB$ product, making it automatically normal ordered. Thus we get
\begin{equation}
\left[\normord{A^i_{m}B^j_{-m}},\normord{A^k_{n}B^l_{-n}}\right] = -2m\left(\normord{A^i_{-m}B^l_{m}}\delta^{jk}-\normord{A^k_{-m}B^j_{m}}\delta^{il}\right)\delta_{m,n}
\end{equation}
Doing this for all four terms,
\begin{equation}\label{SSindflrr}
\begin{split}
\left[S^{ij},S^{kl}\right] =& \sum_{m\neq 0}\frac{1}{2m} \bigg(\left(\normord{A^i_{-m}B^l_{m}}\delta^{jk}-\normord{A^k_{-m}B^j_{m}}\delta^{il}\right)-\left(\normord{A^i_{-m}B^k_{m}}\delta^{jl}-\normord{A^l_{-m}B^j_{m}}\delta^{ik}\right)\\
&-\left(\normord{A^j_{-m}B^l_{m}}\delta^{ik}-\normord{A^k_{-m}B^i_{m}}\delta^{jl}\right)+\left(\normord{A^j_{-m}B^k_{m}}\delta^{il}-\normord{A^l_{-m}B^i_{m}}\delta^{jk}\right)\bigg)\\
=&-\frac{1}2\sum_{m\neq 0}\frac{1}{m} \bigg(\left(\normord{A^l_{-m}B^i_{m}}-\normord{A^i_{-m}B^l_{m}}\right)\delta^{jk}+\left(\normord{A^k_{-m}B^j_{m}}-\normord{A^j_{-m}B^k_{m}}\right)\delta^{il}\\
&+\left(\normord{A^i_{-m}B^k_{m}}-\normord{A^k_{-m}B^i_{m}}\right)\delta^{jl}+\left(\normord{A^j_{-m}B^l_{m}}-\normord{A^l_{-m}B^j_{m}}\right)\delta^{ik}\bigg)\\
=&i\left(S^{li}\delta^{jk}+S^{kj}\delta^{il}+S^{ik}\delta^{jl}+S^{jl}\delta^{ik}\right)
\end{split}
\end{equation}
When we combine \eqref{LLrr}, \eqref{LSindflrr} and \eqref{SSindflrr}, the zero mode parts from \eqref{LLrr} perfectly combine the oscillator parts from \eqref{SSindflrr} to give the full Lorentz generators.
\begin{equation}
\begin{split}
\left[J^{ij},J^{kl}\right] =&\left[L^{ij},L^{kl}\right]+\left[S^{ij},L^{kl}\right]+\left[L^{ij},S^{kl}\right]+\left[S^{ij},S^{kl}\right]
=i\left(J^{li}\delta^{jk}+J^{kj}\delta^{il}+J^{ik}\delta^{jl}+J^{jl}\delta^{ik}\right)
\end{split}
\end{equation}
Thus the spatial rotations in orthogonal directions obey the $SO(D-2)$ Lorentz subalgebra.

\subsubsection*{Computation of $[J^{i-},J^{kl}]$}
As in \eqref{Jsplit}, we separate $J^{i-}$ into zero mode and oscillator modes
$$J^{i-} = L^{i-}+S^{i-},$$
following the definitions \eqref{Ldef} and \eqref{Sdef}. We assume $k\neq l$. Now we compute
\begin{equation}\label{LLbr}
\begin{split}
\left[L^{i-},L^{kl}\right] =& \left[\left(x^ip^--x^-p^i\right),\left(x^kp^l-x^lp^k\right)\right]\\
=& \left[x^i,\left(x^kp^l-x^lp^k\right)\right]p^--x^-\left[p^i,\left(x^kp^l-x^lp^k\right)\right]\\
=& ix^kp^-\delta^{il}-ix^lp^-\delta^{ik}+ix^-p^l\delta^{ik}-ix^-p^k\delta^{il}\\
=& i\left(x^kp^--x^-p^k\right)\delta^{il}-i\left(x^lp^--x^-p^l\right)\delta^{ik}\\
=& i\left(L^{k-}\delta^{il}-L^{l-}\delta^{ik}\right)
\end{split}
\end{equation}
Then we have
\begin{equation}\label{LSbr}
\begin{split}
\left[L^{i-},S^{kl}\right] =& \frac{i}2\sum_{m\neq 0}\frac{1}m\left[\left(x^ip^--x^-p^i\right),\left(\normord{A^k_mB^l_{-m}}-\normord{A^l_mB^k_{-m}}\right)\right] = 0.
\end{split}
\end{equation}
where we have used \eqref{Ldef} for $L^{i-}$. Then we evaluate, making ample use of \eqref{xAB}
\begin{equation}\label{LSrb}
\begin{split}
\left[S^{i-},L^{kl}\right]  &= -\left[L^{kl},S^{i-}\right] = \frac{i}2\sum_{m\neq 0}\frac{1}m\left[\left(x^kp^l-x^lp^k\right),\left(\normord{A^i_{-m}B^-_{m}}+\normord{A^-_mB^i_{-m}}\right)\right]\\
=& -\frac{\sqrt{2c'}}{2B^+_0}\sum_{m\neq 0}\frac{1}m\left(\normord{A^i_{-m}B^k_{m}}+\normord{A^k_mB^i_{-m}}\right)p^l +\frac{\sqrt{2c'}}{2B^+_0}\sum_{m\neq 0}\frac{1}m\left(\normord{A^i_{-m}B^l_{m}}+\normord{A^l_mB^i_{-m}}\right)p^k
\end{split}
\end{equation}
And finally we compute using \eqref{ABij}
\begin{equation}\label{SSbr1}
\begin{split}
\left[S^{i-},S^{kl}\right] 
=& \sum_{\substack{m\neq 0\\n \neq 0}}\frac{1}{4mn}\left[\left(\normord{A^i_{-m}B^-_{m}}+\normord{A^-_mB^i_{-m}}\right),\left(\normord{A^k_nB^l_{-n}}-\normord{A^l_nB^k_{-n}}\right)\right],\\
=& \textcolor[rgb]{1.00,0.00,0.00}{\sum_{m\neq 0}\frac{1}{2m}\left(\normord{A^k_{-m}B^-_{m}}\delta^{il}-\normord{A^-_mB^l_{-m}}\delta^{ik}-\normord{A^l_{-m}B^-_{m}}\delta^{ik}+\normord{A^-_mB^k_{-m}}\delta^{il}\right)}\\
&\textcolor[rgb]{0.00,0.70,0.00}{-\sum_{\substack{m\neq 0\\n \neq 0}}\frac{1}{2mB^+_0}\normord{A^i_{-m}\left(B^k_{m+n}B^l_{-n}-B^l_{m+n}B^k_{-n}\right)}}\\
&\textcolor[rgb]{0.00,0.00,1.00}{-\sum_{\substack{m\neq 0\\n \neq 0\\m\neq  -n}}\frac{1}{2mB^+_0}\normord{\left(A^k_{m+n}B^l_{-n}-A^l_{m+n}B^k_{-n}\right)B^i_{-m}}} \textcolor[rgb]{0.00,0.00,1.00}{+\sum_{\substack{m\neq 0\\n \neq 0}}\frac{1}{2mB^+_0}\normord{\left(A^k_{n}B^l_{m-n}-A^l_{n}B^k_{m-n}\right)B^i_{-m}}}
\end{split}
\end{equation}
Now let's look at the term in \textcolor[rgb]{1.00,0.00,0.00}{red}. These have come from commuting $A^i$s and $B^i$s with $S^{kl}$. It's evident that
\begin{equation}\label{SSbr2}
\begin{split}
&\sum_{m\neq 0}\frac{1}{2m}\left(\normord{A^k_{-m}B^-_{m}}\delta^{il}-\normord{A^-_mB^l_{-m}}\delta^{ik}-\normord{A^l_{-m}B^-_{m}}\delta^{ik}+\normord{A^-_mB^k_{-m}}\delta^{il}\right) =i\left(S^{k-}\delta^{il}-S^{l-}\delta^{ik}\right)
\end{split}
\end{equation}
Then we turn to the terms in \textcolor[rgb]{0.00,0.70,0.00}{green}, whic have resulted from the commutation of $B^-$s with $S^{kl}$. In the second term, we need to shift the dummy index $n \to -(m+n)$, and then a simple algebra gives
\begin{equation}\label{SSbr3}
\begin{split}
\sum_{\substack{m\neq 0\\n \neq 0}}\frac{1}{2mB^+_0}\normord{A^i_{-m}\left(B^k_{m+n}B^l_{-n}-B^l_{m+n}B^k_{-n}\right)}
=&\sum_{m\neq 0}\frac{1}{2mB^+_0}\left(\normord{A^i_{-m}B^l_m}B^k_0-\normord{A^i_{-m}B^k_m}B^l_0\right)\\
=&\frac{\sqrt{2c'}}{2B^+_0}\sum_{m\neq 0}\frac{1}{m}\left(\normord{A^i_{-m}B^l_m}p^k-\normord{A^i_{-m}B^k_m}p^l\right)
\end{split}
\end{equation}
Whereas the remaining terms are indicated in \textcolor[rgb]{0.00,0.00,1.00}{blue}. Notice that the normal ordering for the $A^kB^l$ and $A^lB^k$ terms is removed, since they commute as $k\neq l$. A pretty similar change of dummy index $n$ as in case of the \textcolor[rgb]{0.00,0.70,0.00}{green} terms yields
\begin{equation}\label{SSbr4}
\begin{split}
&\sum_{\substack{m\neq 0\\n \neq 0\\m\neq  -n}}\frac{1}{2mB^+_0}\normord{\left(A^k_{m+n}B^l_{-n}-A^l_{m+n}B^k_{-n}\right)B^i_{-m}} -\sum_{\substack{m\neq 0\\n \neq 0}}\frac{1}{2mB^+_0}\normord{\left(A^k_{n}B^l_{m-n}-A^l_{n}B^k_{m-n}\right)B^i_{-m}}\\
=&-\sum_{m\neq 0}\frac{1}{2mB^+_0}\left(\normord{A^k_mB^i_{-m}}B^l_0-\normord{A^l_mB^i_{-m}}B^k_0\right) = -\frac{\sqrt{2c'}}{2B^+_0}\sum_{m\neq 0}\frac{1}{m}\left(\normord{A^k_mB^i_{-m}}p^l-\normord{A^l_mB^i_{-m}}p^k\right)
\end{split}
\end{equation}
Using \eqref{SSbr2}, \eqref{SSbr3} and \eqref{SSbr4} in \eqref{SSbr1} we get
\begin{equation}\label{SSbr}
\begin{split}
\left[S^{i-},S^{kl}\right]
=&i\left(S^{k-}\delta^{il}-S^{l-}\delta^{ik}\right) +\frac{\sqrt{2c'}}{2B^+_0}\sum_{m\neq 0}\frac{1}m\left(\normord{A^i_{-m}B^k_{m}}+\normord{A^k_mB^i_{-m}}\right)p^l\\
&-\frac{\sqrt{2c'}}{2B^+_0}\sum_{m\neq 0}\frac{1}m\left(\normord{A^i_{-m}B^l_{m}}+\normord{A^l_mB^i_{-m}}\right)p^k\\
\end{split}
\end{equation}
Thus, using \eqref{LLbr}, \eqref{LSbr}, \eqref{LSrb} and \eqref{SSbr}, ultimately we get
\begin{equation}
\begin{split}
\left[J^{i-},J^{kl}\right] =&\left[L^{i-},L^{kl}\right]+\left[S^{i-},L^{kl}\right]+\left[L^{i-},S^{kl}\right]+\left[S^{i-},S^{kl}\right]\\
=& i\left((L^{k-}+S^{k-})\delta^{il}-(L^{l-}+S^{l-})\delta^{ik}\right) = i\left(J^{k-}\delta^{il}-J^{l-}\delta^{ik}\right)
\end{split}
\end{equation}
Thus this component of the Lorentz algebra is also satisfied.

\subsection{Oscillator vacuum}
We treat the oscillator vacuum separately from the induced and fipped vacua because the earlier argument for commutators of normal ordered $AB$ products becomes obscure as we choose to write the quantities in this case in terms of $C$ and $\tilde{C}$ operators. So it is convenient to deal with the explicitly normal ordered expressions.
\subsubsection*{Computation of $[J^{ij},J^{kl}]$}
Here we have, with explicit normal ordering,
\begin{equation}\label{Sijosc}
S^{ij}=i\sum_{n>0}\frac{1}{n}\left(C^j_{-n}C^i_{n}-C^{i}_{-n}C^j_{n}+\tilde{C}^j_{-n}\tilde{C}^i_{n}-\tilde{C}^{i}_{-n}\tilde{C}^j_{n}\right)
\end{equation}
From \eqref{LLrr},
\begin{equation}
[L^{ij},L^{kl}]=i\left(L^{li}\delta^{jk}+L^{kj}\delta^{il}+L^{ik}\delta^{jl}+L^{jl}\delta^{ik}\right)
\end{equation}
Here also it's trivial to see that
\begin{equation}\label{LSoscrr}
\begin{split}
[L^{ij},S^{kl}]=&i\sum_{n>0}\frac{1}{n}\left[\left(x^ip^j-x^jp^i\right),\left(C^j_{-n}C^i_{n}-C^{i}_{-n}C^j_{n}+\tilde{C}^j_{-n}\tilde{C}^i_{n}-\tilde{C}^{i}_{-n}\tilde{C}^j_{n}\right)\right] =0
\end{split}
\end{equation}
Now we have
\begin{equation}
\begin{split}
\left[\frac{1}{n}C^j_{-n}C^i_n,S^{kl}\right]=\sum_{m>0}\left(\frac{i}{nm}\right)& \bigg\{[C^j_{-n}C^i_n,C^l_{-m}C^k_{m}]-[C^j_{-n}C^i_n,C^k_{-m}C^l_{m}] +[C^j_{-n}C^i_n,\tilde{C}^l_{-m}\tilde{C}^k_{m}]-[C^j_{-n}C^i_n,\tilde{C}^k_{-m}\tilde{C}^l_{m}]\bigg\} \\
=\sum_{m>0}\left(\frac{i}{nm}\right)& \bigg\{[C^j_{-n}C^i_n,C^l_{-m}C^k_{m}]-[C^j_{-n}C^i_n,C^k_{-m}C^l_{m}] +[C^j_{-n}C^i_n,\tilde{C}^l_{-m}\tilde{C}^k_{m}]-[C^j_{-n}C^i_n,\tilde{C}^k_{-m}\tilde{C}^l_{m}]\bigg\} \\
=\sum_{m>0}\left(\frac{i}{nm}\right)& \bigg\{[C^j_{-n}C^i_n,C^l_{-m}]C^k_{m}+C^l_{-m}[C^j_{-n}C^i_n,C^k_{m}] -[C^j_{-n}C^i_n,C^k_{-m}]C^l_{m}-C^k_{-m}[C^j_{-n}C^i_n,C^l_{m}]\bigg\} \\
=\left(\frac{i}{n}\right)& \bigg\{C^j_{-n}C^k_{n}\delta^{il}-C^l_{-n}C^i_n\delta^{jk}-C^j_{-n}C^l_{n}\delta^{ik}+C^k_{-n}C^i_n\delta^{jl}\bigg\} \\
\end{split}
\end{equation}
This implies,
\begin{equation}
\begin{split}
\left[\frac{1}{n}C^j_{-n}C^i_n,S^{kl}\right]=\left(\frac{i}{n}\right)& \bigg\{C^j_{-n}C^k_{n}\delta^{il}-C^l_{-n}C^i_n\delta^{jk}-C^j_{-n}C^l_{n}\delta^{ik}+C^k_{-n}C^i_n\delta^{jl}\bigg\}
\end{split}
\end{equation}
Similarly,
\begin{equation}
\begin{split}
\left[\frac{1}{n}\tilde{C}^j_{-n}\tilde{C}^i_n,S^{kl}\right]=\left(\frac{i}{n}\right)& \bigg\{\tilde{C}^j_{-n}\tilde{C}^k_{n}\delta^{il}-\tilde{C}^l_{-n}\tilde{C}^i_n\delta^{jk}-\tilde{C}^j_{-n}\tilde{C}^l_{n}\delta^{ik}+\tilde{C}^k_{-n}\tilde{C}^i_n\delta^{jl}\bigg\}
\end{split}
\end{equation}
This gives us $[S^{ij},S^{kl}]$ to be
\begin{equation}\label{SSoscrr}
\begin{split}
[S^{ij},S^{kl}]=i\sum_{n>0}\frac{1}{n}&[C^j_{-n}C^i_n-C^i_{-n}C^j_n+\tilde{C}^j_{-n}\tilde{C}^i_n-\tilde{C}^i_{-n}\tilde{C}^j_n,S^{kl}] \\
=(-1)\sum_{n>0}\frac{1}{n}&\Bigg[\bigg\{C^j_{-n}C^k_{n}\delta^{il}-C^l_{-n}C^i_n\delta^{jk}-C^j_{-n}C^l_{n}\delta^{ik}+C^k_{-n}C^i_n\delta^{jl}\bigg\} \\
&-\bigg\{C^i_{-n}C^k_{n}\delta^{jl}-C^l_{-n}C^j_n\delta^{ik}-C^i_{-n}C^l_{n}\delta^{jk}+C^k_{-n}C^j_n\delta^{il}\bigg\} \\
&+\bigg\{\tilde{C}^j_{-n}\tilde{C}^k_{n}\delta^{il}-\tilde{C}^l_{-n}\tilde{C}^i_n\delta^{jk}-\tilde{C}^j_{-n}\tilde{C}^l_{n}\delta^{ik}+\tilde{C}^k_{-n}\tilde{C}^i_n\delta^{jl}\bigg\} \\
&-\bigg\{\tilde{C}^i_{-n}\tilde{C}^k_{n}\delta^{jl}-\tilde{C}^l_{-n}\tilde{C}^j_n\delta^{ik}-\tilde{C}^i_{-n}\tilde{C}^l_{n}\delta^{jk}+\tilde{C}^k_{-n}\tilde{C}^j_n\delta^{il}\bigg\} \Bigg]\\
=(-1)\sum_{n>0}\frac{1}{n}&\Bigg[\bigg\{C^j_{-n}C^k_{n}-C^k_{-n}C^j_n+\tilde{C}^j_{-n}\tilde{C}^k_{n}-\tilde{C}^k_{-n}\tilde{C}^j_n \bigg\}\delta^{il} \\
&+\bigg\{-C^l_{-n}C^i_n+C^i_{-n}C^l_{n}-\tilde{C}^l_{-n}\tilde{C}^i_n+\tilde{C}^i_{-n}\tilde{C}^l_{n} \bigg\}\delta^{jk} \\
&+\bigg\{-C^j_{-n}C^l_{n}+C^l_{-n}C^j_n-\tilde{C}^j_{-n}\tilde{C}^l_{n}+\tilde{C}^l_{-n}\tilde{C}^j_n \bigg\}\delta^{ik} \\
&+\bigg\{+C^k_{-n}C^i_n-C^i_{-n}C^k_{n}+\tilde{C}^k_{-n}\tilde{C}^i_n-\tilde{C}^i_{-n}\tilde{C}^k_{n}\bigg\}\delta^{jl}\Bigg] \\
=&i\Bigg(S^{kj}\delta^{il}+S^{li}\delta^{jk}+S^{jl}\delta^{ik}+S^{ik}\delta^{jl}\Bigg)
\end{split}
\end{equation}
Finally, combining \eqref{LLrr}, \eqref{LSoscrr} and \eqref{SSoscrr},
\begin{equation}
\begin{split}
[J^{ij},J^{kl}]=&i\bigg(L^{li}\delta^{jk}+L^{kj}\delta^{il}+L^{ik}\delta^{jl}+L^{jl}\delta^{ik}\bigg)+i\bigg(S^{kj}\delta^{il}+S^{li}\delta^{jk}+S^{jl}\delta^{ik}+S^{ik}\delta^{jl}\bigg) \\
=&i\bigg(J^{kj}\delta^{il}+J^{li}\delta^{jk}+J^{jl}\delta^{ik}+J^{ik}\delta^{jl}\bigg) \\
\end{split}
\end{equation}
Thus here as well the spatial rotations in orthogonal directions obey the $SO(D-2)$ Lorentz subalgebra.

\subsubsection*{Computation of $[J^{ij},J^{k-}]$: Oscillator Case}

Writing the expanded form for the angular momentum generators, we see
\begin{equation}\label{Jijkmin}
\begin{split}
[J^{ij},J^{k-}] = \left[L^{ij}+S^{ij},L^{k-}+S^{k-}\right] = \left[L^{ij},L^{k-}\right]+\left[L^{ij},S^{k-}\right]+\left[S^{ij},L^{k-}\right]+\left[S^{ij},S^{k-}\right]
\end{split}
\end{equation}
where $L^{ij}$ and $S^{ij}$ are defined in \eqref{Lijdef} and \eqref{Sijosc} respectively, while $L^{k-}$ and $S^{k-}$ are defined in \eqref{Ldef} and \eqref{Sosc} respectively. First, we calculate
\begin{equation}\label{LLrbosc}
\begin{split}
[L^{ij},L^{k-}] =& [x^ip^j-x^jp^i, x^kp^--x^-p^k] 
= x^i\left[p^j,x^k\right]p^--x^-\left[x^i,p^k\right]p^j-x^j\left[p^i,x^k\right]p^-+x^-\left[x^j,p^k\right]p^i\\
=& i\left(\delta^{ik}\left(x^jp^--x^-p^j\right)-\delta^{jk}\left(x^ip^--x^-p^i\right)\right) = i\left(\delta^{ik}L^{j-}-\delta^{jk}L^{i-}\right)\\
\end{split}
\end{equation}

Then we turn to $\left[L^{ij},S^{k-}\right]$.
\begin{equation}
\begin{split}
\left[x^ip^j,S^{k-}\right] 
=& -\sum_{n>0}\frac{i}n\left[x^i,\left(C^k_{-n}C^-_n-\tilde{C}^-_{-n}\tilde{C}^k_n\right)\right]p^j -\sum_{n<0}\frac{i}n\left[x^i,\left(C^-_nC^k_{-n}-\tilde{C}^k_{n}\tilde{C}^-_{-n}\right)\right]p^j\\
=& -\sum_{n>0}\frac{i}n\left(C^k_{-n}\left[x^i,C^-_n\right]-\left[x^i,\tilde{C}^-_{-n}\right]\tilde{C}^k_n\right)p^j -\sum_{n<0}\frac{i}n\left(\left[x^i,C^-_n\right]C^k_{-n}-\tilde{C}^k_{n}\left[x^i,\tilde{C}^-_{-n}\right]\right)p^j\\
=& \sum_{n>0}\frac{\sqrt{2c'}}{nB^+_0}\left(C^k_{-n}C^i_n-\tilde{C}^i_{-n}\tilde{C}^k_n\right)p^j +\sum_{n<0}\frac{\sqrt{2c'}}{nB^+_0}\left(C^i_nC^k_{-n}-\tilde{C}^k_{n}\tilde{C}^i_{-n}\right)p^j = -i\frac{\sqrt{2c'}}{B^+_0}S^{ik}p^j
\end{split}
\end{equation}
Therefore
\begin{equation}\label{LSrbosc}
\begin{split}
\left[L^{ij},S^{k-}\right] &= \left[x^ip^j,S^{k-}\right] - \left[x^jp^i,S^{k-}\right] = -i\frac{\sqrt{2c'}}{B^+_0}\left(S^{ik}p^j-S^{jk}p^i\right)
\end{split}
\end{equation}

Next, we can see that $\left[S^{ij},L^{k-}\right]=0$ trivially since $x^-$, $p^k$, $x^k$ and $p^-$ commute with the transverse $C$ and $\tilde{C}$ oscillators and therefore with $S^{ij}$.

Finally, we compute $[S^{ij},S^{k-}]$.
\begin{equation}
\begin{split}
[C^i_{n},S^{k-}]= \frac{i}{2B^+_0}&\bigg[\sum_{m>0}\left(\frac{n}{m}\right)\Big[(B^i_{n-m}+2C^i_{n-m})C^j_m-C^j_{-m}(B^i_{n+m}+2C^i_{n+m})   \\
&+(B^i_{n+m}-2C^i_{n+m})\tilde{C}^j_{m}-\tilde{C}^j_{-m}(B^i_{n-m}-2C^i_{n-m})\Big]-\delta^{ij}C^-_{n} \bigg],
\end{split}
\end{equation}

\begin{equation}
\begin{split}
[\tilde{C}^i_{n},S^{j-}]= \frac{i}{2B^+_0}& \bigg[\sum_{m>0}\left(\frac{n}{m}\right)\Big[(B^i_{-n-m}-2\tilde{C}^i_{n+m})C^j_m-C^j_{-m}(B^i_{m-n}-2\tilde{C}^i_{n-m}) \\
&+(B^i_{m-n}+2\tilde{C}^i_{n-m})\tilde{C}^j_{m}-\tilde{C}^j_{-m}(B^i_{-n-m}+2\tilde{C}^i_{n+m})\Big]-\delta^{ij}\tilde{C}^-_{n} \bigg],
\end{split}
\end{equation}

Let us begin writing the prime commutator now.
\begin{equation}
\begin{split}
[S^{ij},S^{k-}] =& i\sum_{n>0}\frac{1}{n}\left[C^j_{-n}C^i_{n}-C^i_{-n}C^j_{n}+\tilde{C}^j_{-n}\tilde{C}^i_{n}-\tilde{C}^i_{-n}\tilde{C}^j_{n},S^{k-}\right] \\
=& i\sum_{n>0}\frac{1}{n}\bigg\{[C^j_{-n}C^i_{n},S^{k-}]-[C^i_{-n}C^j_{n},S^{k-}]+[\tilde{C}^j_{-n}\tilde{C}^i_{n},S^{k-}]-[\tilde{C}^i_{-n}\tilde{C}^j_{n},S^{k-}]\bigg\} \\
=& i\sum_{n>0}\frac{1}{n}\bigg\{[C^j_{-n},S^{k-}]C^i_{n}+C^j_{-n}[C^i_{n},S^{k-}]-[C^i_{-n},S^{k-}]C^j_{n}-C^i_{-n}[C^j_{n},S^{k-}] \\
& \hspace{1cm}+[\tilde{C}^j_{-n},S^{k-}]\tilde{C}^i_{n}+\tilde{C}^j_{-n}[\tilde{C}^i_{n},S^{k-}]-[\tilde{C}^i_{-n},S^{k-}]\tilde{C}^j_{n}-\tilde{C}^i_{-n}[\tilde{C}^j_{n},S^{k-}]\bigg\} \\
\end{split}
\end{equation}

Expanding the commutators, we get

\begin{equation}
\begin{split}
[S^{ij},S^{k-}] =& \frac{-1}{2B^+_0}\sum_{n,m>0}\frac{1}{m}\bigg\{-\Big[(B^j_{-n-m}+2C^j_{-n-m})C^k_m-C^k_{-m}(B^j_{m-n}+2C^j_{m-n}) \\
				 & \hspace{2cm}+(B^j_{m-n}-2C^j_{m-n})\tilde{C}^k_{m}-\tilde{C}^k_{-m}(B^j_{-n-m}-2C^j_{-n-m})\Big]C^i_{n} \\
				 & \hspace{1cm}+C^j_{-n}\Big[(B^i_{n-m}+2C^i_{n-m})C^k_m-C^k_{-m}(B^i_{n+m}+2C^i_{n+m}) \\
				 & \hspace{2cm}+(B^i_{n+m}-2C^i_{n+m})\tilde{C}^k_{m}-\tilde{C}^k_{-m}(B^i_{n-m}-2C^i_{n-m})\Big] \\
				 & \hspace{1cm}+\Big[(B^i_{-n-m}+2C^i_{-n-m})C^k_m-C^k_{-m}(B^i_{m-n}+2C^i_{m-n}) \\
				 & \hspace{2cm}+(B^i_{m-n}-2C^i_{m-n})\tilde{C}^k_{m}-\tilde{C}^k_{-m}(B^i_{-n-m}-2C^i_{-n-m})\Big]C^j_{n} \\
				 & \hspace{1cm}-C^i_{-n}\Big[(B^j_{n-m}+2C^j_{n-m})C^k_m-C^k_{-m}(B^j_{n+m}+2C^j_{n+m}) \\
				 & \hspace{2cm}+(B^j_{n+m}-2C^j_{n+m})\tilde{C}^k_{m}-\tilde{C}^k_{-m}(B^j_{n-m}-2C^j_{n-m})\Big] \\
				 & \hspace{1cm}-\Big[(B^j_{n-m}-2\tilde{C}^j_{m-n})C^k_m-C^k_{-m}(B^j_{m+n}-2\tilde{C}^j_{-n-m}) \\
				 & \hspace{2cm}+(B^j_{m+n}+2\tilde{C}^j_{-n-m})\tilde{C}^k_{m}-\tilde{C}^k_{-m}(B^j_{n-m}+2\tilde{C}^j_{m-n})\Big]\tilde{C}^i_{n} \\
				 & \hspace{1cm}+\tilde{C}^j_{-n}\Big[(B^i_{-n-m}-2\tilde{C}^i_{n+m})C^k_m-C^k_{-m}(B^i_{m-n}-2\tilde{C}^i_{n-m}) \\
				 & \hspace{2cm}+(B^i_{m-n}+2\tilde{C}^i_{n-m})\tilde{C}^k_{m}-\tilde{C}^k_{-m}(B^i_{-n-m}+2\tilde{C}^i_{n+m})\Big] \\
				 & \hspace{1cm}+\Big[(B^i_{n-m}-2\tilde{C}^i_{m-n})C^k_m-C^k_{-m}(B^i_{m+n}-2\tilde{C}^i_{-n-m}) \\
				 & \hspace{2cm}+(B^i_{m+n}+2\tilde{C}^i_{-n-m})\tilde{C}^k_{m}-\tilde{C}^k_{-m}(B^i_{n-m}+2\tilde{C}^i_{m-n})\Big]\tilde{C}^j_{n} \\
				 & \hspace{1cm}-\tilde{C}^i_{-n}\Big[(B^j_{-n-m}-2\tilde{C}^j_{n+m})C^k_m-C^k_{-m}(B^j_{m-n}-2\tilde{C}^j_{n-m}) \\
				 & \hspace{2cm}+(B^j_{m-n}+2\tilde{C}^j_{n-m})\tilde{C}^k_{m}-\tilde{C}^k_{-m}(B^j_{-n-m}+2\tilde{C}^j_{n+m})\Big]\bigg\} \\
				 & +i\sum_{n>0}\frac{i}{n}\bigg(-C^-_{-n}C^i_{n}\delta^{jk}+C^i_{-n}C^-_{n}\delta^{jk}-\tilde{C}^-_{-n}\tilde{C}^i_{n}\delta^{jk}+\tilde{C}^i_{-n}\tilde{C}^-_{n}\delta^{jk} \\
				 & \hspace{2cm}-C^j_{-n}C^-_{n}\delta^{ik}+C^-_{-n}C^j_{n}\delta^{ik}-\tilde{C}^j_{-n}\tilde{C}^-_{n}\delta^{ik}+\tilde{C}^-_{-n}\tilde{C}^j_{n}\delta^{ik}\bigg).
\end{split}
\end{equation}
Rearranging terms and compiling those that have similar structure to $C^k_m$.
\begin{equation}
\begin{split}
[S^{ij},S^{k-}] =& \frac{-1}{2B^+_0}\sum_{n,m>0}\frac{1}{m}\bigg\{\Big[-(B^j_{-n-m}+2C^j_{-n-m})C^i_{n}+C^j_{-n}(B^i_{n-m}+2C^i_{n-m}) \\
				 & \hspace{2cm}+(B^i_{-n-m}+2C^i_{-n-m})C^j_{n}-C^i_{-n}(B^j_{n-m}+2C^j_{n-m}) \\
				 & \hspace{2cm}-(B^j_{n-m}-2\tilde{C}^j_{m-n})\tilde{C}^i_{n}+\tilde{C}^j_{-n}(B^i_{-n-m}-2\tilde{C}^i_{n+m}) \\
				 & \hspace{2cm}+(B^i_{n-m}-2\tilde{C}^i_{m-n})\tilde{C}^j_{n}-\tilde{C}^i_{-n}(B^j_{-n-m}-2\tilde{C}^j_{n+m})\Big]C^k_m \\
				 & \hspace{1cm}+C^k_{-m}\Big[-(B^i_{m-n}+2C^i_{m-n})C^j_{n}+C^i_{-n}(B^j_{n+m}+2C^j_{n+m}) \\
				 & \hspace{2cm}+(B^j_{m-n}+2C^j_{m-n})C^i_{n}-C^j_{-n}(B^i_{n+m}+2C^i_{n+m}) \\
				 & \hspace{2cm}+(B^j_{m+n}-2\tilde{C}^j_{-n-m})\tilde{C}^i_{n}-\tilde{C}^j_{-n}(B^i_{m-n}-2\tilde{C}^i_{n-m}) \\
				 & \hspace{2cm}-(B^i_{m+n}-2\tilde{C}^i_{-n-m})\tilde{C}^j_{n}+\tilde{C}^i_{-n}(B^j_{m-n}-2\tilde{C}^j_{n-m})\Big] \\
				 & \hspace{1cm}+\Big[(B^i_{m-n}-2C^i_{m-n})C^j_{n}-(B^j_{m-n}-2C^j_{m-n})C^i_{n} \\
				 & \hspace{2cm}-(B^j_{m+n}+2\tilde{C}^j_{-n-m})\tilde{C}^i_{n}+C^j_{-n}(B^i_{n+m}-2C^i_{n+m}) \\
				 & \hspace{2cm}+(B^i_{m+n}+2\tilde{C}^i_{-n-m})\tilde{C}^j_{n}-C^i_{-n}(B^j_{n+m}-2C^j_{n+m}) \\
				 & \hspace{2cm}+\tilde{C}^j_{-n}(B^i_{m-n}+2\tilde{C}^i_{n-m})-\tilde{C}^i_{-n}(B^j_{m-n}+2\tilde{C}^j_{n-m})\Big]\tilde{C}^k_{m} \\
				 & \hspace{1cm}+\tilde{C}^k_{-m}\Big[(B^j_{n-m}+2\tilde{C}^j_{m-n})\tilde{C}^i_{n}+(B^j_{-n-m}-2C^j_{-n-m})C^i_{n} \\
				 & \hspace{2cm}-(B^i_{n-m}+2\tilde{C}^i_{m-n})\tilde{C}^j_{n}-(B^i_{-n-m}-2C^i_{-n-m})C^j_{n} \\
				 & \hspace{2cm}-C^j_{-n}(B^i_{n-m}-2C^i_{n-m})+C^i_{-n}(B^j_{n-m}-2C^j_{n-m}) \\
				 & \hspace{2cm}+\tilde{C}^i_{-n}(B^j_{-n-m}+2\tilde{C}^j_{n+m})-\tilde{C}^j_{-n}(B^i_{-n-m}+2\tilde{C}^i_{n+m})\Big]\bigg\} \\
				 & +i\sum_{n>0}\frac{i}{n}\bigg(-C^-_{-n}C^i_{n}\delta^{jk}+C^i_{-n}C^-_{n}\delta^{jk}-\tilde{C}^-_{-n}\tilde{C}^i_{n}\delta^{jk}+\tilde{C}^i_{-n}\tilde{C}^-_{n}\delta^{jk} \\
				 & \hspace{2cm}-C^j_{-n}C^-_{n}\delta^{ik}+C^-_{-n}C^j_{n}\delta^{ik}-\tilde{C}^j_{-n}\tilde{C}^-_{n}\delta^{ik}+\tilde{C}^-_{-n}\tilde{C}^j_{n}\delta^{ik}\bigg).
\end{split}
\end{equation}
First, we expand the $B$'s in terms of $C,\tilde{C}$'s. To get more cancellations and in trying to bring the expression in a familiar form, we shift $(C/\tilde{C})^i$'s indices to "$n$ or $-n$".

\begin{align}
[S^{ij},S^{k-}] =& \frac{-1}{2B^+_0}\sum_{n,m>0}\frac{1}{m}\bigg\{\Big[-(\tilde{C}^j_{n+m}+3C^j_{-n-m})C^i_{n}-C^i_{-n}(\tilde{C}^j_{m-n}+3C^j_{n-m}) \nonumber \\
				 & \hspace{2cm}-(C^j_{n-m}-\tilde{C}^j_{m-n})\tilde{C}^i_{n}-\tilde{C}^i_{-n}(C^j_{-n-m}-\tilde{C}^j_{n+m})\Big]C^k_m \nonumber \\
				 & \hspace{1cm}+C^k_{-m}\Big[(\tilde{C}^j_{n-m}+3C^j_{m-n})C^i_{n}+C^i_{-n}(\tilde{C}^j_{-n-m}+3C^j_{n+m}) \nonumber \\
				 & \hspace{2cm}+(C^j_{m+n}-\tilde{C}^j_{-n-m})\tilde{C}^i_{n}+\tilde{C}^i_{-n}(C^j_{m-n}-\tilde{C}^j_{n-m})\Big] \nonumber\\
				 & \hspace{1cm}+\Big[-(C^j_{m+n}+3\tilde{C}^j_{-n-m})\tilde{C}^i_{n}-(C^j_{m-n}-C^j_{m-n})C^i_{n} \nonumber \\
				 & \hspace{2cm}-\tilde{C}^i_{-n}(C^j_{m-n}+3\tilde{C}^j_{n-m})-C^i_{-n}(\tilde{C}^j_{-n-m}-C^j_{n+m})\Big]\tilde{C}^k_{m} \nonumber \\
				 & \hspace{1cm}+\tilde{C}^k_{-m}\Big[(C^j_{n-m}+3\tilde{C}^j_{m-n})\tilde{C}^i_{n}+(\tilde{C}^j_{n+m}-C^j_{-n-m})C^i_{n} \nonumber\\
				 & \hspace{2cm}+\tilde{C}^i_{-n}(C^j_{-n-m}+3\tilde{C}^j_{n+m})+C^i_{-n}(\tilde{C}^j_{m-n}-C^j_{n-m})\Big]\bigg\} \nonumber \\
				 & +\frac{-1}{2B^+_0}\sum_{n,m>0}\frac{1}{m}\bigg\{\Big[(\tilde{C}^i_{n}+3C^i_{-n})C^j_{n-m}+\tilde{C}^j_{m-n}(C^i_{-n}-\tilde{C}^i_{n})\Big]C^k_m \nonumber\\
				 & \hspace{2cm}+C^k_{-m}\Big[-C^j_{m-n}(\tilde{C}^i_{-n}+3C^i_{n})-(C^i_{n}-\tilde{C}^i_{-n})\tilde{C}^j_{n-m}\Big] \nonumber \\
				 & \hspace{2cm}+\Big[C^j_{m-n}(\tilde{C}^i_{-n}-C^i_{n})+(C^i_{n}+3\tilde{C}^i_{-n})\tilde{C}^j_{n-m}\Big]\tilde{C}^k_{m} \nonumber \\
				 & \hspace{2cm}+\tilde{C}^k_{-m}\Big[-\tilde{C}^j_{m-n}(C^i_{-n}+3\tilde{C}^i_{n})-(\tilde{C}^i_{n}-C^i_{-n})C^j_{n-m}\Big] \nonumber \\
				 & \hspace{2cm}+\Big[(C^i_{n}-\tilde{C}^i_{-n})\tilde{C}^j_{n+m}+C^j_{-n-m}(\tilde{C}^i_{-n}+3C^i_{n})\Big]C^k_m \nonumber \\
				 & \hspace{2cm}+C^k_{-m}\Big[-\tilde{C}^j_{-n-m}(C^i_{-n}-\tilde{C}^i_{n})-(\tilde{C}^i_{n}+3C^i_{-n})C^j_{n+m}\Big] \nonumber \\
				 & \hspace{2cm}+\Big[\tilde{C}^j_{-n-m}(C^i_{-n}+3\tilde{C}^i_{n})+(\tilde{C}^i_{n}-C^i_{-n})C^j_{n+m}\Big]\tilde{C}^k_{m} \nonumber \\
				 & \hspace{2cm}+\tilde{C}^k_{-m}\Big[-C^j_{-n-m}(\tilde{C}^i_{-n}-C^i_{n})-(C^i_{n}+3\tilde{C}^i_{-n})\tilde{C}^j_{n+m}\Big]\bigg\} \nonumber \\
				 & -\frac{-1}{2B^+_0}\sum^{n\leq m}_{n,m>0}\frac{1}{m}\bigg\{\Big[(\tilde{C}^i_{n}+3C^i_{-n})C^j_{n-m}+\tilde{C}^j_{m-n}(C^i_{-n}-\tilde{C}^i_{n})\Big]C^k_m  \nonumber \\
				 & \hspace{2cm}+C^k_{-m}\Big[-C^j_{m-n}(\tilde{C}^i_{-n}+3C^i_{n})-(C^i_{n}-\tilde{C}^i_{-n})\tilde{C}^j_{n-m}\Big] \nonumber \\
				 & \hspace{2cm}+\Big[+C^j_{m-n}(\tilde{C}^i_{-n}-C^i_{n})+(C^i_{n}+3\tilde{C}^i_{-n})\tilde{C}^j_{n-m}\Big]\tilde{C}^k_{m} \nonumber \\
				 & \hspace{2cm}+\tilde{C}^k_{-m}\Big[-\tilde{C}^j_{m-n}(C^i_{-n}+3\tilde{C}^i_{n})-(\tilde{C}^i_{n}-C^i_{-n})C^j_{n-m}\Big]\bigg\} \nonumber \\
				 & +\frac{-1}{2B^+_0}\sum^{n\leq 0}_{m>0,n>-m}\frac{1}{m}\bigg\{\Big[(C^i_{n}-\tilde{C}^i_{-n})\tilde{C}^j_{n+m}+C^j_{-n-m}(\tilde{C}^i_{-n}+3C^i_{n})\Big]C^k_m \nonumber \\
				 & \hspace{2cm}+C^k_{-m}\Big[-\tilde{C}^j_{-n-m}(C^i_{-n}-\tilde{C}^i_{n})-(\tilde{C}^i_{n}+3C^i_{-n})C^j_{n+m}\Big] \nonumber \\
				 & \hspace{2cm}+\Big[+\tilde{C}^j_{-n-m}(C^i_{-n}+3\tilde{C}^i_{n})+(\tilde{C}^i_{n}-C^i_{-n})C^j_{n+m}\Big]\tilde{C}^k_{m} \nonumber \\
				 & \hspace{2cm}+\tilde{C}^k_{-m}\Big[-C^j_{-n-m}(\tilde{C}^i_{-n}-C^i_{n})-(C^i_{n}+3\tilde{C}^i_{-n})\tilde{C}^j_{n+m}\Big]\bigg\} \nonumber \\
				 & +i\sum_{n>0}\frac{i}{n}\bigg(-C^-_{-n}C^i_{n}\delta^{jk}+C^i_{-n}C^-_{n}\delta^{jk}-\tilde{C}^-_{-n}\tilde{C}^i_{n}\delta^{jk}+\tilde{C}^i_{-n}\tilde{C}^-_{n}\delta^{jk} \nonumber \\
				 & \hspace{2cm}-C^j_{-n}C^-_{n}\delta^{ik}+C^-_{-n}C^j_{n}\delta^{ik}-\tilde{C}^j_{-n}\tilde{C}^-_{n}\delta^{ik}+\tilde{C}^-_{-n}\tilde{C}^j_{n}\delta^{ik}\bigg).
\end{align}

First two major brackets cancel with each other exactly. The result is the remainder. After changing the sign of '$n$' in the second bracket so that that it takes only positive values, we get

\begin{equation}
\begin{split}
[S^{ij},S^{k-}] =& -\frac{-1}{2B^+_0}\sum^{n\leq m}_{n,m>0}\frac{1}{m}\bigg\{\Big[(\tilde{C}^i_{n}+3C^i_{-n})C^j_{n-m}+\tilde{C}^j_{m-n}(C^i_{-n}-\tilde{C}^i_{n})\Big]C^k_m \\
				 & \hspace{2cm}+C^k_{-m}\Big[-C^j_{m-n}(\tilde{C}^i_{-n}+3C^i_{n})-(C^i_{n}-\tilde{C}^i_{-n})\tilde{C}^j_{n-m}\Big] \\
				 & \hspace{2cm}+\Big[C^j_{m-n}(\tilde{C}^i_{-n}-C^i_{n})+(C^i_{n}+3\tilde{C}^i_{-n})\tilde{C}^j_{n-m}\Big]\tilde{C}^k_{m} \\
				 & \hspace{2cm}+\tilde{C}^k_{-m}\Big[-\tilde{C}^j_{m-n}(C^i_{-n}+3\tilde{C}^i_{n})-(\tilde{C}^i_{n}-C^i_{-n})C^j_{n-m}\Big]\bigg\} \\
				 & +\frac{-1}{2B^+_0}\sum^{n<m}_{m>0,n\geq 0}\frac{1}{m}\bigg\{\Big[(C^i_{-n}-\tilde{C}^i_{n})\tilde{C}^j_{m-n}+C^j_{n-m}(\tilde{C}^i_{n}+3C^i_{-n})\Big]C^k_m \\
				 & \hspace{2cm}+C^k_{-m}\Big[-\tilde{C}^j_{n-m}(C^i_{n}-\tilde{C}^i_{-n})-(\tilde{C}^i_{-n}+3C^i_{n})C^j_{m-n}\Big] \\
				 & \hspace{2cm}+\Big[\tilde{C}^j_{n-m}(C^i_{n}+3\tilde{C}^i_{-n})+(\tilde{C}^i_{-n}-C^i_{n})C^j_{m-n}\Big]\tilde{C}^k_{m} \\
				 & \hspace{2cm}+\tilde{C}^k_{-m}\Big[-C^j_{n-m}(\tilde{C}^i_{n}-C^i_{-n})-(C^i_{-n}+3\tilde{C}^i_{n})\tilde{C}^j_{m-n}\Big]\bigg\} \\
				 & +i\sum_{n>0}\frac{i}{n}\bigg\{-C^-_{-n}C^i_{n}\delta^{jk}+C^i_{-n}C^-_{n}\delta^{jk}-\tilde{C}^-_{-n}\tilde{C}^i_{n}\delta^{jk}+\tilde{C}^i_{-n}\tilde{C}^-_{n}\delta^{jk} \\
				 & \hspace{2cm}-C^j_{-n}C^-_{n}\delta^{ik}+C^-_{-n}C^j_{n}\delta^{ik}-\tilde{C}^j_{-n}\tilde{C}^-_{n}\delta^{ik}+\tilde{C}^-_{-n}\tilde{C}^j_{n}\delta^{ik}\bigg\}
\end{split}
\end{equation}

As is easily seen, the terms in the two brackets gives commutator of the terms inside the  square brackets apart from the $n=m$-component of the first and $n=0$-component of the second. Further, all commutators of first bracket are identically "zero". Once all that's taken into account, we get:

\begin{equation}
\begin{split}
[S^{ij},S^{k-}] =& +\frac{-1}{2B^+_0}\sum_{m>0}\frac{1}{m}\bigg\{\Big[-(\tilde{C}^i_{m}+3C^i_{-m})C^j_{0}-\tilde{C}^j_{0}(C^i_{-m}-\tilde{C}^i_{m})\Big]C^k_m \\
				 & \hspace{2cm}+C^k_{-m}\Big[C^j_{0}(\tilde{C}^i_{-m}+3C^i_{m})+(C^i_{m}-\tilde{C}^i_{-m})\tilde{C}^j_{0}\Big] \\
				 & \hspace{2cm}+\Big[-C^j_{0}(\tilde{C}^i_{-m}-C^i_{m})-(C^i_{m}+3\tilde{C}^i_{-m})\tilde{C}^j_{0}\Big]\tilde{C}^k_{m} \\
				 & \hspace{2cm}+\tilde{C}^k_{-m}\Big[\tilde{C}^j_{0}(C^i_{-m}+3\tilde{C}^i_{m})+(\tilde{C}^i_{m}-C^i_{-m})C^j_{0}\Big]\bigg\} \\
				 & +\frac{-1}{2B^+_0}\sum_{m>0}\frac{1}{m}\bigg\{\Big[(C^i_{0}-\tilde{C}^i_{0})\tilde{C}^j_{m}+C^j_{-m}(\tilde{C}^i_{0}+3C^i_{0})\Big]C^k_m \\
				 & \hspace{2cm}+C^k_{-m}\Big[-\tilde{C}^j_{-m}(C^i_{0}-\tilde{C}^i_{0})-(\tilde{C}^i_{0}+3C^i_{0})C^j_{m}\Big] \\
				 & \hspace{2cm}+\Big[\tilde{C}^j_{-m}(C^i_{0}+3\tilde{C}^i_{0})+(\tilde{C}^i_{0}-C^i_{0})C^j_{m}\Big]\tilde{C}^k_{m} \\
				 & \hspace{2cm}+\tilde{C}^k_{-m}\Big[-C^j_{-m}(\tilde{C}^i_{0}-C^i_{0})-(C^i_{0}+3\tilde{C}^i_{0})\tilde{C}^j_{m}\Big]\bigg\} \\
				 & +i\sum_{n>0}\frac{i}{n}\bigg(-C^-_{-n}C^i_{n}\delta^{jk}+C^i_{-n}C^-_{n}\delta^{jk}-\tilde{C}^-_{-n}\tilde{C}^i_{n}\delta^{jk}+\tilde{C}^i_{-n}\tilde{C}^-_{n}\delta^{jk} \\
				 & \hspace{2cm}-C^j_{-n}C^-_{n}\delta^{ik}+C^-_{-n}C^j_{n}\delta^{ik}-\tilde{C}^j_{-n}\tilde{C}^-_{n}\delta^{ik}+\tilde{C}^-_{-n}\tilde{C}^j_{n}\delta^{ik}\bigg)
\end{split}
\end{equation}

Since, the terms commutes in the first bracket for "$m\neq 0$", the first bracket is basically zero.

\begin{equation}
\begin{split}
[S^{ij},S^{k-}] =& +\frac{-1}{2B^+_0}\sum_{m>0}\frac{1}{m}\bigg\{\Big[-(\tilde{C}^i_{m}+3C^i_{-m})C^j_{0}-\tilde{C}^j_{0}(C^i_{-m}-\tilde{C}^i_{m})\Big]C^k_m \\
				 & +C^k_{-m}\Big[C^j_{0}(\tilde{C}^i_{-m}+3C^i_{m})+(C^i_{m}-\tilde{C}^i_{-m})\tilde{C}^j_{0}\Big] +\Big[-C^j_{0}(\tilde{C}^i_{-m}-C^i_{m})-(C^i_{m}+3\tilde{C}^i_{-m})\tilde{C}^j_{0}\Big]\tilde{C}^k_{m} \\
				 & +\tilde{C}^k_{-m}\Big[\tilde{C}^j_{0}(C^i_{-m}+3\tilde{C}^i_{m})+(\tilde{C}^i_{m}-C^i_{-m})C^j_{0}\Big] +\Big[(C^i_{0}-\tilde{C}^i_{0})\tilde{C}^j_{m}+C^j_{-m}(\tilde{C}^i_{0}+3C^i_{0})\Big]C^k_m \\
				 & +C^k_{-m}\Big[-\tilde{C}^j_{-m}(C^i_{0}-\tilde{C}^i_{0})-(\tilde{C}^i_{0}+3C^i_{0})C^j_{m}\Big] +\Big[+\tilde{C}^j_{-m}(C^i_{0}+3\tilde{C}^i_{0})+(\tilde{C}^i_{0}-C^i_{0})C^j_{m}\Big]\tilde{C}^k_{m} \\
				 & +\tilde{C}^k_{-m}\Big[-C^j_{-m}(\tilde{C}^i_{0}-C^i_{0})-(C^i_{0}+3\tilde{C}^i_{0})\tilde{C}^j_{m}\Big]\bigg\} \\
				 & +i\sum_{n>0}\frac{i}{n}\bigg(-C^-_{-n}C^i_{n}\delta^{jk}+C^i_{-n}C^-_{n}\delta^{jk}-\tilde{C}^-_{-n}\tilde{C}^i_{n}\delta^{jk}+\tilde{C}^i_{-n}\tilde{C}^-_{n}\delta^{jk} \\
				 & \hspace{2cm}-C^j_{-n}C^-_{n}\delta^{ik}+C^-_{-n}C^j_{n}\delta^{ik}-\tilde{C}^j_{-n}\tilde{C}^-_{n}\delta^{ik}+\tilde{C}^-_{-n}\tilde{C}^j_{n}\delta^{ik}\bigg)
\end{split}
\end{equation}

Using the condition that $C_0=\tilde{C}_0$, we get

\begin{equation}
\begin{split}
[S^{ij},S^{k-}] =& +\frac{-1}{2B^+_0}\sum_{m>0}\frac{1}{m}\bigg\{4\left(-C^i_{-m}C^k_m+C^k_{-m}C^i_{m}-\tilde{C}^i_{-m}\tilde{C}^k_{m}+\tilde{C}^k_{-m}\tilde{C}^i_{m}\right)C^j_{0} \\
				 & \hspace{2cm}+4\left(C^j_{-m}C^k_m-C^k_{-m}C^j_{m}+\tilde{C}^j_{-m}\tilde{C}^k_{m}-\tilde{C}^k_{-m}\tilde{C}^j_{m}\right)C^i_{0}\bigg\} \\
				 & +i\sum_{n>0}\frac{i}{n}\bigg\{\left(-C^-_{-n}C^i_{n}+C^i_{-n}C^-_{n}-\tilde{C}^-_{-n}\tilde{C}^i_{n}+\tilde{C}^i_{-n}\tilde{C}^-_{n}\right)\delta^{jk} \\
				 & \hspace{2cm}+\left(-C^j_{-n}C^-_{n}+C^-_{-n}C^j_{n}-\tilde{C}^j_{-n}\tilde{C}^-_{n}+\tilde{C}^-_{-n}\tilde{C}^j_{n}\right)\delta^{ik}\bigg\}
\end{split}
\end{equation}

\begin{equation}
\begin{split}
[S^{ij},S^{k-}] =& +\frac{i}{B^+_0}\bigg\{S^{ik}B^j_{0}+S^{kj}B^i_{0}+B^+_0S^{-i}+B^+_0S^{j-}\bigg\} \\
				=& +\frac{i}{p^+_0}\bigg\{S^{ik}p^j_{0}+S^{kj}p^i_{0}+p^+S^{-i}\delta^{jk}+p^+S^{j-}\delta^{ik}\bigg\} \\
\end{split}
\end{equation}

So, the final expression of the $[J^{ij},J^{k-}]$ looks like,

\begin{equation}
\begin{split}
[J^{ij},J^{k-}] =& [L^{ij}+S^{ij},L^{k-}+S^{k-}] = [L^{ij},L^{k-}]+[L^{ij},S^{k-}]+[S^{ij},L^{k-}]+[S^{ij},S^{k-}] \\
				=& iL^{-i}\delta^{jk}+iL^{j-}\delta^{ik} +\frac{i}{p^+}\bigg\{S^{ki}p^j+S^{jk}p^i \bigg\} +\frac{i}{p^+}\bigg\{S^{ik}p^j+S^{kj}p^i+p^+S^{-i}\delta^{jk}+p^+S^{j-}\delta^{ik}\bigg\} \\
				=& i(L^{-i}\delta^{jk}+L^{j-}\delta^{ik}) +\frac{i}{p^+}\bigg\{p^+S^{-i}\delta^{jk}+p^+S^{j-}\delta^{ik} \bigg\} \\
				=& i\left(\left(L^{-i}+S^{-i}\right)\delta^{jk}+\left(L^{j-}+S^{j-}\right)\delta^{ik}\right) = i\left(J^{-i}\delta^{jk}+J^{j-}\delta^{ik}\right) \\
\end{split}
\end{equation}

So this part of the Lorentz algebra also closes without any trouble. 

\newpage

\section{Detailed analysis: Induced vacuum}\label{Ivac}
This appendix involves the detailed calculations relevent to \ref{Indsec}.
\subsection*{Central Algebra}
Using the normal ordered expressions \eqref{minusdefind}, we evaluate
\begin{equation}\label{AA}
\begin{split}
\left[A^-_m,A^-_n\right] &= \frac{1}{B^+_0}\sum_{p\neq 0}\left[A^i_p B^j_{m-p},A^-_n\right]\\
&= \frac{1}{\left(B^+_0\right)^2}\left(\sum_{p\neq -n}2pA^i_{p+n} B^j_{m-p}+\sum_{p\neq 0}2(m-p)A^i_p B^j_{m+n-p}\right)\\
&= \frac{1}{\left(B^+_0\right)^2}\left(\sum_{p\neq 0}2(p-n)A^i_{p} B^j_{m+n-p}+\sum_{p\neq 0}2(m-p)A^i_p B^j_{m+n-p}\right)\\
&= \frac{2(m-n)}{\left(B^+_0\right)^2}\sum_{p\neq 0}A^i_p B^j_{m+n-p}\\
&= \frac{2(m-n)}{B^+_0}A_{m+n},
\end{split}
\end{equation}
and
\begin{equation}\label{BA}
\begin{split}
\left[B^-_m,A^-_n\right] &= \frac{1}{2B^+_0}\sum_p\left[B^i_p B^j_{m-p},A^-_n\right]\\
&= \frac{1}{2\left(B^+_0\right)^2}\left(\sum_p 2pB^i_{p+n} B^j_{m-p}+\sum_p 2(m-p)B^i_p B^j_{m+n-p}\right)\\
&= \frac{1}{2\left(B^+_0\right)^2}\left(\sum_p 2(p-n)B^i_{p} B^j_{m+n-p}+\sum_p 2(m-p)B^i_p B^j_{m+n-p}\right)\\
&= \frac{2(m-n)}{2\left(B^+_0\right)^2}\sum_p B^i_p B^j_{m+n-p}\\
&= \frac{2(m-n)}{B^+_0}B_{m+n}.
\end{split}
\end{equation}
In the third step of both \eqref{AA} and \eqref{BA} in the first term the dummy index $p$ has been shifted.

\subsection*{Calculating $\left[L^{i-},S^{j-}\right]$}
We have from \eqref{xAB}
\begin{equation}
\begin{split}
\left[x^ip^-,S^{j-}\right] &= -\frac{i}2\sum_{m\neq 0}\frac{1}m\left[x^ip^-,\left(A^j_{-m}B^-_m + B^j_{-m}A^-_m\right)\right]\\
&= \frac{\sqrt{2c'}}{2B^+_0}\sum_{m\neq 0}\frac{1}m\left(A^j_{-m}B^i_m + B^j_{-m}A^i_m\right)p^-.
\end{split}
\end{equation}
and
\begin{equation}
\begin{split}
\left[x^-p^i,S^{j-}\right] &= -\frac{i}2\sum_{m\neq 0}\frac{1}m\left[x^-,\left(A^j_{-m}B^-_m + B^j_{-m}A^-_m\right)\right]p^i\\
&= \frac{\sqrt{2c'}}{2B^+_0}\sum_{m\neq 0}\frac{1}m\left(A^j_{-m}B^-_m + B^j_{-m}A^-_m\right)p^i~.
\end{split}
\end{equation}
So,
\begin{equation}
\begin{split}
\left[L^{i-},S^{j-}\right] = \frac{\sqrt{2c'}}{2B^+_0}\sum_{m\neq 0}\frac{1}m\left(A^j_{-m}B^i_m + B^j_{-m}A^i_m\right)p^- -\frac{\sqrt{2c'}}{2B^+_0}\sum_{m\neq 0}\frac{1}m\left(A^j_{-m}B^-_m + B^j_{-m}A^-_m\right)p^i~.
\end{split} 
\end{equation}
Similarly,
\begin{equation}
\begin{split}
\left[S^{i-},L^{j-}\right] = -\frac{\sqrt{2c'}}{2B^+_0}\sum_{m\neq 0}\frac{1}m\left(A^i_{-m}B^j_m + B^i_{-m}A^j_m\right)p^- +\frac{\sqrt{2c'}}{2B^+_0}\sum_{m\neq 0}\frac{1}m\left(A^i_{-m}B^-_m + B^i_{-m}A^-_m\right)p^j~.
\end{split} 
\end{equation}

Therefore
\begin{equation}
\begin{split}
\left[L^{i-},S^{j-}\right]+\left[L^{i-},S^{j-}\right]
=& -\sqrt{2c'}\sum_{m\neq 0}\frac{1}{mB^+_0}\left(A^i_{-m}B^j_m + B^i_{-m}A^j_m\right)p^-\\
&+\frac{\sqrt{2c'}}{2}\sum_{m\neq 0}\frac{1}{mB^+_0}\left(A^i_{-m}p^j-p^iA^j_{-m}\right)B^-_m\\
&+\frac{\sqrt{2c'}}{2}\sum_{m\neq 0}\frac{1}{mB^+_0}\left(B^i_{-m}p^j-p^iB^j_{-m}\right)A^-_m~.
\end{split}
\end{equation}

\subsection*{Calculation of $\left[S^{i-},S^{j-}\right]$}
Now we turn to the commutator of the oscillator contributions:
\begin{equation}
\left[S^{i-},S^{j-}\right] = -\frac{1}4\sum_{\substack{m\neq 0\\n\neq 0}}\frac{1}{mn}\left[\left(A^i_{-m}B^-_m + B^i_{-m}A^-_m\right),\left(A^j_{-n}B^-_n + B^j_{-n}A^-_n\right)\right].
\end{equation}
The commutator of the first terms in both the brackets is evaluated as follows.
\begin{equation}\label{11}
\begin{split}
\sum_{\substack{m\neq 0\\n\neq 0}}\frac{1}{mn}\left[A^i_{-m}B^-_m,A^j_{-n}B^-_n\right] 
&= -\sum_{\substack{m\neq 0\\n\neq 0}}\frac{2}{nB^+_0} B^i_{n-m}A^j_{-n}B^-_m+\sum_{\substack{m\neq 0\\n\neq 0}}\frac{2}{mB^+_0} A^i_{-m}B^j_{m-n}B^-_m\\
&= -\sum_{\substack{n\neq 0\\m\neq -n}}\frac{2}{nB^+_0} B^i_{-m}A^j_{-n}B^-_{m+n}+\sum_{\substack{m\neq 0\\n\neq -m}}\frac{2}{mB^+_0} A^i_{-m}B^j_{-n}B^-_{m+n}.
\end{split}
\end{equation}
Notice the change of dummy indices. Ultimately all the terms have to be brought in the following form. Each of the term in $\left[S^{i-},S^{j-}\right]$ is a product of three operators. The operator with the spacetime index $i$ has index $-m$, the operator with the spacetime index $j$ has index $-n$ and the operator with the spacetime index $-$ has index $m+n$. This will be the same form we would like to bring all the terms into in the flipped vacuum case, and with some modifications in the oscillator vacuum case. Following this pattern, the other terms give
\begin{equation}\label{12}
\begin{split}
\sum_{\substack{m\neq 0\\n\neq 0}}\frac{1}{mn}\left[A^i_{-m}B^-_m,B^j_{-n}A^-_n\right] 
&= 2\sum_{\substack{m\neq 0\\n\neq 0}}\left(\frac{1}n-\frac{1}m\right)\frac{1}{B^+_0} A^i_{-m}B^j_{-n}B^-_{m+n}-\sum_{\substack{n\neq 0\\m\neq -n\\m\neq 0}}\frac{2}{nB^+_0} A^i_{-m}B^j_{-n}B^-_{m+n}\\
&= -\sum_{\substack{m\neq 0\\n\neq 0}}\frac{2}{mB^+_0} A^i_{-m}B^j_{-n}B^-_{m+n}+\sum_{n\neq 0} \frac{2}{nB^+_0} A^i_n B^j_{-n}B^-_0.
\end{split}
\end{equation}
and
\begin{equation}\label{21}
\begin{split}
\sum_{\substack{m\neq 0\\n\neq 0}}\frac{1}{mn}\left[B^i_{-m}A^-_m,A^j_{-n}B^-_n\right]
&= 2\sum_{\substack{m\neq 0\\n\neq 0}}\left(\frac{1}n-\frac{1}m\right)\frac{1}{B^+_0} B^i_{-m}A^j_{-n}B^-_{m+n}+\sum_{\substack{m\neq 0\\n\neq -m\\n\neq 0}}\frac{2}{mB^+_0} B^i_{-m}A^j_{-n}B^-_{m+n}\\
&= \sum_{\substack{m\neq 0\\n\neq 0}}\frac{2}{nB^+_0} B^i_{-m}A^j_{-n}B^-_{m+n}-\sum_{m\neq 0} \frac{2}{mB^+_0} B^i_{-m} A^j_m B^-_0.
\end{split}
\end{equation}
and
\begin{equation}\label{22}
\begin{split}
\sum_{\substack{m\neq 0\\n\neq 0}}\frac{1}{mn}\left[B^i_{-m}A^-_m,B^j_{-n}A^-_n\right] =& 2\sum_{\substack{m\neq 0\\n\neq 0}}\left(\frac{1}n-\frac{1}m\right)\frac{1}{B^+_0} B^i_{-m}B^j_{-n}A^-_{m+n}
-\sum_{\substack{n\neq 0\\m\neq -n}}\frac{2}{nB^+_0} B^i_{-m}B^j_{-n}A^-_{m+n} \\ & +\sum_{\substack{m\neq 0\\n\neq -m}}\frac{2}{mB^+_0} B^i_{-m}B^j_{-n}A^-_{m+n}\\
=& \sum_{m\neq 0}\frac{2}{mB^+_0}\left(B^i_{-m}B^j_0A^-_m-B^i_{-m}B^j_mA^-_0\right) - \sum_{n\neq 0}\frac{2}{nB^+_0}\left(B^i_0B^j_{-n}A^-_n-B^i_nB^j_{-n}A^-_0\right)\\
=& -4\sum_{m\neq 0}\frac{1}{mB^+_0}B^i_{-m}B^j_mA^-_0+2\sum_{m\neq 0}\frac{1}{mB^+_0}\left(B^i_{-m}B^j_0-B^i_0B^j_{-m}\right)A^-_m.
\end{split}
\end{equation}

Notice that the terms in \eqref{11}, \eqref{12} and \eqref{21} have similar structure, namely they all involve $B^-_{\dots}$, hence they should be combined together. The terms in \eqref{22} have different form and they combine among themselves. This pattern is useful again in the flipped vacuum and oscillator vacuum cases.\\

The combination of \eqref{11}, \eqref{12} and \eqref{21} is simplified to
\begin{equation}\label{combo1}
\begin{split}
&\sum_{m\neq 0}\frac{2}{mB^+_0}\left(A^i_{-m}B^j_0B^-_m-A^i_{-m}B^j_mB^-_m\right) +\sum_{n\neq 0}\frac{2}{nB^+_0}\left(B^i_nA^j_{-n}B^-_0-B^i_0A^j_{-n}B^-_n\right)\\
&+\sum_{n\neq 0}\frac{2}{nB^+_0}A^i_nB^j_{-n}B^-_0-\sum_{m\neq 0}\frac{2}{mB^+_0}B^i_{-m}A^j_mB^-_0\\
=&-4\sum_{m\neq 0}\frac{1}{mB^+_0}\left(A^i_{-m}B^j_m+B^i_{-m}A^j_m\right)B^-_0 +2\sum_{m\neq 0}\frac{1}{mB^+_0}\left(A^i_{-m}B^j_0-B^i_0A^j_{-m}\right)B^-_m.
\end{split}
\end{equation}

Then, using \eqref{22} and \eqref{combo1},
\begin{equation}
\begin{split}
\left[S^{i-},S^{j-}\right] 
=& \sum_{m\neq 0}\frac{1}{mB^+_0}\left(A^i_{-m}B^j_m+B^i_{-m}A^j_m\right)B^-_0\\
&-\frac{\sqrt{2c'}}2\sum_{m\neq 0}\frac{1}{mB^+_0}\left(A^i_{-m}p^j-p^iA^j_{-m}\right)B^-_m\\
&-\frac{\sqrt{2c'}}2\sum_{m\neq 0}\frac{1}{mB^+_0}\left(B^i_{-m}p^j-p^iB^j_{-m}\right)A^-_m\\
&+ \sum_{m\neq 0}\frac{1}{mB^+_0}B^i_{-m}B^j_mA^-_0.\\
\end{split}
\end{equation}

\newpage

\section{Detailed analysis: Flipped vacuum}\label{FVac}
This appendix involves the detailed calculations relevent to \ref{Flsec}.
\subsection*{Central algebra}
We use the normal ordered expressions \eqref{minusdeffl} to obtain the central algebra for the flipped vacuum.\\
Without loss of generality, let's assume $n<0$ for the following
\begin{equation}\label{AAflip}
\begin{split}
\left[A^-_m,A^-_n\right] =& \frac{1}{B_0^+}\sum_{p<0}\left[A^i_p B^j_{m-p},A^-_n\right]+\frac{1}{B_0^+}\sum_{p>0}\left[B^j_{m-p}A^i_p,A^-_n\right]\\
=& \frac{2}{(B_0^+)^2}\left(\sum_{p<0}pA^i_{p+n}B^i_{m-p}+\sum_{p<0}(m-p)A^i_{p}B^i_{m+n-p}\right)\\
&+\frac{2}{(B_0^+)^2}\left(\sum_{\substack{p > 0\\p\neq -n}}pB^i_{m-p}A^i_{p+n}+\sum_{p>0}(m-p)B^i_{m+n-p}A^i_{p}\right)\\
=&\frac{2}{(B_0^+)^2}\left(\sum_{p<n}(p-n)A^i_pB^i_{m+n-p}+\sum_{p<0}(m-p)A^i_{p}B^i_{m+n-p}\right)\\
&+\frac{2}{(B_0^+)^2}\left(\sum_{\substack{p > n\\p\neq 0}}(p-n)B^i_{m+n-p}A^i_p+\sum_{p>0}(m-p)B^i_{m+n-p}A^i_{p}\right)\\
=& \frac{2(m-n)}{(B_0^+)^2}\left(\sum_{p<0}A^i_{p}B^i_{m+n-p}+\sum_{p>0}B^i_{m+n-p}A^i_{p}\right)\\
&- \frac{2}{(B_0^+)^2}\sum_{p=n+1}^{-1}(p-n)\left[A^i_p,B^i_{m+n-p}\right]\\
=& \frac{2(m-n)}{B_0^+}A^-_{m+n}-\frac{4}{(B_0^+)^2}(D-2)\delta_{m+n,0}\sum_{p=n+1}^{-1}p(p-n)\\
=& \frac{2(m-n)}{B_0^+}A^-_{m+n}-\frac{4}{(B_0^+)^2}(D-2)\delta_{m+n,0}\sum_{p=1}^{m-1}p(p-m)\\
&= \frac{2(m-n)}{B_0^+}A^-_{m+n}+\frac{2(D-2)}{3(B_0^+)^2}(m^3-m)\delta_{m+n,0}~.
\end{split}
\end{equation}
Similarly,
\begin{equation}\label{BAflip}
\begin{split}
\left[B^-_m,A^-_n\right] =& \frac{1}{B_0^+}\sum_p\left[B^i_p B^j_{m-p},A^-_n\right]\\
=& \frac{2}{(B_0^+)^2}\left(\sum_{p}pB^i_{p+n}B^i_{m-p}+\sum_{p}(m-p)B^i_{p}B^i_{m+n-p}\right)\\
=&\frac{2}{(B_0^+)^2}\left(\sum_{p}(p-n)A^i_pB^i_{m+n-p}+\sum_{p}(m-p)A^i_{p}B^i_{m+n-p}\right)\\
=& \frac{2(m-n)}{(B_0^+)^2}\sum_{p}B^i_{p}B^i_{m+n-p}\\
=& \frac{2(m-n)}{B_0^+}B^-_{m+n}~.
\end{split}
\end{equation}

\subsection*{Calculation of $\left[S^{i-},S^{j-}\right]$}

This calculation is divided into five sectors, taking into consideration the lessons about which terms combine nicely from the $\left[S^{i-},S^{j-}\right]$ calculation in the induced vacuum case.
\begin{equation}\label{sectintro}
\left[S^{i-},S^{j-}\right] = S_1 + S_2 + S_3 + S_4 + S_5~.
\end{equation}

\subsubsection*{Sector $1$}
\begin{equation}\label{S1def}
S_1 = -\sum_{\substack{m > 0\\n > 0}}\frac{1}{mn}
\left(\left[A^i_{-m}B^-_m,A^j_{-n}B^-_n\right]+\left[A^i_{-m}B^-_m,B^j_{-n}A^-_n\right]+\left[B^i_{-m}A^-_m,A^j_{-n}B^-_n\right]\right).
\end{equation}
Note that these are three terms out of four in the $m > 0,~n > 0$ portion of the calculation. The remaining term is a part of $S_5$. This pattern will repeat for $S_2$, $S_3$ and $S_4$ too, just for different signs of $m$ and $n$.\\
Now look at the three terms in $S_1$ one by one.
\begin{equation}\label{S1T1}
\begin{split}
\sum_{\substack{m > 0\\n > 0}}\frac{1}{mn}\left[A^i_{-m}B^-_m,A^j_{-n}B^-_n\right] =& \sum_{\substack{m > 0\\n > 0}}\frac{2}{mB^+_0}A^i_{-m}B^j_{m-n}B^-_n-\sum_{\substack{m > 0\\n > 0}}\frac{2}{nB^+_0}B^i_{n-m}A^j_{-n}B^-_m\\
=& \sum_{\substack{m > 0\\n > -m}}\frac{2}{mB^+_0}A^i_{-m}B^j_{-n}B^-_{m+n}-\sum_{\substack{n > 0\\m > -n}}\frac{2}{nB^+_0}B^i_{-m}A^j_{-n}B^-_{m+n}.
\end{split}
\end{equation}
Note the shift of dummy indices performed in the second step in order to conform with the convention as discussed in the induced vacuum case.\\
Similarly with other terms,
\begin{equation}\label{S1T2}
\begin{split}
\sum_{\substack{m > 0\\n > 0}}\frac{1}{mn}\left[A^i_{-m}B^-_m,B^j_{-n}A^-_n\right]=& 2\sum_{\substack{m > 0\\n > 0}}\left(\frac{1}{n}-\frac{1}{m}\right)\frac{1}{B^+_0}A^i_{-m}B^j_{-n}B^-_{m+n} -\sum_{\substack{n > 0\\m > -n\\m\neq 0}}\frac{2}{nB^+_0}A^i_{-m}B^j_{-n}B^-_{m+n}\\
=& -2\sum_{\substack{m > 0\\n > 0}}\frac{1}{mB^+_0}A^i_{-m}B^j_{-n}B^-_{m+n} -\sum_{n > 0}\frac{2}{nB^+_0}\sum_{m=-n+1}^{-1}A^i_{-m}B^j_{-n}B^-_{m+n}.
\end{split}
\end{equation}
and
\begin{equation}\label{S1T3}
\begin{split}
\sum_{\substack{m > 0\\n > 0}}\frac{1}{mn}\left[B^i_{-m}A^-_m,A^j_{-n}B^-_n\right]=& 2\sum_{\substack{m > 0\\n > 0}}\left(\frac{1}{n}-\frac{1}{m}\right)\frac{1}{B^+_0}B^i_{-m}A^j_{-n}B^-_{m+n}+\sum_{\substack{m > 0\\n > -m\\n\neq 0}}\frac{2}{mB^+_0}B^i_{-m}A^j_{-n}B^-_{m+n}\\
=& 2\sum_{\substack{m > 0\\n > 0}}\frac{1}{nB^+_0}B^i_{-m}A^j_{-n}B^-_{m+n} +\sum_{m > 0}\frac{2}{mB^+_0}\sum_{n=-m+1}^{-1}B^i_{-m}A^j_{-n}B^-_{m+n}.
\end{split}
\end{equation}
Similar looking terms in \eqref{S1T1}, \eqref{S1T2} and \eqref{S1T3} can now be grouped together and combined easily to get
\begin{equation}\label{S1tot}
\begin{split}
S_1 =& \sum_{n > 0}\frac{2}{nB^+_0}\sum_{m=-n+1}^{-1}A^i_{-m}B^j_{-n}B^-_{m+n}+\sum_{n > 0}\frac{2}{nB^+_0}\sum_{m=-n+1}^{0}B^i_{-m}A^j_{-n}B^-_{m+n}\\
&-\sum_{m > 0}\frac{2}{mB^+_0}\sum_{n=-m+1}^{0}A^i_{-m}B^j_{-n}B^-_{m+n}
-\sum_{m > 0}\frac{2}{mB^+_0}\sum_{n=-m+1}^{-1}B^i_{-m}A^j_{-n}B^-_{m+n}.
\end{split}
\end{equation}

\subsubsection*{Sector $2$}
\begin{equation}\label{S2def}
\begin{split}
S_2 =& -\sum_{\substack{m > 0\\n < 0}}\frac{1}{mn} \left(\left[A^i_{-m}B^-_m,B^-_nA^j_{-n}\right]+\left[A^i_{-m}B^-_m,A^-_nB^j_{-n}\right]+\left[B^i_{-m}A^-_m,B^-_nA^j_{-n}\right]\right).
\end{split}
\end{equation}
Similar to sector $1$, the terms are calculated individually and combined.
\begin{equation}\label{S2totnq}
\begin{split}
S_2 =& \sum_{m > 0}\frac{2}{mB^+_0}\sum_{n=-m}^{-1}\left(A^i_{-m}B^j_{-n}B^-_{m+n}+B^-_{m+n}B^i_{-m}A^j_{-n}\right)\\
&- \sum_{n < 0}\frac{2}{nB^+_0}\sum_{m=1}^{-n}A^i_{-m}B^j_{-n}B^-_{m+n}\\
&+ \sum_{\substack{n < 0\\m > -n}}\frac{2}{nB^+_0}B^i_{-m}A^j_{-n}B^-_{m+n}-\sum_{\substack{n < 0\\ m > 0}}\frac{2}{nB^+_0}B^-_{m+n}B^i_{-m}A^j_{-n}.
\end{split}
\end{equation}
Note that the last two terms appear with different ordering, and they are combined as follows.
\begin{equation}\label{commord}
\begin{split}
&\sum_{\substack{n < 0\\m > -n}}\frac{2}{nB^+_0}B^i_{-m}A^j_{-n}B^-_{m+n}-\sum_{\substack{n < 0\\ m > 0}}\frac{2}{nB^+_0}B^-_{m+n}B^i_{-m}A^j_{-n}\\
=& -\sum_{n<0}\frac{2}{nB^+_0}\sum_{m=1}^{-n}B^-_{m+n}B^i_{-m}A^j_{-n}+\sum_{\substack{n < 0\\ m > -n}}\frac{2}{nB^+_0}B^i_{-m}\left[A^j_{-n},B^-_{m+n}\right]\\
=& -\sum_{n<0}\frac{2}{nB^+_0}\sum_{m=1}^{-n}B^-_{m+n}B^i_{-m}A^j_{-n} - \frac{4}{(B^+_0)^2}\sum_{\substack{n < 0\\ m > -n}}B^i_{-m}B^j_m~.
\end{split}
\end{equation}
It is easy to show that
\begin{equation}\label{trick}
\sum_{\substack{n < 0\\ m > -n}}B^i_{-m}B^j_m = \sum_{m > 1}\sum_{n=-m+1}^{-1}B^i_{-m}B^j_m = \sum_{m>0}(m-1)B^i_{-m}B^j_m~.
\end{equation}
Then using \eqref{commord} and \eqref{trick} in \eqref{S2totnq}
\begin{equation}\label{S2tot}
\begin{split}
S_2 =& \sum_{m > 0}\frac{2}{mB^+_0}\sum_{n=-m}^{-1}\left(A^i_{-m}B^j_{-n}B^-_{m+n}+B^-_{m+n}B^i_{-m}A^j_{-n}\right)\\
&- \sum_{n < 0}\frac{2}{nB^+_0}\sum_{m=1}^{-n}\left(A^i_{-m}B^j_{-n}B^-_{m+n}+B^-_{m+n}B^i_{-m}A^j_{-n}\right)\\
&-\frac{4}{(B^+_0)^2}\sum_{m>0}(m-1)B^i_{-m}B^j_m~.
\end{split}
\end{equation}

\subsubsection*{Sectors $3$ and $4$}
\begin{equation}\label{S3def}
S_3 = -\sum_{\substack{m < 0\\n > 0}}\frac{1}{mn} \left(\left[B^-_mA^i_{-m},A^j_{-n}B^-_n\right]+\left[B^-_mA^i_{-m},B^j_{-n}A^-_n\right]+\left[A^-_mB^i_{-m},A^j_{-n}B^-_n\right]\right).
\end{equation}

\begin{equation}\label{S4def}
S_4 =-\sum_{\substack{m < 0\\n < 0}}\frac{1}{mn} \left(\left[B^-_mA^i_{-m},B^-_nA^j_{-n}\right]+\left[B^-_mA^i_{-m},A^-_nB^j_{-n}\right]+\left[A^-_mB^i_{-m},B^-_nA^j_{-n}\right]\right).
\end{equation}
Computations of sectors $3$ and $4$ follow those of sectors $2$ and $1$ respectively. The final expressions:
\begin{equation}\label{S3tot}
\begin{split}
S_3 =& \sum_{m < 0}\frac{2}{mB^+_0}\sum_{n=1}^{-m}\left(B^-_{m+n}A^i_{-m}B^j_{-n}+B^i_{-m}A^j_{-n}B^-_{m+n}\right)\\
&- \sum_{n > 0}\frac{2}{nB^+_0}\sum_{m=-n}^{-1}\left(B^-_{m+n}A^i_{-m}B^j_{-n}+B^i_{-m}A^j_{-n}B^-_{m+n}\right)\\
&-\frac{4}{(B^+_0)^2}\sum_{m<0}(m+1)B^i_{-m}B^j_m~.
\end{split}
\end{equation}
and
\begin{equation}\label{S4tot}
\begin{split}
S_4 =& \sum_{n < 0}\frac{2}{nB^+_0}\sum_{m=1}^{-n-1}B^-_{m+n}A^i_{-m}B^j_{-n}+\sum_{n < 0}\frac{2}{nB^+_0}\sum_{m=0}^{-n-1}B^-_{m+n}B^i_{-m}A^j_{-n}\\
&-\sum_{m < 0}\frac{2}{mB^+_0}\sum_{n=0}^{-m-1}B^-_{m+n}A^i_{-m}B^j_{-n}
-\sum_{m < 0}\frac{2}{mB^+_0}\sum_{n=1}^{-m-1}B^-_{m+n}B^i_{-m}A^j_{-n}~.
\end{split}
\end{equation}
Now let's combine the sectors $1$ to $4$ (\eqref{S1tot}, \eqref{S2tot},\eqref{S3tot} and \eqref{S4tot}) by collecting similar terms. First notice that, for example, the first term in second line of \eqref{S1tot} and the first term in first line of \eqref{S2tot} have the same structure and the same ordering, and they combine to give
\begin{equation}\label{samp1}
\begin{split}
&-\sum_{m > 0}\frac{2}{mB^+_0}\sum_{n=-m+1}^{0}A^i_{-m}B^j_{-n}B^-_{m+n}+\sum_{m > 0}\frac{2}{mB^+_0}\sum_{n=-m}^{-1}A^i_{-m}B^j_{-n}B^-_{m+n}\\
=& \sum_{m > 0}\frac{2}{mB^+_0}A^i_{-m}B^j_m B^-_0 - \sum_{m > 0}\frac{2}{mB^+_0}A^i_{-m}B^j_0 B^-_m~.
\end{split}
\end{equation} 
while the second term in second line of \eqref{S1tot} and the second term in first line of \eqref{S2tot} have the same structure but {\it not} the same ordering, so combining them proceeds as follows.
\begin{equation}\label{samp2}
\begin{split}
&-\sum_{m > 0}\frac{2}{mB^+_0}\sum_{n=-m+1}^{-1}B^i_{-m}A^j_{-n}B^-_{m+n}+\sum_{m > 0}\frac{2}{mB^+_0}\sum_{n=-m}^{-1}B^-_{m+n}B^i_{-m}A^j_{-n}\\
=& \sum_{m > 0}\frac{2}{mB^+_0}B^i_{-m}A^j_m B^-_0 - \sum_{m > 0}\frac{2}{mB^+_0}\sum_{n=-m}^{-1}B^i_{-m}\left[A^j_{-n},B^-_{m+n}\right]\\
=& \sum_{m > 0}\frac{2}{mB^+_0}B^i_{-m}A^j_m B^-_0 + \sum_{m > 0}\frac{4}{m(B^+_0)^2}\left(\sum_{n=-m}^{-1}n\right)B^i_{-m}B^j_m\\
=& \sum_{m > 0}\frac{2}{mB^+_0}B^i_{-m}A^j_m B^-_0 - \sum_{m > 0}\frac{2(m+1)}{(B^+_0)^2}B^i_{-m}B^j_m~.
\end{split}
\end{equation} 
Doing this for all the terms\footnote{Some terms have $n$ as the summed index, that shold be transformed to $m$ depending on the desired final form as in \eqref{S1234}.}, we get
\begin{equation}\label{S1234}
\begin{split}
 S_1+S_2+S_3&+S_4 
= \sum_{m\neq 0}\frac{1}{mB^+_0}\left(A^i_{-m}B^j_m + B^i_{-m}A^j_m\right)B^-_0 -\frac{\sqrt{2c'}}2\sum_{m > 0}\frac{1}{mB^+_0}\left(A^i_{-m}p^j-p^iA^j_{-m}\right)B^-_m\\
&-\frac{\sqrt{2c'}}2\sum_{m < 0}\frac{1}{mB^+_0}B^-_m\left(A^i_{-m}p^j-p^iA^j_{-m}\right) -2\sum_{m>0}\frac{m}{(B^+_0)^2}B^i_{-m}B^j_m-2\sum_{m<0}\frac{m}{(B^+_0)^2}B^i_{-m}B^j_m~.
\end{split}
\end{equation}

\subsubsection*{Sector $5$}
\begin{equation}\label{S5def}
\begin{split}
S_5 =& -\sum_{\substack{m > 0\\n > 0}}\frac{1}{mn}\left[B^i_{-m}A^-_m,B^j_{-n}A^-_n\right]-\sum_{\substack{m > 0\\n < 0}}\frac{1}{mn}\left[B^i_{-m}A^-_m,A^-_nB^j_{-n}\right]\\
&-\sum_{\substack{m < 0\\n > 0}}\frac{1}{mn}\left[A^-_mB^i_{-m},B^j_{-n}A^-_n\right]-\sum_{\substack{m < 0\\n < 0}}\frac{1}{mn}\left[A^-_mB^i_{-m},A^-_nB^j_{-n}\right]~.
\end{split}
\end{equation}
The manipulations involved in combining the terms are similar to those used in sectors $1$ to $4$ even though the terms themselves are of a different form. for example, consider the first term in \eqref{S5def}
\begin{equation}\label{S5T1}
\begin{split}
&\sum_{\substack{m > 0\\n > 0}}\frac{1}{mn}\left[B^i_{-m}A^-_m,B^j_{-n}A^-_n\right] = \sum_{\substack{m > 0\\n > 0}}\left(\frac{1}n-\frac{1}m\right)\frac{2}{B^+_0}B^i_{-m}B^j_{-n}A^-_{m+n}\\
&-\sum_{\substack{n > 0\\m > -n}}\frac{2}{nB^+_0}B^i_{-m}B^j_{-n}A^-_{m+n}+\sum_{\substack{m > 0\\n > -m}}\frac{2}{mB^+_0}B^i_{-m}B^j_{-n}A^-_{m+n}\\
=& \sum_{m>0}\frac{2}{mB^+_0}\sum_{n=-m+1}^{0}B^i_{-m}B^j_{-n}A^-_{m+n}-\sum_{n>0}\frac{2}{nB^+_0}\sum_{m=-n+1}^{0}B^i_{-m}B^j_{-n}A^-_{m+n}~.
\end{split}
\end{equation}
The fourth term is similarly straightforward
\begin{equation}\label{S5T4}
\begin{split}
&\sum_{\substack{m < 0\\n < 0}}\frac{1}{mn}\left[B^i_{-m}A^-_m,B^j_{-n}A^-_n\right]\\
=& \sum_{m<0}\frac{2}{mB^+_0}\sum_{n=0}^{-m-1}B^i_{-m}B^j_{-n}A^-_{m+n}-\sum_{n<0}\frac{2}{nB^+_0}\sum_{m=0}^{-n-1}B^i_{-m}B^j_{-n}A^-_{m+n}.
\end{split}
\end{equation}
Second and third terms are little more complicated due to mismatched ordering similar to the sectors $2$ and $3$, and also due to the presence of central terms. Let's calculate for example the second term.
\begin{equation}\label{S5T2}
\begin{split}
&\sum_{\substack{m > 0\\n < 0}}\frac{1}{mn}\left[B^i_{-m}A^-_m,A^-_nB^j_{-n}\right]\\
=& \sum_{\substack{m > 0\\n < 0}}\left(\frac{1}n-\frac{1}m\right)\frac{2}{B^+_0}B^i_{-m}A^-_{m+n}B^j_{-n}-\sum_{m>0}\frac{2(D-2)}{3m^2(B^+_0)^2}(m^3-m)B^i_{-m}B^j_m\\
&-\sum_{\substack{n < 0\\m > -n}}\frac{2}{nB^+_0}B^i_{-m}B^j_{-n}A^-_{m+n}+\sum_{\substack{m > 0\\n < -m}}\frac{2}{mB^+_0}B^i_{-m}A^-_{m+n}B^j_{-n}\\
=&-\sum_{m>0}\frac{2}{mB^+_0}\sum_{n=-m}^{-1}B^i_{-m}A^-_{m+n}B^j_{-n}-\sum_{n<0}\frac{2}{nB^+_0}\sum_{m=1}^{-n}B^i_{-m}A^-_{m+n}B^j_{-n}\\
&+4\sum_{m>0}\frac{m-1}{(B^+_0)^2}B^i_{-m}B^j_m-\frac{2(D-2)}{3(B^+_0)^2}\sum_{m>0}\left(m-\frac{1}m\right)B^i_{-m}B^j_m~.
\end{split}
\end{equation}
The third term in the second step has resulted from the mismatch between the ordering of the first term in the first line and the first term in the second line of the first step. This calculation is similar to the sector $2$. Likewise the third term in \eqref{S5def} gives
\begin{equation}\label{S5T3}
\begin{split}
&\sum_{\substack{m > 0\\n < 0}}\frac{1}{mn}\left[A^-_m B^i_{-m},B^j_{-n}A^-_n\right]\\
=&-\sum_{m<0}\frac{2}{mB^+_0}\sum_{n=1}^{-m}B^j_{-n}A^-_{m+n}B^i_{-m}-\sum_{n>0}\frac{2}{nB^+_0}\sum_{m=-n}^{-1}B^j_{-n}A^-_{m+n}B^i_{-m}\\
&+4\sum_{m<0}\frac{m-1}{(B^+_0)^2}B^i_{-m}B^j_m-\frac{2(D-2)}{3(B^+_0)^2}\sum_{m<0}\left(m-\frac{1}m\right)B^i_{-m}B^j_m~.
\end{split}
\end{equation}
Combining these four terms in the same way the sectors $1$ to $4$ were combined,
\begin{equation}\label{S5tot}
\begin{split}
S_5 =& \sum_{m\neq 0}\frac{1}{mB^+_0}B^i_{-m}B^j_mA^-_0\\
&-\frac{\sqrt{2c'}}2\sum_{m > 0}\frac{1}{mB^+_0}\left(B^i_{-m}p^j-p^iB^j_{-m}\right)A^-_m\\
&-\frac{\sqrt{2c'}}2\sum_{m < 0}\frac{1}{mB^+_0}A^-_m\left(B^i_{-m}p^j-p^iB^j_{-m}\right)\\
&-2\sum_{m>0}\frac{m}{(B^+_0)^2}B^i_{-m}B^j_m-2\sum_{m<0}\frac{m}{(B^+_0)^2}B^i_{-m}B^j_m\\
&+\frac{(D-2)}{6(B^+_0)^2}\left(\sum_{m>0}\left(m-\frac{1}m\right)B^i_{-m}B^j_m+\sum_{m<0}\left(m-\frac{1}m\right)B^i_{-m}B^j_m\right)~.
\end{split}
\end{equation}
And then from \eqref{sectintro}, \eqref{S1234} and \eqref{S5tot},
\begin{equation}
\begin{split}
\left[S^{i-},S^{j-}\right]
=& \sum_{m\neq 0}\frac{1}{mB^+_0}\left(A^i_{-m}B^j_m + B^i_{-m}A^j_m\right)B^-_0 -\frac{\sqrt{2c'}}2\sum_{m > 0}\frac{1}{mB^+_0}\left(A^i_{-m}p^j-p^iA^j_{-m}\right)B^-_m\\
&-\frac{\sqrt{2c'}}2\sum_{m < 0}\frac{1}{mB^+_0}B^-_m\left(A^i_{-m}p^j-p^iA^j_{-m}\right) -\frac{\sqrt{2c'}}2\sum_{m > 0}\frac{1}{mB^+_0}\left(B^i_{-m}p^j-p^iB^j_{-m}\right)A^-_m\\
&-\frac{\sqrt{2c'}}2\sum_{m < 0}\frac{1}{mB^+_0}A^-_m\left(B^i_{-m}p^j-p^iB^j_{-m}\right) -4\sum_{m>0}\frac{m}{(B^+_0)^2}B^i_{-m}B^j_m-4\sum_{m<0}\frac{m}{(B^+_0)^2}B^i_{-m}B^j_m\\
&+\frac{(D-2)}{6(B^+_0)^2}\left(\sum_{m>0}\left(m-\frac{1}m\right)B^i_{-m}B^j_m+\sum_{m<0}\left(m-\frac{1}m\right)B^i_{-m}B^j_m\right) + \sum_{m\neq 0}\frac{1}{mB^+_0}B^i_{-m}B^j_mA^-_0~.
\end{split}
\end{equation}

\newpage 

\section{Detailed analysis: Oscillator vacuum}\label{Cvac}
This appendix involves the detailed calculations relevent to \ref{Oscsec}. In addition to \eqref{Ccomm} we also have
\begin{equation}\label{xC}
\left[x^i,C^-_m\right] = \frac{i\sqrt{2c'}}{B^+_0}C^i_m,\quad
\left[x^-,C^-_m\right] = \frac{i\sqrt{2c'}}{B^+_0}C^-_m. 
\end{equation}
and identical relations for $\tilde{C}^-$. 
Some more useful commutators are
\begin{equation}\label{CABcomm}
\begin{split}
\left[C^i_m,A^-_n\right] &= \frac{2m}{B^+_0}C^i_{m+n}~~~\forall m\neq -n,\quad \left[C^i_{-n},A^-_n\right] = -\frac{n}{B^+_0}B^i_0,\\
\left[\tilde{C}^i_m,A^-_n\right] &= -\frac{2m}{B^+_0}\tilde{C}^i_{m-n}~~~\forall m\neq n,\quad
\left[\tilde{C}^i_n,A^-_n\right] = -\frac{n}{B^+_0}B^i_0,\\
\left[C^i_m,B^-_n\right] &= \frac{m}{B^+_0}\left(C^i_{m+n}+\tilde{C}^i_{-m-n}\right)~~~\forall m\neq -n,\quad
\left[C^i_{-n},B^-_n\right] = -\frac{n}{B^+_0}B^i_0,\\
\left[\tilde{C}^i_m,B^-_n\right] &= \frac{m}{B^+_0}\left(C^i_{n-m}+\tilde{C}^i_{m-n}\right)~~~\forall m\neq n,\quad
\left[\tilde{C}^i_n,B^-_n\right] = \frac{n}{B^+_0}B^i_0~.
\end{split}
\end{equation}

\subsection*{Central algebra}
According to the normal ordering suited for this vacuum,
\begin{equation}\label{Aosc}
\begin{split}
A^-_n =& \frac{1}{B^+_0}\sum_{m\neq 0}\normord{A^i_mB^i_{n-m}}\\
=& \frac{1}{B^+_0}\sum_{m>0}\left(C^i_{n-m}C^i_m - \tilde{C}^i_{-m}\tilde{C}^i_{m-n}\right)+\frac{1}{B^+_0}\sum_{m<0}\left(C^i_mC^i_{n-m} -\tilde{C}^i_{m-n}\tilde{C}^i_{-m} \right)+ \frac{1}{B^+_0}A^i_nB^i_0~.
\end{split}
\end{equation}

Without loss of generality, let's assume $n < 0$.
\begin{equation}\label{AAcommOsc}
\begin{split}
\hspace{-1cm} \left[A^-_m,A^-_n\right] =& \frac{1}{B^+_0} \left\{ \sum_{p>0}\left(\left[C^i_{m-p}C^i_p,A^-_n\right]-\left[\tilde{C}^i_{-p}\tilde{C}^i_{p-m},A^-_n\right]  \right)+ \sum_{p<0}\left(\left[C^i_pC^i_{m-p},A^-_n\right]  -\left[\tilde{C}^i_{p-m}\tilde{C}^i_{-p},A^-_n\right]\right) + \left[A^i_mB^i_0,A^-_n\right] \right\} \\
=& \frac{2}{(B^+_0)^2}\sum_{\substack{p>0\\p\neq -n}}p\left(C^i_{m-p}C^i_{p+n}-\tilde{C}^i_{-p-n}\tilde{C}^i_{p-m}\right) + \frac{2}{(B^+_0)^2}\sum_{\substack{p>0\\p\neq m+n}}(m-p)\left(C^i_{m+n-p}C^i_p-\tilde{C}^i_{-p}\tilde{C}^i_{p-m-n}\right) \\ &+ \frac{2}{(B^+_0)^2}\sum_{p<0}p\left(C^i_{p+n}C^i_{m-p}-\tilde{C}^i_{p-m}\tilde{C}^i_{-p-n}\right)+ \frac{2}{(B^+_0)^2}\sum_{\substack{p<0\\p\neq m+n}}(m-p)\left(C^i_pC^i_{m+n-p}-\tilde{C}^i_{p-m-n}\tilde{C}^i_{-p}\right)\\
&-\frac{2n}{(B^+_0)^2}A^i_{m+n}B^i_0 +\frac{2m}{(B^+_0)^2}A^i_{m+n}B^i_0\\
=& \frac{2}{(B^+_0)^2}\sum_{\substack{p>n\\p\neq 0}}(p-n)\left(C^i_{m+n-p}C^i_p-\tilde{C}^i_{-p}\tilde{C}^i_{p-m-n}\right)+ \frac{2}{(B^+_0)^2}\sum_{\substack{p>0\\p\neq m+n}}(m-p)\left(C^i_{m+n-p}C^i_p-\tilde{C}^i_{-p}\tilde{C}^i_{p-m-n}\right)\\
&+ \frac{2}{(B^+_0)^2}\sum_{p<n}(p-n)\left(C^i_pC^i_{m+n-p}-\tilde{C}^i_{p-m-n}\tilde{C}^i_{-p}\right) + \frac{2}{(B^+_0)^2}\sum_{\substack{p<0\\p\neq m+n}}(m-p)\left(C^i_pC^i_{m+n-p}-\tilde{C}^i_{p-m-n}\tilde{C}^i_{-p}\right)\\ &+ 2\frac{(m-n)}{(B^+_0)^2}A^i_{m+n}B^i_0\\
=& 2\frac{(m-n)}{(B^+_0)^2}\bigg(\sum_{\substack{p>0\\p\neq m+n}}\left(C^i_{m+n-p}C^i_p-\tilde{C}^i_{-p}\tilde{C}^i_{p-m-n}\right) + \sum_{\substack{p<0\\p\neq m+n}}\left(C^i_pC^i_{m+n-p}-\tilde{C}^i_{p-m-n}\tilde{C}^i_{-p}\right)+A^i_{m+n}B^i_0\bigg)\\
&+ \frac{2}{(B^+_0)^2}\sum_{p=n}^{-1}(p-n)\left(\left[C^i_{m+n-p},C^i_p\right]-\left[\tilde{C}^i_{-p},\tilde{C}^i_{p-m-n}\right]\right) = \frac{2(m-n)}{B^+_0}A^-_{m+n}~.
\end{split}
\end{equation}
While $B^-_n$ is given according to this normal ordering by
\begin{equation}\label{Bosc}
\begin{split}
B^-_n =& \frac{1}{2B^+_0}\sum_{m}\normord{B^i_mB^i_{n-m}} \,
= \frac{1}{2B^+_0}\sum_{m>0}\left(C^i_{n-m}C^i_m + \tilde{C}^i_{-m}\tilde{C}^i_{m-n}\right) \\ &+\frac{1}{2B^+_0}\sum_{m<0}\left(C^i_mC^i_{n-m}+\tilde{C}^i_{m-n}\tilde{C}^i_{-m} \right) +\frac{1}{B^+_0}\sum_{\substack{m\neq 0\\m\neq n}}C^i_m\tilde{C}^i_{m-n}+ \frac{1}{B^+_0}B^i_nB^i_0~.
\end{split}
\end{equation}

Again assuming $n < 0$, and making rearrangements very similar to those in \eqref{AAcommOsc},
\begin{equation}\label{ABcommOsc}
\begin{split}
\left[B^-_m,A^-_n\right] =& \frac{(m-n)}{(B^+_0)^2}\bigg(\sum_{\substack{p>0\\p\neq m+n}}\left(C^i_{m+n-p}C^i_p+\tilde{C}^i_{-p}\tilde{C}^i_{p-m-n}\right) + \sum_{\substack{p<0\\p\neq m+n}}\left(C^i_pC^i_{m+n-p}+\tilde{C}^i_{p-m-n}\tilde{C}^i_{-p}\right) \\
&+2\sum_{\substack{p\neq 0\\p\neq m+n}}C^i_p\tilde{C}^i_{p-m-n} +2B^i_{m+n}B^i_0\bigg)+\frac{1}{(B^+_0)^2}\sum_{p=n}^{-1}(p-n)\left(\left[C^i_{m+n-p},C^i_p\right]+\left[\tilde{C}^i_{-p},\tilde{C}^i_{p-m-n}\right]\right)\\
=& \frac{2(m-n)}{B^+_0}B^-_{m+n}+\frac{(D-2)}{3(B^+_0)^2}(m^3-m)\delta_{n,-m}~.
\end{split}
\end{equation}

Therefore
\begin{equation}\label{CC1}
\begin{split}
\left[C^-_m,C^-_n\right] =&\frac{1}{4}\left(\left[B^-_m,A^-_n\right]+\left[A^-_m,B^-_n\right]+\left[A^-_m,A^-_n\right]\right)\\
=& \frac{m-n}{2B^+_0}\left(2B^-_{m+n}+A^-_{m+n}\right)+\frac{(D-2)}{12(B^+_0)^2}\left((m^3-m)-(n^3-n)\right)\delta_{n,-m}\\
=& \frac{m-n}{2B^+_0}\left(3C^-_{m+n}+\tilde{C}^-_{-m-n}\right) +\frac{(D-2)}{6(B^+_0)^2}(m^3-m)\delta_{n,-m}~,
\end{split}
\end{equation}

Similarly,
\begin{equation}\label{CC2}
\left[C^-_m,\tilde{C}^-_{-n}\right] = -\frac{(m-n)}{2B^+_0}\left(C^-_{m+n}-\tilde{C}^-_{-m-n}\right)~,
\end{equation}
and
\begin{equation}\label{CC3}
\left[\tilde{C}^-_{-m},\tilde{C}^-_{-n}\right] = -\frac{(m-n)}{2B^+_0}\left(C^-_{m+n}+3\tilde{C}^-_{-m-n}\right)-\frac{(D-2)}{6(B^+_0)^2}(m^3-m)\delta_{n,-m}.
\end{equation}

Also, from \eqref{CABcomm}
\begin{equation}\label{CiCcomm}
\begin{split}
&\left[C^i_{-m},C^-_n\right] = -\frac{m}{2B^+_0}\left(3C^i_{n-m}+\tilde{C}^i_{m-n}\right)~~~\forall m\neq n,\quad \textcolor[rgb]{1.00,0.00,1.00}{\left[C^i_{-m},C^-_m\right] = -\frac{m}{B^+_0}B^i_0 = -\frac{m\sqrt{2c'}}{B^+_0}p^i},\\
&\left[\tilde{C}^i_m,C^-_n\right] = \frac{m}{2B^+_0}\left(C^i_{n-m}-\tilde{C}^i_{m-n}\right)~~~\forall m\neq n,\quad \left[\tilde{C}^i_m,C^-_m\right] = 0,\\
&\left[C^i_{-m},\tilde{C}^-_{-n}\right] = \frac{m}{2B^+_0}\left(C^i_{n-m}-\tilde{C}^i_{m-n}\right)~~~\forall m\neq n,\quad \left[C^i_{-m},\tilde{C}^-_{-m}\right] = 0,\\
&\left[\tilde{C}^i_m,\tilde{C}^-_{-n}\right] = \frac{m}{2B^+_0}\left(C^i_{n-m}+3\tilde{C}^i_{m-n}\right)~~~\forall m\neq n,\quad \textcolor[rgb]{1.00,0.00,1.00}{\left[\tilde{C}^i_m,\tilde{C}^-_{-m}\right] = \frac{m}{B^+_0}B^i_0 = \frac{m\sqrt{2c'}}{B^+_0}p^i}.
\end{split}
\end{equation}

\subsection*{Calculation of  $\left[L^{i-},S^{j-}\right]+\left[S^{i-},L^{j-}\right]$}

We have from \eqref{xC}
\begin{equation}
\begin{split}
\left[x^ip^-,S^{j-}\right] =& -i\sum_{m > 0}\frac{1}m\left[x^i,\left(C^j_{-m}C^-_m - \tilde{C}^-_{-m}\tilde{C}^j_m\right)\right]p^- -i\sum_{m < 0}\frac{1}m\left[x^i,\left(C^-_mC^j_{-m} - \tilde{C}^j_m\tilde{C}^-_{-m}\right)\right]p^-\\
=& \frac{\sqrt{2c'}}{B^+_0}\sum_{m\neq 0}\frac{1}m\left(C^j_{-m}C^i_m - \tilde{C}^i_{-m}\tilde{C}^j_m\right)p^-,
\end{split}
\end{equation}
and
\begin{equation}
\begin{split}
\left[x^-p^i,S^{j-}\right] =& -i\sum_{m > 0}\frac{1}m\left[x^-,\left(C^j_{-m}C^-_m - \tilde{C}^-_{-m}\tilde{C}^j_m\right)\right]p^i -i\sum_{m < 0}\frac{1}m\left[x^-,\left(C^-_mC^j_{-m} - \tilde{C}^j_m\tilde{C}^-_{-m}\right)\right]p^i\\
=& \frac{\sqrt{2c'}}{B^+_0}\sum_{m> 0}\frac{1}m \left(C^j_{-m}C^-_m - \tilde{C}^-_{-m}\tilde{C}^j_m\right)p^i +\frac{\sqrt{2c'}}{B^+_0}\sum_{m < 0}\frac{1}m \left(C^-_mC^j_{-m} - \tilde{C}^j_m\tilde{C}^-_{-m}\right)p^i,
\end{split}
\end{equation}

So
\begin{equation}
\begin{split}
\left[L^{i-},S^{j-}\right] =& \frac{\sqrt{2c'}}{B^+_0}\sum_{m\neq 0}\frac{1}m\left(C^j_{-m}C^i_m - \tilde{C}^i_{-m}\tilde{C}^j_m\right)p^-\\
&-\frac{\sqrt{2c'}}{B^+_0}\sum_{m> 0}\frac{1}m \left(C^j_{-m}C^-_m - \tilde{C}^-_{-m}\tilde{C}^j_m\right)p^i\\
&-\frac{\sqrt{2c'}}{B^+_0}\sum_{m < 0}\frac{1}m \left(C^-_mC^j_{-m} - \tilde{C}^j_m\tilde{C}^-_{-m}\right)p^i.
\end{split} 
\end{equation}

Similarly
\begin{equation}
\begin{split}
\left[S^{i-},L^{j-}\right] =& -\frac{\sqrt{2c'}}{B^+_0}\sum_{m\neq 0}\frac{1}m\left(C^i_{-m}C^j_m - \tilde{C}^j_{-m}\tilde{C}^i_m\right)p^-\\
&\frac{\sqrt{2c'}}{B^+_0}\sum_{m> 0}\frac{1}m \left(C^i_{-m}C^-_m - \tilde{C}^-_{-m}\tilde{C}^i_m\right)p^j\\
&+\frac{\sqrt{2c'}}{B^+_0}\sum_{m < 0}\frac{1}m \left(C^-_mC^i_{-m} - \tilde{C}^i_m\tilde{C}^-_{-m}\right)p^j.
\end{split} 
\end{equation}

Therefore
\begin{equation}
\begin{split}
\left[L^{i-},S^{j-}\right]+\left[S^{i-},L^{j-}\right]
&= -\frac{2\sqrt{2c'}}{B^+_0}\sum_{m\neq 0}\frac{1}m\left(C^i_{-m}C^j_m - \tilde{C}^i_m\tilde{C}^j_{-m}\right)p^-\\
&-\frac{\sqrt{2c'}}{B^+_0}\sum_{m > 0}\frac{1}m \left(\left(C^j_{-m}C^-_m - \tilde{C}^-_{-m}\tilde{C}^j_m\right)p^i- \left(C^i_{-m}C^-_m - \tilde{C}^-_{-m}\tilde{C}^i_m\right)p^j\right)\\
&-\frac{\sqrt{2c'}}{B^+_0}\sum_{m < 0}\frac{1}m \left(\left(C^-_mC^j_{-m} - \tilde{C}^j_m\tilde{C}^-_{-m}\right)p^i - \left(C^-_mC^i_{-m} - \tilde{C}^i_m\tilde{C}^-_{-m}\right)p^j\right).
\end{split}
\end{equation}

\subsection*{Calculation of  $\left[S^{i-},S^{j-}\right]$}
The sector structure here is easier than that in the case of the flipped vacuum.

\begin{equation}\label{sectintroosc}
\left[S^{i-},S^{j-}\right] = \mathcal{S}_1 + \mathcal{S}_2 + \mathcal{S}_3 + \mathcal{S}_4.
\end{equation}
We now proceed to calculate these sectors one by one.

\subsubsection*{Sector $1$}
\begin{equation}\label{S1defosc}
\begin{split}
-\mathcal{S}_1 = \sum_{\substack{m > 0\\n > 0}}\frac{1}{mn}\bigg(\left[C^i_{-m}C^-_m,C^j_{-n}C^-_n\right]-\left[C^i_{-m}C^-_m,\tilde{C}^-_{-n}\tilde{C}^j_n\right] -\left[\tilde{C}^-_{-m}\tilde{C}^i_m,C^j_{-n}C^-_n\right]+\left[\tilde{C}^-_{-m}\tilde{C}^i_m,\tilde{C}^-_{-n}\tilde{C}^j_n\right]\bigg).
\end{split}
\end{equation}
We evaluate \eqref{S1defosc} as following. First, let's collect the terms on the RHS resulting from $[-,-]$ type commutators, namely $\left[C^-_m,C^-_n\right]$, $\left[C^-_m,\tilde{C}^-_{-n}\right]$, $\left[\tilde{C}^-_{-m},C^-_n\right]$ and $\left[\tilde{C}^-_{-m},\tilde{C}^-_{-n}\right]$ 
\begin{equation}\label{S1T1osc}
\begin{split}
&\sum_{\substack{m > 0\\n > 0}}\left(\frac{1}{n}-\frac{1}{m}\right)\frac{1}{2B^+_0}C^i_{-m}C^j_{-n}\left(3C^-_{m+n}+\tilde{C}^-_{-m-n}\right)\\
&+\sum_{\substack{m > 0\\n > 0}}\left(\frac{1}{n}-\frac{1}{m}\right)\frac{1}{2B^+_0}C^i_{-m}\left(C^-_{m+n}-\tilde{C}^-_{-m-n}\right)\tilde{C}^j_n\\
&+\sum_{\substack{m > 0\\n > 0}}\left(\frac{1}{n}-\frac{1}{m}\right)\frac{1}{2B^+_0}C^j_{-n}\left(C^-_{m+n}-\tilde{C}^-_{-m-n}\right)\tilde{C}^i_m\\
&-\sum_{\substack{m > 0\\n > 0}}\left(\frac{1}{n}-\frac{1}{m}\right)\frac{1}{2B^+_0}\left(C^-_{m+n}+3\tilde{C}^-_{-m-n}\right)\tilde{C}^i_m\tilde{C}^j_n~.\\
\end{split}
\end{equation}
Note the operator ordering that the individual product terms have. Corresponding products in all other terms in this sector will be brought in this order. For example, the $[i,-]$ type terms give
\begin{equation}\label{S1T2osc}
\begin{split}
&\sum_{\substack{n > 0\\m > -n\\m\neq 0}}\frac{1}{2nB^+_0}	\bigg(-\left(3C^i_{-m}+\tilde{C}^i_m\right)C^j_{-n}C^-_{m+n}-\left(C^i_{-m}-\tilde{C}^i_m\right)\tilde{C}^j_nC^-_{m+n}\\
&-\tilde{C}^-_{-m-n}\left(C^i_{-m}-\tilde{C}^i_m\right)C^j_{-n}+\tilde{C}^-_{-m-n}\left(C^i_{-m}+3\tilde{C}^i_m\right)\tilde{C}^j_n\bigg) \textcolor[rgb]{1.00,0.00,1.00}{-\frac{\sqrt{2c'}}{B^+_0}\sum_{m > 0}\frac{1}m\left(C^j_{-m}C^-_m - \tilde{C}^-_{-m}\tilde{C}^j_m\right)p^i}\\
=&\sum_{\substack{n > 0\\m > -n\\m\neq 0}}\frac{1}{2nB^+_0}	\bigg(-3C^i_{-m}C^j_{-n}C^-_{m+n}-C^j_{-n}C^-_{m+n}\tilde{C}^i_m-C^j_{-n}\left[\tilde{C}^i_m,C^-_{m+n}\right] -C^i_{-m}C^-_{m+n}\tilde{C}^j_n  -C^i_{-m}\left[\tilde{C}^j_n,C^-_{m+n}\right] \\ &
+C^-_{m+n}\tilde{C}^i_m\tilde{C}^j_n+\left[\tilde{C}^i_m\tilde{C}^j_n,C^-_{m+n}\right] -C^i_{-m}C^j_{-n}\tilde{C}^-_{-m-n}+\left[C^i_{-m}C^j_{-n},\tilde{C}^-_{-m-n}\right] +C^j_{-n}\tilde{C}^-_{-m-n}\tilde{C}^i_m
-\left[C^j_{-n},\tilde{C}^-_{-m-n}\right]\tilde{C}^i_m \\ & +C^i_{-m}\tilde{C}^-_{-m-n}\tilde{C}^j_n-\left[C^i_{-m},\tilde{C}^-_{-m-n}\right]\tilde{C}^j_n+3\tilde{C}^-_{-m-n}\tilde{C}^i_m\tilde{C}^j_n\bigg)-\textcolor[rgb]{1.00,0.00,1.00}{\frac{\sqrt{2c'}}{B^+_0}\sum_{m > 0}\frac{1}m\left(C^j_{-m}C^-_m - \tilde{C}^-_{-m}\tilde{C}^j_m\right)p^i}\\
=& \sum_{\substack{n > 0\\m > -n\\m\neq 0}}\frac{1}{2nB^+_0}\bigg(-C^i_{-m}C^j_{-n}\left(3C^-_{m+n}+\tilde{C}^-_{-m-n}\right)-C^i_{-m}\left(C^-_{m+n}-\tilde{C}^-_{-m-n}\right)\tilde{C}^j_n\\
&-C^j_{-n}\left(C^-_{m+n}-\tilde{C}^-_{-m-n}\right)\tilde{C}^i_m+\left(C^-_{m+n}+3\tilde{C}^-_{-m-n}\right)\tilde{C}^i_m\tilde{C}^j_n\bigg)\textcolor[rgb]{1.00,0.00,1.00}{-\frac{\sqrt{2c'}}{B^+_0}\sum_{m > 0}\frac{1}m\left(C^j_{-m}C^-_m - \tilde{C}^-_{-m}\tilde{C}^j_m\right)p^i}~.
\end{split}
\end{equation}
In the first step, the dummy index $m$ is shifted so that the index structure is brought in the form of \eqref{S1T1osc}, on similar lines as discussed in the induced vacuum case. The \textcolor[rgb]{1.00,0.00,1.00}{magenta} terms involving $p^i$ come from the $m=n$ contribution (\textcolor[rgb]{1.00,0.00,1.00}{magenta} equations) from \eqref{CiCcomm}. In the second step, the terms are brought into the order same as that in \eqref{S1T1osc}, resulting into commutator terms. While going from the second step to the last step, notice that the commutator terms cancel out.

And similarly the $[-,j]$ type terms give
\begin{equation}\label{S1T3osc}
\begin{split}
&\sum_{\substack{m > 0\\n > -m\\n\neq 0}}\frac{1}{2mB^+_0}	\bigg(C^i_{-m}\left(3C^j_{-n}+\tilde{C}^j_n\right)C^-_{m+n}+C^i_{-m}\tilde{C}^-_{-m-n}\left(C^j_{-n}-\tilde{C}^j_n\right)\\
&+\left(C^j_{-n}-\tilde{C}^j_n\right)C^-_{m+n}\tilde{C}^i_m-\tilde{C}^-_{-m-n}\tilde{C}^i_m\left(C^j_{-n}+3\tilde{C}^j_n\right)\bigg)\\
&\textcolor[rgb]{1.00,0.00,1.00}{+\frac{\sqrt{2c'}}{B^+_0}\sum_{n > 0}\frac{1}n\left(C^i_{-n}C^-_n - \tilde{C}^-_{-n}\tilde{C}^i_n\right)p^j}\\
=&\sum_{\substack{m > 0\\n > -m\\n\neq 0}}\frac{1}{2mB^+_0}	\bigg(3C^i_{-m}C^j_{-n}C^-_{m+n}+C^i_{-m}C^-_{m+n}\tilde{C}^j_n+C^i_{-m}\left[\tilde{C}^j_n,C^-_{m+n}\right]\\
&+C^i_{-m}C^j_{-n}\tilde{C}^-_{-m-n}-C^i_{-m}\left[C^j_{-n},\tilde{C}^-_{-m-n}\right]-C^i_{-m}\tilde{C}^-_{-m-n}\tilde{C}^j_n\\
&+C^j_{-n}C^-_{m+n}\tilde{C}^i_m-C^-_{m+n}\tilde{C}^i_m\tilde{C}^j_n-\left[\tilde{C}^j_n,C^-_{m+n}\right]\tilde{C}^i_m\\
&-C^j_{-n}\tilde{C}^-_{-m-n}\tilde{C}^i_m+\left[C^j_{-n},\tilde{C}^-_{-m-n}\right]\tilde{C}^i_m-3\tilde{C}^-_{-m-n}\tilde{C}^i_m\tilde{C}^j_n\bigg)\textcolor[rgb]{1.00,0.00,1.00}{+\frac{\sqrt{2c'}}{B^+_0}\sum_{n > 0}\frac{1}n\left(C^i_{-n}C^-_n - \tilde{C}^-_{-n}\tilde{C}^i_n\right)p^j}\\
=& \sum_{\substack{m > 0\\n > -m\\n\neq 0}}\frac{1}{2mB^+_0}\bigg(C^i_{-m}C^j_{-n}\left(3C^-_{m+n}+\tilde{C}^-_{-m-n}\right)+C^i_{-m}\left(C^-_{m+n}-\tilde{C}^-_{-m-n}\right)\tilde{C}^j_n\\
&+C^j_{-n}\left(C^-_{m+n}-\tilde{C}^-_{-m-n}\right)\tilde{C}^i_m-\left(C^-_{m+n}+3\tilde{C}^-_{-m-n}\right)\tilde{C}^i_m\tilde{C}^j_n\bigg)\textcolor[rgb]{1.00,0.00,1.00}{+\frac{\sqrt{2c'}}{B^+_0}\sum_{m > 0}\frac{1}m\left(C^i_{-m}C^-_m - \tilde{C}^-_{-m}\tilde{C}^i_m\right)p^j}~.
\end{split}
\end{equation}
Here, too, the commutators arising from rearranging the terms cancel out.\\
Combining \eqref{S1T1osc}, \eqref{S1T2osc} and \eqref{S1T3osc}, we get
\begin{equation}\label{S1totosc}
\begin{split}
\mathcal{S}_1 =& \sum_{n > 0}\frac{1}{2nB^+_0}\sum_{m=-n+1}^{-1}\bigg(C^i_{-m}C^j_{-n}\left(3C^-_{m+n}+\tilde{C}^-_{-m-n}\right)+C^i_{-m}\left(C^-_{m+n}-\tilde{C}^-_{-m-n}\right)\tilde{C}^j_n\\
&+C^j_{-n}\left(C^-_{m+n}-\tilde{C}^-_{-m-n}\right)\tilde{C}^i_m-\left(C^-_{m+n}+3\tilde{C}^-_{-m-n}\right)\tilde{C}^i_m\tilde{C}^j_n\bigg)\\
&-\sum_{m > 0}\frac{1}{2mB^+_0}\sum_{n=-m+1}^{-1}\bigg(C^i_{-m}C^j_{-n}\left(3C^-_{m+n}+\tilde{C}^-_{-m-n}\right)+C^i_{-m}\left(C^-_{m+n}-\tilde{C}^-_{-m-n}\right)\tilde{C}^j_n\\
&+C^j_{-n}\left(C^-_{m+n}-\tilde{C}^-_{-m-n}\right)\tilde{C}^i_m-\left(C^-_{m+n}+3\tilde{C}^-_{-m-n}\right)\tilde{C}^i_m\tilde{C}^j_n\bigg)\\
&\textcolor[rgb]{1.00,0.00,1.00}{+\frac{\sqrt{2c'}}{B^+_0}\sum_{m > 0}\frac{1}m\left(\left(C^j_{-m}C^-_m - \tilde{C}^-_{-m}\tilde{C}^j_m\right)p^i-\left(C^i_{-m}C^-_m - \tilde{C}^-_{-m}\tilde{C}^i_m\right)p^j\right)}
\end{split}
\end{equation}

\begin{equation}\label{S1totosc1}
\begin{split}
\mathcal{S}_1=& \sum_{n > 0}\frac{1}{2nB^+_0}\sum_{m=\textcolor[rgb]{0.00,1.00,0.00}{-n}}^{-1}\bigg(C^i_{-m}C^j_{-n}\left(3C^-_{m+n}+\tilde{C}^-_{-m-n}\right)+C^i_{-m}\left(C^-_{m+n}-\tilde{C}^-_{-m-n}\right)\tilde{C}^j_n\\
&+C^j_{-n}\left(C^-_{m+n}-\tilde{C}^-_{-m-n}\right)\tilde{C}^i_m-\left(C^-_{m+n}+3\tilde{C}^-_{-m-n}\right)\tilde{C}^i_m\tilde{C}^j_n\bigg)\\
&\textcolor[rgb]{0.50,0.00,0.00}{-\sum_{m > 0}\frac{1}{2mB^+_0}\sum_{n=\textcolor[rgb]{0.00,1.00,0.00}{-m}}^{-1}\bigg(C^i_{-m}C^j_{-n}\left(3C^-_{m+n}+\tilde{C}^-_{-m-n}\right)+C^i_{-m}\left(C^-_{m+n}-\tilde{C}^-_{-m-n}\right)\tilde{C}^j_n}\\
&\textcolor[rgb]{0.50,0.00,0.00}{+C^j_{-n}\left(C^-_{m+n}-\tilde{C}^-_{-m-n}\right)\tilde{C}^i_m-\left(C^-_{m+n}+3\tilde{C}^-_{-m-n}\right)\tilde{C}^i_m\tilde{C}^j_n\bigg)}\\
&\textcolor[rgb]{0.00,0.70,0.00}{-\sum_{n > 0}\frac{1}{2nB^+_0}\bigg(C^i_{n}C^j_{-n}\left(2B^-_0+A^-_0\right)+C^i_{n}A^-_0\tilde{C}^j_n} \textcolor[rgb]{0.00,0.70,0.00}{+C^j_{-n}A^-_0\tilde{C}^i_{-n}-\left(2B^-_0-A^-_0\right)\tilde{C}^i_{-n}\tilde{C}^j_n\bigg)}\\
&\textcolor[rgb]{0.00,0.70,0.00}{+\sum_{m > 0}\frac{1}{2mB^+_0}\bigg(C^i_{-m}C^j_{m}\left(2B^-_0+A^-_0\right)+C^i_{-m}A^-_0\tilde{C}^j_{-m}}
\textcolor[rgb]{0.00,0.70,0.00}{+C^j_{m}A^-_0\tilde{C}^i_m-\left(2B^-_0-A^-_0\right)\tilde{C}^i_m\tilde{C}^j_{-m}\bigg)}\\
&\textcolor[rgb]{1.00,0.00,1.00}{+\frac{\sqrt{2c'}}{B^+_0}\sum_{m > 0}\frac{1}m\left(\left(C^j_{-m}C^-_m - \tilde{C}^-_{-m}\tilde{C}^j_m\right)p^i-\left(C^i_{-m}C^-_m - \tilde{C}^-_{-m}\tilde{C}^i_m\right)p^j\right)}.
\end{split}
\end{equation}
In the last step, terms involving $C^-_0$ and $\tilde{C}^-_0$ (in \textcolor[rgb]{0.00,0.70,0.00}{green}) are added and subtracted, and written in terms of $A^-_0$ and $B^-_0$.

\subsubsection*{Sector 2}
\begin{equation}\label{S2defosc}
\begin{split}
-\mathcal{S}_2 = \sum_{\substack{m > 0\\n < 0}}\frac{1}{mn}\bigg(\left[C^i_{-m}C^-_m,C^-_nC^j_{-n}\right]-\left[C^i_{-m}C^-_m,\tilde{C}^j_n\tilde{C}^-_{-n}\right] -\left[\tilde{C}^-_{-m}\tilde{C}^i_m,C^-_nC^j_{-n}\right]+\left[\tilde{C}^-_{-m}\tilde{C}^i_m,\tilde{C}^j_n\tilde{C}^-_{-n}\right]\bigg).
\end{split}
\end{equation}
This sector allows $m = -n$ since $m$ and $n$ run over opposite signed integers. So the $[-,-]$ part gives central terms (in \textcolor[rgb]{1.00,0.00,0.00}{red}).
\begin{equation}\label{S2T1osc}
\begin{split}
&\sum_{\substack{m > 0\\n < 0}}\left(\frac{1}{n}-\frac{1}{m}\right)\frac{1}{2B^+_0}C^i_{-m}\left(3C^-_{m+n}+\tilde{C}^-_{-m-n}\right)C^j_{-n}\\
&+\sum_{\substack{m > 0\\n < 0}}\left(\frac{1}{n}-\frac{1}{m}\right)\frac{1}{2B^+_0}C^i_{-m}\tilde{C}^j_n\left(C^-_{m+n}-\tilde{C}^-_{-m-n}\right)\\
&+\sum_{\substack{m > 0\\n < 0}}\left(\frac{1}{n}-\frac{1}{m}\right)\frac{1}{2B^+_0}\left(C^-_{m+n}-\tilde{C}^-_{-m-n}\right)\tilde{C}^i_mC^j_{-n}\\
&-\sum_{\substack{m > 0\\n < 0}}\left(\frac{1}{n}-\frac{1}{m}\right)\frac{1}{2B^+_0}\tilde{C}^j_n\left(C^-_{m+n}+3\tilde{C}^-_{-m-n}\right)\tilde{C}^i_m\\
&\textcolor[rgb]{1.00,0.00,0.00}{-\frac{(D-2)}{6\left(B^+_0\right)^2}\sum_{m>0}\left(m-\frac{1}m\right)\left(C^i_{-m}C^j_m-\tilde{C}^i_m\tilde{C}^j_{-m}\right)}.
\end{split}
\end{equation}
However, since $m=n$ is not allowed, the terms involving $p^i$ and $p^j$ won't appear unlike sector $1$.\\
The $[i,-]$ type terms give
\begin{equation}\label{S2T2osc}
\begin{split}
&\sum_{\substack{n < 0\\m > -n}}\frac{1}{2nB^+_0}	\bigg(-\left(3C^i_{-m}+\tilde{C}^i_m\right)C^j_{-n}C^-_{m+n}-\left(C^i_{-m}-\tilde{C}^i_m\right)\tilde{C}^j_nC^-_{m+n}\\
&-\tilde{C}^-_{-m-n}\left(C^i_{-m}-\tilde{C}^i_m\right)C^j_{-n}+\tilde{C}^-_{-m-n}\left(C^i_{-m}+3\tilde{C}^i_m\right)\tilde{C}^j_n\bigg)\\
=&\sum_{\substack{n < 0\\m > -n}}\frac{1}{2nB^+_0}	\bigg(-3C^i_{-m}C^-_{m+n}C^j_{-n}-3C^i_{-m}\left[C^j_{-n},C^-_{m+n}\right]-C^-_{m+n}\tilde{C}^i_mC^j_{-n}\\
&-\left[\tilde{C}^i_mC^j_{-n},C^-_{m+n}\right]-C^i_{-m}\tilde{C}^j_nC^-_{m+n}+\tilde{C}^j_nC^-_{m+n}\tilde{C}^i_m+\tilde{C}^j_n\left[\tilde{C}^i_m,C^-_{m+n}\right]\\
&-C^i_{-m}\tilde{C}^-_{-m-n}C^j_{-n}+\left[C^i_{-m},\tilde{C}^-_{-m-n}\right]C^j_{-n}+\tilde{C}^-_{-m-n}\tilde{C}^i_mC^j_{-n}\\
&+C^i_{-m}\tilde{C}^j_n\tilde{C}^-_{-m-n}-\left[C^i_{-m}\tilde{C}^j_n,\tilde{C}^-_{-m-n}\right]+3\tilde{C}^j_n\tilde{C}^-_{-m-n}\tilde{C}^i_m-3\left[\tilde{C}^j_n,\tilde{C}^-_{-m-n}\right]\tilde{C}^i_m\bigg)\\
=& \sum_{\substack{n < 0\\m > -n}}\frac{1}{2nB^+_0}\bigg(-C^i_{-m}\left(3C^-_{m+n}+\tilde{C}^-_{-m-n}\right)C^j_{-n}-C^i_{-m}\tilde{C}^j_n\left(C^-_{m+n}-\tilde{C}^-_{-m-n}\right)\\
&-\left(C^-_{m+n}-\tilde{C}^-_{-m-n}\right)\tilde{C}^i_mC^j_{-n}+\tilde{C}^j_n\left(C^-_{m+n}+3\tilde{C}^-_{-m-n}\right)\tilde{C}^i_m\bigg)\\
&\textcolor[rgb]{0.00,0.00,1.00}{+\frac{2}{\left(B^+_0\right)^2}\sum_{\substack{n < 0\\m > -n}}\left(C^i_{-m}C^j_m - \tilde{C}^i_m\tilde{C}^j_{-m}\right)}.
\end{split}
\end{equation}
Notice that another difference from sector $1$ is that in this sector the commutators arising from rearrangements don't cancel out, resulting into the \textcolor[rgb]{0.00,0.00,1.00}{blue} terms. The explicit calculation of the commutator contribution is as follows.
\begin{equation}
\begin{split}
&\sum_{\substack{n < 0\\m > -n}}\frac{1}{2nB^+_0}	\bigg(-3C^i_{-m}\left[C^j_{-n},C^-_{m+n}\right]-\left[\tilde{C}^i_mC^j_{-n},C^-_{m+n}\right]+\tilde{C}^j_n\left[\tilde{C}^i_m,C^-_{m+n}\right]\\
&+\left[C^i_{-m},\tilde{C}^-_{-m-n}\right]C^j_{-n}-\left[C^i_{-m}\tilde{C}^j_n,\tilde{C}^-_{-m-n}\right]-3\left[\tilde{C}^j_n,\tilde{C}^-_{-m-n}\right]\tilde{C}^i_m\bigg)\\
=&\sum_{\substack{n < 0\\m > -n}}\frac{1}{4\left(B^+_0\right)^2}	\bigg(3C^i_{-m}\left(3C^j_m+\tilde{C}^j_{-m}\right)+\tilde{C}^i_m\left(3C^j_m+\tilde{C}^j_{-m}\right)\\
&-\frac{m}n\left(C^i_n-\tilde{C}^i_{-n}\right)C^j_{-n}+\frac{m}n\left(C^i_n-\tilde{C}^i_{-n}\right)\tilde{C}^j_n+\frac{m}n\left(C^i_n-\tilde{C}^i_{-n}\right)C^j_{-n}\\
&-C^i_{-m}\left(C^j_m+3\tilde{C}^j_{-m}\right)-\frac{m}n\left(C^i_n-\tilde{C}^i_{-n}\right)\tilde{C}^j_n-3\tilde{C}^i_m\left(C^j_m+3\tilde{C}^j_{-m}\right)\bigg)\\
=&\textcolor[rgb]{0.00,0.00,1.00}{\frac{2}{\left(B^+_0\right)^2}\sum_{\substack{n < 0\\m > -n}}\left(C^i_{-m}C^j_m - \tilde{C}^i_m\tilde{C}^j_{-m}\right)}.
\end{split}
\end{equation}
And the $[-,j]$ type terms give
\begin{equation}\label{S2T3osc}
\begin{split}
&\sum_{\substack{m > 0\\n < -m}}\frac{1}{2mB^+_0}	\bigg(C^i_{-m}C^-_{m+n}\left(3C^j_{-n}+\tilde{C}^j_n\right)+C^i_{-m}\left(C^j_{-n}-\tilde{C}^j_n\right)\tilde{C}^-_{-m-n}\\
&+C^-_{m+n}\tilde{C}^i_m\left(C^j_{-n}-\tilde{C}^j_n\right)-\left(C^j_{-n}+3\tilde{C}^j_n\right)\tilde{C}^-_{-m-n}\tilde{C}^i_m\bigg)\\
=&\sum_{\substack{m > 0\\n < -m}}\frac{1}{2mB^+_0}	\bigg(3C^i_{-m}C^-_{m+n}C^j_{-n}+C^i_{-m}\tilde{C}^j_nC^-_{m+n}-C^i_{-m}\left[\tilde{C}^j_n,C^-_{m+n}\right]\\
&+C^i_{-m}\tilde{C}^-_{-m-n}C^j_{-n}+C^i_{-m}\left[C^j_{-n},\tilde{C}^-_{-m-n}\right]-C^i_{-m}\tilde{C}^j_n\tilde{C}^-_{-m-n}\\
&+C^-_{m+n}\tilde{C}^i_mC^j_{-n}-\tilde{C}^j_nC^-_{m+n}\tilde{C}^i_m+\left[\tilde{C}^j_n,C^-_{m+n}\right]\tilde{C}^i_m\\
&-\tilde{C}^-_{-m-n}\tilde{C}^i_mC^j_{-n}-\left[C^j_{-n},\tilde{C}^-_{-m-n}\right]\tilde{C}^i_m-3\tilde{C}^j_n\tilde{C}^-_{-m-n}\tilde{C}^i_m\bigg)\\
=& \sum_{\substack{m > 0\\n < -m}}\frac{1}{2mB^+_0}\bigg(C^i_{-m}\left(3C^-_{m+n}+\tilde{C}^-_{-m-n}\right)C^j_{-n}+C^i_{-m}\tilde{C}^j_n\left(C^-_{m+n}-\tilde{C}^-_{-m-n}\right)\\
&+\left(C^-_{m+n}-\tilde{C}^-_{-m-n}\right)\tilde{C}^i_mC^j_{-n}-\tilde{C}^j_n\left(C^-_{m+n}+3\tilde{C}^-_{-m-n}\right)\tilde{C}^i_m\bigg).
\end{split}
\end{equation}
The final result for this sector
\begin{equation}\label{S2totosc}
\begin{split}
\mathcal{S}_2
=& -\sum_{n < 0}\frac{1}{2nB^+_0}\sum_{m=1}^{-n}\bigg(C^i_{-m}\left(3C^-_{m+n}+\tilde{C}^-_{-m-n}\right)C^j_{-n}+C^i_{-m}\tilde{C}^j_n\left(C^-_{m+n}-\tilde{C}^-_{-m-n}\right)\\
&+\left(C^-_{m+n}-\tilde{C}^-_{-m-n}\right)\tilde{C}^i_mC^j_{-n}-\tilde{C}^j_n\left(C^-_{m+n}+3\tilde{C}^-_{-m-n}\right)\tilde{C}^i_m\bigg)\\
&\textcolor[rgb]{0.50,0.00,0.00}{+\sum_{m > 0}\frac{1}{2mB^+_0}\sum_{n=-m}^{-1}\bigg(C^i_{-m}\left(3C^-_{m+n}+\tilde{C}^-_{-m-n}\right)C^j_{-n}+C^i_{-m}\tilde{C}^j_n\left(C^-_{m+n}-\tilde{C}^-_{-m-n}\right)}\\
&\textcolor[rgb]{0.50,0.00,0.00}{+\left(C^-_{m+n}-\tilde{C}^-_{-m-n}\right)\tilde{C}^i_mC^j_{-n}-\tilde{C}^j_n\left(C^-_{m+n}+3\tilde{C}^-_{-m-n}\right)\tilde{C}^i_m\bigg)}\\
&\textcolor[rgb]{1.00,0.00,0.00}{+\frac{(D-2)}{6\left(B^+_0\right)^2}\sum_{m>0}\left(m-\frac{1}m\right)\left(C^i_{-m}C^j_m-\tilde{C}^i_m\tilde{C}^j_{-m}\right)} \textcolor[rgb]{0.00,0.00,1.00}{-\frac{2}{\left(B^+_0\right)^2}\sum_{m>0}(m-1)\left(C^i_{-m}C^j_m-\tilde{C}^i_m\tilde{C}^j_{-m}\right)}.
\end{split}
\end{equation}
For the blue term we have used the change of order of summation
\begin{equation}\label{sumordmillionaire}
\begin{split}
\sum_{\substack{n<0\\m>-n}}\left(C^i_{-m}C^j_m-\tilde{C}^i_m\tilde{C}^j_{-m}\right) &= \sum_{m>0}\sum_{n=-m+1}^{-1}\left(C^i_{-m}C^j_m-\tilde{C}^i_m\tilde{C}^j_{-m}\right)\\
&= \sum_{m>0}(m-1)\left(C^i_{-m}C^j_m-\tilde{C}^i_m\tilde{C}^j_{-m}\right).
\end{split}
\end{equation}

\subsubsection*{Sectors 3 and 4}
The calculations for sectors $3$ and $4$ proceed similarly to those for sectors $2$ and $1$ respectively.
\begin{equation}\label{S3totosc}
\begin{split}
\mathcal{S}_3 &=-\sum_{\substack{m < 0\\n > 0}}\frac{1}{mn}\bigg(\left[C^-_mC^i_{-m},C^j_{-n}C^-_n\right]-\left[C^-_mC^i_{-m},\tilde{C}^-_{-n}\tilde{C}^j_n\right] -\left[\tilde{C}^i_m\tilde{C}^-_{-m},C^j_{-n}C^-_n\right]+\left[\tilde{C}^i_m\tilde{C}^-_{-m},\tilde{C}^-_{-n}\tilde{C}^j_n\right]\bigg)\\
=& -\sum_{n > 0}\frac{1}{2nB^+_0}\sum_{m=-n}^{-1}\bigg(C^j_{-n}\left(3C^-_{m+n}+\tilde{C}^-_{-m-n}\right)C^i_{-m}+\left(C^-_{m+n}-\tilde{C}^-_{-m-n}\right)C^i_{-m}\tilde{C}^j_n\\
&+\tilde{C}^i_mC^j_{-n}\left(C^-_{m+n}-\tilde{C}^-_{-m-n}\right)-\tilde{C}^i_m\left(C^-_{m+n}+3\tilde{C}^-_{-m-n}\right)\tilde{C}^j_n\bigg)\\
&+\sum_{m < 0}\frac{1}{2mB^+_0}\sum_{n=1}^{-m}\bigg(C^j_{-n}\left(3C^-_{m+n}+\tilde{C}^-_{-m-n}\right)C^i_{-m}+\left(C^-_{m+n}-\tilde{C}^-_{-m-n}\right)C^i_{-m}\tilde{C}^j_n\\
&+\tilde{C}^i_mC^j_{-n}\left(C^-_{m+n}-\tilde{C}^-_{-m-n}\right)-\tilde{C}^i_m\left(C^-_{m+n}+3\tilde{C}^-_{-m-n}\right)\tilde{C}^j_n\bigg)\\
&\textcolor[rgb]{1.00,0.00,0.00}{+\frac{(D-2)}{6\left(B^+_0\right)^2}\sum_{m<0}\left(m-\frac{1}m\right)\left(C^i_{-m}C^j_m-\tilde{C}^i_m\tilde{C}^j_{-m}\right)} \textcolor[rgb]{0.00,0.00,1.00}{-\frac{2}{\left(B^+_0\right)^2}\sum_{m<0}(m-1)\left(C^i_{-m}C^j_m-\tilde{C}^i_m\tilde{C}^j_{-m}\right)}.
\end{split}
\end{equation}
and
\begin{equation}\label{S4totosc}
\begin{split}
\mathcal{S}_4 
=& -\sum_{\substack{m < 0\\n < 0}}\frac{1}{mn}\bigg(\left[C^-_mC^i_{-m},C^-_nC^j_{-n}\right]-\left[C^-_mC^i_{-m},\tilde{C}^j_n\tilde{C}^-_{-n}\right] 
-\left[\tilde{C}^i_m\tilde{C}^-_{-m},C^-_nC^j_{-n}\right]+\left[\tilde{C}^i_m\tilde{C}^-_{-m},\tilde{C}^j_n\tilde{C}^-_{-n}\right]\bigg)\\
=& \sum_{n < 0}\frac{1}{2nB^+_0}\sum_{m=1}^{-n}\bigg(\left(3C^-_{m+n}+\tilde{C}^-_{-m-n}\right)C^i_{-m}C^j_{-n}+\tilde{C}^j_n\left(C^-_{m+n}-\tilde{C}^-_{-m-n}\right)C^i_{-m}\\
&+\tilde{C}^i_m\left(C^-_{m+n}-\tilde{C}^-_{-m-n}\right)C^j_{-n}-\tilde{C}^i_m\tilde{C}^j_n\left(C^-_{m+n}+3\tilde{C}^-_{-m-n}\right)\bigg)\\
&-\sum_{m < 0}\frac{1}{2mB^+_0}\sum_{n=1}^{-m}\bigg(\left(3C^-_{m+n}+\tilde{C}^-_{-m-n}\right)C^i_{-m}C^j_{-n}+\tilde{C}^j_n\left(C^-_{m+n}-\tilde{C}^-_{-m-n}\right)C^i_{-m}\\
&+\tilde{C}^i_m\left(C^-_{m+n}-\tilde{C}^-_{-m-n}\right)C^j_{-n}-\tilde{C}^i_m\tilde{C}^j_n\left(C^-_{m+n}+3\tilde{C}^-_{-m-n}\right)\bigg)\\
&\textcolor[rgb]{0.00,0.70,0.00}{-\sum_{n < 0}\frac{1}{2nB^+_0}\bigg(\left(2B^-_0+A^-_0\right)C^i_{n}C^j_{-n}+\tilde{C}^j_nA^-_0C^i_{n}}\textcolor[rgb]{0.00,0.70,0.00}{+\tilde{C}^i_{-n}A^-_0C^j_{-n}-\tilde{C}^i_{-n}\tilde{C}^j_n\left(2B^-_0-A^-_0\right)\bigg)}\\
&\textcolor[rgb]{0.00,0.70,0.00}{+\sum_{m < 0}\frac{1}{2mB^+_0}\bigg(\left(2B^-_0+A^-_0\right)C^i_{-m}C^j_{m}+\tilde{C}^j_{-m}A^-_0C^i_{-m}}\textcolor[rgb]{0.00,0.70,0.00}{+\tilde{C}^i_mA^-_0C^j_{m}-\tilde{C}^i_m\tilde{C}^j_{-m}\left(2B^-_0-A^-_0\right)\bigg)}\\
&\textcolor[rgb]{1.00,0.00,1.00}{-\frac{\sqrt{2c'}}{2\pi B^+_0}\sum_{m < 0}\frac{1}m\left(\left(C^j_{-m}C^-_m - \tilde{C}^-_{-m}\tilde{C}^j_m\right)p^i-\left(C^i_{-m}C^-_m - \tilde{C}^-_{-m}\tilde{C}^i_m\right)p^j\right)}.
\end{split}
\end{equation}
Now it's time to combine the four sectors. The black (and \textcolor[rgb]{0.50,0.00,0.00}{brown}) terms in \eqref{S1totosc}, \eqref{S2totosc}, \eqref{S3totosc} and \eqref{S4totosc} can be organised in equal and opposite pairs, but these pairs have dissimilar ordering. For example, consider the \textcolor[rgb]{0.50,0.00,0.00}{brown} terms in \eqref{S1totosc} and \eqref{S2totosc}.
\begin{equation}\label{combexosc}
\begin{split}
&-\sum_{m > 0}\frac{1}{2mB^+_0}\sum_{n=-m}^{-1}\bigg(C^i_{-m}C^j_{-n}\left(3C^-_{m+n}+\tilde{C}^-_{-m-n}\right)+C^i_{-m}\left(C^-_{m+n}-\tilde{C}^-_{-m-n}\right)\tilde{C}^j_n\\
&+C^j_{-n}\left(C^-_{m+n}-\tilde{C}^-_{-m-n}\right)\tilde{C}^i_m-\left(C^-_{m+n}+3\tilde{C}^-_{-m-n}\right)\tilde{C}^i_m\tilde{C}^j_n\bigg)\\
&+\sum_{m > 0}\frac{1}{2mB^+_0}\sum_{n=-m}^{-1}\bigg(C^i_{-m}\left(3C^-_{m+n}+\tilde{C}^-_{-m-n}\right)C^j_{-n}+C^i_{-m}\tilde{C}^j_n\left(C^-_{m+n}-\tilde{C}^-_{-m-n}\right)\\
&+\left(C^-_{m+n}-\tilde{C}^-_{-m-n}\right)\tilde{C}^i_mC^j_{-n}-\tilde{C}^j_n\left(C^-_{m+n}+3\tilde{C}^-_{-m-n}\right)\tilde{C}^i_m\bigg)\\
=&\sum_{m > 0}\frac{1}{2mB^+_0}\sum_{n=-m}^{-1} \bigg(-C^i_{-m}\left[C^j_{-n},\left(3C^-_{m+n}+\tilde{C}^-_{-m-n}\right)\right]+C^i_{-m}\left[\tilde{C}^j_n,\left(C^-_{m+n}-\tilde{C}^-_{-m-n}\right)\right]\\
&\left[C^j_{-n},\left(C^-_{m+n}-\tilde{C}^-_{-m-n}\right)\right]\tilde{C}^i_m-\left[\tilde{C}^j_n,\left(C^-_{m+n}+3\tilde{C}^-_{-m-n}\right)\right]\tilde{C}^i_m\bigg)\\
=&\sum_{m > 0}\frac{2}{m\left(B^+_0\right)^2}\left(\sum_{n=-m}^{-1}n\right)\left(C^i_{-m}C^j_m-\tilde{C}^i_m\tilde{C}^j_{-m}\right) 
=\textcolor[rgb]{0.50,0.00,0.00}{-\sum_{m > 0}\frac{(m+1)}{\left(B^+_0\right)^2}\left(C^i_{-m}C^j_m-\tilde{C}^i_m\tilde{C}^j_{-m}\right)}.
\end{split}
\end{equation}
Similarly combining other black terms, we get one more $m > 0$ and two $m<0$ summations. So the total sum reads
\begin{equation}
\begin{split}
\left[S^{i-},S^{j-}\right]
=& \textcolor[rgb]{0.50,0.00,0.00}{-\sum_{m \neq 0}\frac{2(m+1)}{\left(B^+_0\right)^2}\left(C^i_{-m}C^j_m-\tilde{C}^i_m\tilde{C}^j_{-m}\right)}
\textcolor[rgb]{0.00,0.00,1.00}{-\sum_{m \neq 0}\frac{2(m-1)}{\left(B^+_0\right)^2}\left(C^i_{-m}C^j_m-\tilde{C}^i_m\tilde{C}^j_{-m}\right)}\\
&\textcolor[rgb]{1.00,0.00,0.00}{+\frac{(D-2)}{6\left(B^+_0\right)^2}\sum_{m<0}\left(m-\frac{1}m\right)\left(C^i_{-m}C^j_m-\tilde{C}^i_m\tilde{C}^j_{-m}\right)}\\
&\textcolor[rgb]{0.00,0.70,0.00}{+\frac{1}{2B^+_0}\sum_{m \neq 0}\frac{1}m \left(C^i_{-m}C^j_m+C^i_{-m}\tilde{C}^j_{-m}+\tilde{C}^i_mC^j_m+\tilde{C}^i_m\tilde{C}^j_{-m}\right)A^-_0}\\
&\textcolor[rgb]{0.00,0.70,0.00}{+\frac{1}{2B^+_0}\sum_{m \neq 0}\frac{1}m A^-_0\left(C^i_{-m}C^j_m+C^i_{-m}\tilde{C}^j_{-m}+\tilde{C}^i_mC^j_m+\tilde{C}^i_m\tilde{C}^j_{-m}\right)}\\
&\textcolor[rgb]{0.00,0.70,0.00}{+\frac{1}{B^+_0}\sum_{m > 0}\frac{1}m\left(C^i_{-m}C^j_mB^-_0 - B^-_0\tilde{C}^i_m\tilde{C}^j_{-m}+B^-_0C^i_{-m}C^j_m - \tilde{C}^i_m\tilde{C}^j_{-m}B^-_0\right)}\\
&\textcolor[rgb]{1.00,0.00,1.00}{+\frac{\sqrt{2c'}}{B^+_0}\sum_{m > 0}\frac{1}m \left(\left(C^j_{-m}C^-_m - \tilde{C}^-_{-m}\tilde{C}^j_m\right)p^i- \left(C^i_{-m}C^-_m - \tilde{C}^-_{-m}\tilde{C}^i_m\right)p^j\right)}\\
&\textcolor[rgb]{1.00,0.00,1.00}{+\frac{\sqrt{2c'}}{B^+_0}\sum_{m < 0}\frac{1}m \left(\left(C^-_mC^j_{-m} - \tilde{C}^j_m\tilde{C}^-_{-m}\right)p^i - \left(C^-_mC^i_{-m} - \tilde{C}^i_m\tilde{C}^-_{-m}\right)p^j\right)}\\
\end{split}
\end{equation}
This gives us the result for this part of the calculation as,
\begin{equation}
\begin{split}
\left[S^{i-},S^{j-}\right]
=&\frac{1}{\left(B^+_0\right)^2}\left(\frac{(D-2)}{6}-4\right)\sum_{m<0}m\left(C^i_{-m}C^j_m-\tilde{C}^i_m\tilde{C}^j_{-m}\right) -\frac{(D-2)}{6\left(B^+_0\right)^2}\sum_{m\neq 0}\frac{1}m\left(C^i_{-m}C^j_m-\tilde{C}^i_m\tilde{C}^j_{-m}\right)\\
&+\frac{1}{B^+_0}\sum_{m \neq 0}\frac{1}m B^i_{-m}B^j_mA^-_0+\frac{2}{B^+_0}\sum_{m \neq 0}\frac{1}m\left(C^i_{-m}C^j_m - \tilde{C}^i_m\tilde{C}^j_{-m}\right)B^-_0\\
&+\frac{\sqrt{2c'}}{B^+_0}\sum_{m > 0}\frac{1}m \left(\left(C^j_{-m}C^-_m - \tilde{C}^-_{-m}\tilde{C}^j_m\right)p^i- \left(C^i_{-m}C^-_m - \tilde{C}^-_{-m}\tilde{C}^i_m\right)p^j\right)\\
&+\frac{\sqrt{2c'}}{B^+_0}\sum_{m < 0}\frac{1}m \left(\left(C^-_mC^j_{-m} - \tilde{C}^j_m\tilde{C}^-_{-m}\right)p^i - \left(C^-_mC^i_{-m} - \tilde{C}^i_m\tilde{C}^-_{-m}\right)p^j\right).
\end{split}
\end{equation}

\pagebreak

\newpage

\bibliographystyle{JHEP}
\bibliography{ref}

\providecommand{\href}[2]{#2}\begingroup\raggedright\begin{thebibliography}{10}

\bibitem{Schild:1976vq}
A.~Schild, \emph{{Classical Null Strings}},
  \href{https://doi.org/10.1103/PhysRevD.16.1722}{\emph{Phys. Rev.} {\bfseries
  D16} (1977) 1722}.

\bibitem{Pisarski:1982cn}
R.~D. Pisarski and O.~Alvarez, \emph{{Strings at Finite Temperature and
  Deconfinement}}, \href{https://doi.org/10.1103/PhysRevD.26.3735}{\emph{Phys.
  Rev.} {\bfseries D26} (1982) 3735}.

\bibitem{Olesen:1985ej}
P.~Olesen, \emph{{Strings, Tachyons and Deconfinement}},
  \href{https://doi.org/10.1016/0370-2693(85)90010-3}{\emph{Phys. Lett.}
  {\bfseries 160B} (1985) 408}.

\bibitem{Atick:1988si}
J.~J. Atick and E.~Witten, \emph{{The Hagedorn Transition and the Number of
  Degrees of Freedom of String Theory}},
  \href{https://doi.org/10.1016/0550-3213(88)90151-4}{\emph{Nucl. Phys.}
  {\bfseries B310} (1988) 291}.

\bibitem{Gross:1987kza}
D.~J. Gross and P.~F. Mende, \emph{{The High-Energy Behavior of String
  Scattering Amplitudes}},
  \href{https://doi.org/10.1016/0370-2693(87)90355-8}{\emph{Phys. Lett.}
  {\bfseries B197} (1987) 129}.

\bibitem{Gross:1987ar}
D.~J. Gross and P.~F. Mende, \emph{{String Theory Beyond the Planck Scale}},
  \href{https://doi.org/10.1016/0550-3213(88)90390-2}{\emph{Nucl. Phys.}
  {\bfseries B303} (1988) 407}.

\bibitem{Gross:1988ue}
D.~J. Gross, \emph{{High-Energy Symmetries of String Theory}},
  \href{https://doi.org/10.1103/PhysRevLett.60.1229}{\emph{Phys. Rev. Lett.}
  {\bfseries 60} (1988) 1229}.

\bibitem{Isberg:1993av}
J.~Isberg, U.~Lindstrom, B.~Sundborg and G.~Theodoridis, \emph{{Classical and
  quantized tensionless strings}},
  \href{https://doi.org/10.1016/0550-3213(94)90056-6}{\emph{Nucl. Phys.}
  {\bfseries B411} (1994) 122}
  [\href{https://arxiv.org/abs/hep-th/9307108}{{\ttfamily hep-th/9307108}}].

\bibitem{Bagchi:2013bga}
A.~Bagchi, \emph{{Tensionless Strings and Galilean Conformal Algebra}},
  \href{https://doi.org/10.1007/JHEP05(2013)141}{\emph{JHEP} {\bfseries 05}
  (2013) 141} [\href{https://arxiv.org/abs/1303.0291}{{\ttfamily 1303.0291}}].

\bibitem{Bagchi:2015nca}
A.~Bagchi, S.~Chakrabortty and P.~Parekh, \emph{{Tensionless Strings from
  Worldsheet Symmetries}},
  \href{https://doi.org/10.1007/JHEP01(2016)158}{\emph{JHEP} {\bfseries 01}
  (2016) 158} [\href{https://arxiv.org/abs/1507.04361}{{\ttfamily
  1507.04361}}].

\bibitem{Bondi:1962px}
H.~Bondi, M.~G.~J. van~der Burg and A.~W.~K. Metzner, \emph{{Gravitational
  waves in general relativity. 7. Waves from axisymmetric isolated systems}},
  \href{https://doi.org/10.1098/rspa.1962.0161}{\emph{Proc. Roy. Soc. Lond.}
  {\bfseries A269} (1962) 21}.

\bibitem{Sachs:1962wk}
R.~K. Sachs, \emph{{Gravitational waves in general relativity. 8. Waves in
  asymptotically flat space-times}},
  \href{https://doi.org/10.1098/rspa.1962.0206}{\emph{Proc. Roy. Soc. Lond.}
  {\bfseries A270} (1962) 103}.

\bibitem{Barnich:2006av}
G.~Barnich and G.~Compere, \emph{{Classical central extension for asymptotic
  symmetries at null infinity in three spacetime dimensions}},
  \href{https://doi.org/10.1088/0264-9381/24/5/F01,
  10.1088/0264-9381/24/11/C01}{\emph{Class. Quant. Grav.} {\bfseries 24} (2007)
  F15} [\href{https://arxiv.org/abs/gr-qc/0610130}{{\ttfamily gr-qc/0610130}}].

\bibitem{Bagchi:2010eg}
A.~Bagchi, \emph{{Correspondence between Asymptotically Flat Spacetimes and
  Nonrelativistic Conformal Field Theories}},
  \href{https://doi.org/10.1103/PhysRevLett.105.171601}{\emph{Phys. Rev. Lett.}
  {\bfseries 105} (2010) 171601}
  [\href{https://arxiv.org/abs/1006.3354}{{\ttfamily 1006.3354}}].

\bibitem{Bagchi:2012xr}
A.~Bagchi, S.~Detournay, R.~Fareghbal and J.~Sim{\'o}n, \emph{{Holography of 3D
  Flat Cosmological Horizons}},
  \href{https://doi.org/10.1103/PhysRevLett.110.141302}{\emph{Phys. Rev. Lett.}
  {\bfseries 110} (2013) 141302}
  [\href{https://arxiv.org/abs/1208.4372}{{\ttfamily 1208.4372}}].

\bibitem{Bagchi:2012yk}
A.~Bagchi, S.~Detournay and D.~Grumiller, \emph{{Flat-Space Chiral Gravity}},
  \href{https://doi.org/10.1103/PhysRevLett.109.151301}{\emph{Phys. Rev. Lett.}
  {\bfseries 109} (2012) 151301}
  [\href{https://arxiv.org/abs/1208.1658}{{\ttfamily 1208.1658}}].

\bibitem{Bagchi:2014iea}
A.~Bagchi, R.~Basu, D.~Grumiller and M.~Riegler, \emph{{Entanglement entropy in
  Galilean conformal field theories and flat holography}},
  \href{https://doi.org/10.1103/PhysRevLett.114.111602}{\emph{Phys. Rev. Lett.}
  {\bfseries 114} (2015) 111602}
  [\href{https://arxiv.org/abs/1410.4089}{{\ttfamily 1410.4089}}].

\bibitem{Bagchi:2019cay}
A.~Bagchi, A.~Banerjee and P.~Parekh, \emph{{Tensionless Path from Closed to
  Open Strings}},
  \href{https://doi.org/10.1103/PhysRevLett.123.111601}{\emph{Phys. Rev. Lett.}
  {\bfseries 123} (2019) 111601}
  [\href{https://arxiv.org/abs/1905.11732}{{\ttfamily 1905.11732}}].

\bibitem{Bagchi:2020fpr}
A.~Bagchi, A.~Banerjee, S.~Chakrabortty, S.~Dutta and P.~Parekh, \emph{{A tale
  of three \textemdash{} tensionless strings and vacuum structure}},
  \href{https://doi.org/10.1007/JHEP04(2020)061}{\emph{JHEP} {\bfseries 04}
  (2020) 061} [\href{https://arxiv.org/abs/2001.00354}{{\ttfamily
  2001.00354}}].

\bibitem{Bagchi:2020ats}
A.~Bagchi, A.~Banerjee and S.~Chakrabortty, \emph{{Rindler Physics on the
  String Worldsheet}},
  \href{https://doi.org/10.1103/PhysRevLett.126.031601}{\emph{Phys. Rev. Lett.}
  {\bfseries 126} (2021) 031601}
  [\href{https://arxiv.org/abs/2009.01408}{{\ttfamily 2009.01408}}].

\bibitem{Bagchi:2016yyf}
A.~Bagchi, S.~Chakrabortty and P.~Parekh, \emph{{Tensionless Superstrings: View
  from the Worldsheet}},
  \href{https://doi.org/10.1007/JHEP10(2016)113}{\emph{JHEP} {\bfseries 10}
  (2016) 113} [\href{https://arxiv.org/abs/1606.09628}{{\ttfamily
  1606.09628}}].

\bibitem{Bagchi:2017cte}
A.~Bagchi, A.~Banerjee, S.~Chakrabortty and P.~Parekh, \emph{{Inhomogeneous
  Tensionless Superstrings}},
  \href{https://doi.org/10.1007/JHEP02(2018)065}{\emph{JHEP} {\bfseries 02}
  (2018) 065} [\href{https://arxiv.org/abs/1710.03482}{{\ttfamily
  1710.03482}}].

\bibitem{Bagchi:2018wsn}
A.~Bagchi, A.~Banerjee, S.~Chakrabortty and P.~Parekh, \emph{{Exotic Origins of
  Tensionless Superstrings}},
  \href{https://doi.org/10.1016/j.physletb.2019.135139}{\emph{Phys. Lett. B}
  (2018) } [\href{https://arxiv.org/abs/1811.10877}{{\ttfamily 1811.10877}}].

\bibitem{Karlhede:1986wb}
A.~Karlhede and U.~Lindstrom, \emph{{The Classical Bosonic String in the Zero
  Tension Limit}},
  \href{https://doi.org/10.1088/0264-9381/3/4/002}{\emph{Class. Quant. Grav.}
  {\bfseries 3} (1986) L73}.

\bibitem{Lizzi:1986nv}
F.~Lizzi, B.~Rai, G.~Sparano and A.~Srivastava, \emph{{Quantization of the Null
  String and Absence of Critical Dimensions}},
  \href{https://doi.org/10.1016/0370-2693(86)90101-2}{\emph{Phys. Lett.}
  {\bfseries B182} (1986) 326}.

\bibitem{Gamboa:1989zc}
J.~Gamboa, C.~Ramirez and M.~Ruiz-Altaba, \emph{{NULL SPINNING STRINGS}},
  \href{https://doi.org/10.1016/0550-3213(90)90627-P}{\emph{Nucl. Phys.}
  {\bfseries B338} (1990) 143}.

\bibitem{Gamboa:1989px}
J.~Gamboa, C.~Ramirez and M.~Ruiz-Altaba, \emph{{QUANTUM NULL (SUPER)STRINGS}},
  \href{https://doi.org/10.1016/0370-2693(89)90578-9}{\emph{Phys. Lett.}
  {\bfseries B225} (1989) 335}.

\bibitem{Gustafsson:1994kr}
H.~Gustafsson, U.~Lindstrom, P.~Saltsidis, B.~Sundborg and R.~van Unge,
  \emph{{Hamiltonian BRST quantization of the conformal string}},
  \href{https://doi.org/10.1016/0550-3213(95)00051-S}{\emph{Nucl. Phys.}
  {\bfseries B440} (1995) 495}
  [\href{https://arxiv.org/abs/hep-th/9410143}{{\ttfamily hep-th/9410143}}].

\bibitem{Lindstrom:2003mg}
U.~Lindstrom and M.~Zabzine, \emph{{Tensionless strings, WZW models at critical
  level and massless higher spin fields}},
  \href{https://doi.org/10.1016/j.physletb.2004.01.035}{\emph{Phys. Lett.}
  {\bfseries B584} (2004) 178}
  [\href{https://arxiv.org/abs/hep-th/0305098}{{\ttfamily hep-th/0305098}}].

\bibitem{Casali:2016atr}
E.~Casali and P.~Tourkine, \emph{{On the null origin of the ambitwistor
  string}}, \href{https://doi.org/10.1007/JHEP11(2016)036}{\emph{JHEP}
  {\bfseries 11} (2016) 036}
  [\href{https://arxiv.org/abs/1606.05636}{{\ttfamily 1606.05636}}].

\bibitem{Harmark:2017rpg}
T.~Harmark, J.~Hartong and N.~A. Obers, \emph{{Nonrelativistic strings and
  limits of the AdS/CFT correspondence}},
  \href{https://doi.org/10.1103/PhysRevD.96.086019}{\emph{Phys. Rev.}
  {\bfseries D96} (2017) 086019}
  [\href{https://arxiv.org/abs/1705.03535}{{\ttfamily 1705.03535}}].

\bibitem{Harmark:2018cdl}
T.~Harmark, J.~Hartong, L.~Menculini, N.~A. Obers and Z.~Yan, \emph{{Strings
  with Non-Relativistic Conformal Symmetry and Limits of the AdS/CFT
  Correspondence}}, \href{https://doi.org/10.1007/JHEP11(2018)190}{\emph{JHEP}
  {\bfseries 11} (2018) 190}
  [\href{https://arxiv.org/abs/1810.05560}{{\ttfamily 1810.05560}}].

\bibitem{Harmark:2019upf}
T.~Harmark, J.~Hartong, L.~Menculini, N.~A. Obers and G.~Oling, \emph{{Relating
  non-relativistic string theories}},
  \href{https://doi.org/10.1007/JHEP11(2019)071}{\emph{JHEP} {\bfseries 11}
  (2019) 071} [\href{https://arxiv.org/abs/1907.01663}{{\ttfamily
  1907.01663}}].

\bibitem{Bergshoeff:2019pij}
E.~A. Bergshoeff, J.~Gomis, J.~Rosseel, C.~\c{S}im\c{s}ek and Z.~Yan,
  \emph{{String Theory and String Newton-Cartan Geometry}},
  \href{https://doi.org/10.1088/1751-8121/ab56e9}{\emph{J. Phys. A} {\bfseries
  53} (2020) 014001} [\href{https://arxiv.org/abs/1907.10668}{{\ttfamily
  1907.10668}}].

\bibitem{Gomis:2019zyu}
J.~Gomis, J.~Oh and Z.~Yan, \emph{{Nonrelativistic String Theory in Background
  Fields}}, \href{https://doi.org/10.1007/JHEP10(2019)101}{\emph{JHEP}
  {\bfseries 10} (2019) 101}
  [\href{https://arxiv.org/abs/1905.07315}{{\ttfamily 1905.07315}}].

\bibitem{Gallegos:2019icg}
A.~D. Gallegos, U.~G\"ursoy and N.~Zinnato, \emph{{Torsional Newton Cartan
  gravity from non-relativistic strings}},
  \href{https://doi.org/10.1007/JHEP09(2020)172}{\emph{JHEP} {\bfseries 09}
  (2020) 172} [\href{https://arxiv.org/abs/1906.01607}{{\ttfamily
  1906.01607}}].

\bibitem{Green:1987sp}
M.~B. Green, J.~H. Schwarz and E.~Witten, \emph{{SUPERSTRING THEORY. VOL. 1:
  INTRODUCTION}}, Cambridge Monographs on Mathematical Physics. 1988.

\bibitem{Gamboa:1991nj}
J.~Gamboa, \emph{{BRST quantization of null spinning p-branes}},
  \href{https://doi.org/10.1142/S0217732392000495}{\emph{Mod. Phys. Lett. A}
  {\bfseries 7} (1992) 533}.

\bibitem{Bozhilov:1997xq}
P.~Bozhilov, \emph{{Tensionless branes and the null string critical
  dimension}}, \href{https://doi.org/10.1142/S0217732398002734}{\emph{Mod.
  Phys. Lett. A} {\bfseries 13} (1998) 2571}
  [\href{https://arxiv.org/abs/hep-th/9711136}{{\ttfamily hep-th/9711136}}].

\bibitem{Duval:2014uoa}
C.~Duval, G.~W. Gibbons, P.~A. Horvathy and P.~M. Zhang, \emph{{Carroll versus
  Newton and Galilei: two dual non-Einsteinian concepts of time}},
  \href{https://doi.org/10.1088/0264-9381/31/8/085016}{\emph{Class. Quant.
  Grav.} {\bfseries 31} (2014) 085016}
  [\href{https://arxiv.org/abs/1402.0657}{{\ttfamily 1402.0657}}].

\bibitem{Cariglia:2016oft}
M.~Cariglia, C.~Duval, G.~W. Gibbons and P.~A. Horvathy, \emph{{Eisenhart lifts
  and symmetries of time-dependent systems}},
  \href{https://doi.org/10.1016/j.aop.2016.07.033}{\emph{Annals Phys.}
  {\bfseries 373} (2016) 631}
  [\href{https://arxiv.org/abs/1605.01932}{{\ttfamily 1605.01932}}].

\end{thebibliography}\endgroup

\end{document}